\documentclass[fleqn,usenatbib]{mnras}

\usepackage{newtxtext,newtxmath}
\usepackage{longtable}

\usepackage[T1]{fontenc}

\DeclareRobustCommand{\VAN}[3]{#2}
\let\VANthebibliography\thebibliography
\def\thebibliography{\DeclareRobustCommand{\VAN}[3]{##3}\VANthebibliography}

\usepackage{graphicx}	
\usepackage{amsmath}	
\usepackage{gensymb}
\usepackage{tablefootnote}
\usepackage{url} 
\usepackage{booktabs}
\defcitealias{Elias2021}{EC21}
\defcitealias{Elias2024}{EC24}

\newcommand{\tm}[1]{\color{purple} #1 \color{black}}

\title[X-ray properties in the Lockman-SpReSO]{The Lockman-SpReSO Project: A Deep X-ray Spectral View of a FIR-selected AGN Population}

\author[Elías-Chávez et al.]{Mauricio Elías-Chávez,$^{1}$\thanks{E-mail: melias@astro.unam.mx}
Takamitsu Miyaji,$^{1}$
Martín Herrera-Endoqui,$^{2}$
Irene Cruz-González,$^{2}$
\newauthor
Yair Krongold,$^{2}$
Héctor Hernández-Toledo,$^{2}$
Mirjana Povic,$^{3,4,5}$
Bereket Assefa,$^{3,6}$
\newauthor
Mauro González-Otero,$^{4,7}$
Castalia Alenka Negrete,$^{2}$
Miguel Cerviño,$^{8}$
J. Antonio de Diego,$^{2}$
\newauthor
Miguel Sánchez-Portal,$^{9}$
Erika Benítez,$^{2}$
Jordi Cepa,$^{10,11}$
José Antonio Vázquez Mata,$^{2}$
\newauthor
Monica I. Rodriguez,$^{12}$
Emilio J. Alfaro,$^{4}$
J. Jesús González,$^{2}$
Carmen P. Padilla-Torres,$^{10,11,7,14}$
\newauthor
J. Ignacio González-Serrano,$^{13,7}$
Vladimir Ávila-Reese,$^{2}$
Ángel Bongiovanni,$^{12,7}$
A. M. Perez Garcia$^{7,8}$
\\
$^{1}$ Universidad Nacional Autónoma de México. Instituto de Astronomía. A.P. 106, 22800. Ensenada, B.C., México\\
$^{2}$ Universidad Nacional Autónoma de México. Instituto de Astronomía. A.P. 70-264, 04510. Ciudad de México, México\\
$^{3}$ Space Science and Geospatial Institute (SSGI), Entoto Observatory and Research Center (EORC),\\ Astronomy and Astrophysics Research Division, PO Box 33679, Addis Ababa, Ethiopia\\
$^{4}$ Instituto de Astrofísica de Andalucía, CSIC, Glorieta de la Astronomía, s/n 18008, Granada, Spain\\
$^{5}$ Physics Department, Mbarara University of Science and Technology (MUST), Mbarara, Uganda\\
$^{6}$ Debre Berhan University, Debre Berhan, Ethiopia\\
$^{7}$ Asociación Astrofísica para la Promoción de la Investigación, Instrumentación y su Desarrollo, ASPID, 38205 La Laguna, Tenerife, Spain\\
$^{8}$ Centro de Astrobiología (CAB), CSIC-INTA, Camino Bajo del Castillo s/n, 28692, Villanueva de la Cañada, Madrid, Spain\\
$^{9}$ Institut de Radioastronomie Millimétrique, Avenida Divina Pastora, 7, Local 20, E 18012 Granada, Spain\\
$^{10}$ Instituto de Astrofísica de Canarias, E-38205 La Laguna, Tenerife, Spain\\
$^{11}$ Departamento de Astrofísica, Universidad de La Laguna (ULL), E-38205 La Laguna, Tenerife, Spain\\
$^{12}$ Institut de Radioastronomie Millimétrique (IRAM), Av. Divina Pastora 7, Núcleo Central 18012, Granada, Spain\\
$^{13}$ Instituto de Física de Cantabria (CSIC-Universidad de Cantabria), 39005 Santander, Spain\\
$^{14}$ Fundación Galileo Galilei-INAF Rambla José Ana Fernández Pérez, 7, 38712 Breña Baja, Tenerife, Spain\\
}

\date{Accepted XXX. Received YYY; in original form ZZZ}

\pubyear{2025}

\begin{document}
\label{firstpage}
\pagerange{\pageref{firstpage}--\pageref{lastpage}}
\maketitle

\begin{abstract}
We present a detailed X-ray spectral analysis of the active galactic nucleus (AGN) population in the Lockman-SpReSO project, a multiwavelength campaign of FIR sources in the Lockman Hole field. Using deep \textit{XMM-Newton} observations cross-matched with FIR-selected galaxies, we characterize 94 AGNs based on their X-ray spectral properties. The sample is distributed over a large redshift range of $z = 0.07-5$, and reaches a flux limit of $5 \times 10^{-16}\, \mathrm{erg\, s^{-1}\, cm^{-2}}$ in the $0.3-10\, \mathrm{keV}$ band. We model the X-ray spectra using absorbed power-law, reflection, and soft excess components to estimate intrinsic column densities ($N_{\mathrm{H}}$), rest-frame $2-10\, \mathrm{keV}$ luminosities, and iron line equivalent widths $(EW)$. Additionally, we included an advanced model fitting for Compton-thick AGNs (CT-AGNs) using the XCLUMPY model. Our results show an increase in the fraction of obscured AGNs toward higher redshifts, including the identification of one strong and two borderline CT-AGN candidates with $N_{\mathrm{H}} \gtrsim  10^{24}\, \mathrm{cm^{-2}}$. Soft excess emission is detected in 10 AGNs, with an average blackbody temperature of $0.12 \pm 0.02\, \mathrm{keV}$. We also detect the X-ray Baldwin effect in both obscured and unobscured populations, and we found a strong correlation between the torus angular width $(\sigma_\mathrm{tor})$, dust covering factor $(f_{\mathrm{cov}})$, and X-ray luminosity, described by $f_{\mathrm{cov}} = (-0.1 \pm 0.01) \times \log(L_{2-10\, \mathrm{keV}}/(10^{44}\, \mathrm{erg\, s^{-1}})) + (0.65 \pm 0.01)$,  supporting the receding torus scenario. While consistent with deep X-ray surveys, the FIR selection favours the identification of dusty star-forming host galaxies and heavily obscured AGNs.

\end{abstract}



\tm{}
\section{Introduction}

X-ray surveys trace some of the most energetic phenomena in the universe, and offer one of the most reliable methods for identifying active galactic nuclei (AGNs) powered by accreting supermassive black holes (SMBHs) across a broad range of redshifts. Studying AGNs is essential for understanding the cosmic co-evolution of SMBHs and galaxies, characterizing the evolution of accretion power in the universe. In particular, the detection of obscured AGNs is found to be crucial to estimate its contribution to the cosmic X-ray background \citep[CXRB,][]{Gilli2007}.

The intrinsic X-ray emission detected in AGNs originates in their innermost regions, primarily through the Comptonization of optical-UV disc seed photons from the accretion disc by a hot corona of highly energetic electrons  \citep{George1991, Haardt1991, Matt1997}. Depending on the level of obscuration, quantified by the hydrogen column density $(N_{\mathrm{H}})$, a significant fraction of the observed X-ray radiation may result from reflection or scattering interactions between the primary X-ray emission and the surrounding nuclear material. In X-ray astronomy, AGNs are typically classified based on their column density absorption, such as unobscured AGNs with  $N_{\mathrm{H}} < 10^{22}\, \mathrm{cm}^{-2}$, while obscured AGNs have $N_{\mathrm{H}} \geq 10^{22}\, \mathrm{cm}^{-2}$. Obscured AGNs are further divided into Compton-thin by moderate to high obscuration, i.e. $N_{\mathrm{H}} = 10^{22} - 10^{24}\, \mathrm{cm}^{-2}$ and Compton-thick AGNs (CT-AGNs), which are extremely obscured sources with  $N_{\mathrm{H}} \gtrsim 10^{24}\, \mathrm{cm}^{-2}$, where the reflected or scattered component dominates the observed luminosity \citep[e.g.,][]{Comastri2004,Hickox2018}.    

Deep X-ray surveys performed with \textit{Chandra} and \textit{XMM-Newton} observations have revealed an evolutionary trend in the obscured AGN population, showing that the fraction of obscured AGNs increases with redshift. This trend has been confirmed by multiple deep-field studies \citep[e.g.,][]{Iwasawa2012, Liu2017, Vito2018, Iwasawa2020, Peca2023}. For instance, \citet{Peca2023} reported a significant increase in the obscured AGN fraction ($f_{\mathrm{obs}}$), reaching $64 \pm 12$ per cent at $z > 3$. At redshifts $z > 3-4$, this effect becomes even more pronounced, with CT-AGNs expected to constitute an important fraction of the AGN population \citep{Vito2018,Iwasawa2020}. 

Supporting evidence for these results was previously provided in \citet{Iwasawa2012},  a study focused on heavily obscured AGNs at $ z > 1.7$  based on \textit{XMM-Newton} CDFS. They identified several highly absorbed sources, including two  CT-AGN candidates. Later, \citet{Liu2017} analysed data from the 7 Ms \textit{Chandra} Deep Field-South Survey (CDFS) and found that the increase in the obscured AGN fraction with redshift seemed to be independent of luminosity.  Their results are well described by the relation $f_{\mathrm{obs}} \approx 0.42(1 + z)^{0.6}$. 

These results are consistent with population synthesis models, which require an increasing fraction of obscured AGNs with redshift to reproduce the observed X-ray background spectrum \citep{Gilli2007, Ananna2019}. For example, \citet{Gilli2007} showed that a redshift-dependent rise in the obscured AGN fraction is necessary to match the spectral shape and intensity of the cosmic X-ray background. More recently, \citet{Ananna2019} refined these models using multi-wavelength data, providing further support for the scenario in which heavily obscured AGNs (including CT-AGNs), become increasingly dominant in the early universe. The agreement between deep X-ray observations and theoretical models suggests that AGN obscuration is closely linked to the gas-rich environments of early galaxies, where black hole growth and galaxy evolution proceed via a co-evolutionary process. 

In the present work, we identified and characterized the X-ray AGN population with \textit{Herschel} far-infrared (FIR) counterparts detected in the Lockman Hole (LH) field as part of the Lockman-SpReSO project. Since X-ray features can be hidden in CT-AGNs, the reprocessed FIR emission arising from hot dust grains in the torus plays a key role in detecting and characterizing these objects. For instance, \citet{KilerciEser2020} found that CT-AGNs typically exhibit a strong FIR emission. Moreover, thanks to the availability of infrared (IR) emission, in addition to identifying highly obscured AGNs, the SED fits in our sample can be used to constrain torus properties and estimate parameters such as the torus angular width ($\sigma_\mathrm{tor}$) and inclination ($i$) \citep[e.g.,][]{Yang2020}.

The paper is organized as follows: In Section \ref{LH}, we describe the Lockman-SpReSO project, the deep \textit{XMM-Newton} observations, and the statistical cross-correlation method used to associate X-ray and FIR sources. Section \ref{Xspectral} presents a detailed overview of our X-ray spectral analysis, including model fitting procedures and AGN classification criteria. The spectral fitting results are presented in Section~\ref{SpecResuts}, where we discuss the detection of soft excess emission, the properties of the obscured AGN population, and the identification of Compton-thick sources. In Section \ref{torus}, we investigate the properties of the dusty torus, focusing on the obscured fraction and the dust covering factor $f_{\mathrm{cov}}$ as a function of the X-ray luminosity. In Section \ref{Discussion}, we discuss the detection of CT-AGNs, the role of X-ray luminosity in shaping the torus structure, the star-forming properties of the FIR-selected AGN population, and implications for the receding torus model. Finally, Section \ref{summary} summarizes our main results.

Throughout this work, we adopted the cosmological parameters $H_0 = 70\, \mathrm{km}\, \mathrm{s}^{-1}\, \mathrm{Mpc}^{-1}$, $\Omega_\mathrm{m} = 0.3$, and $\Omega_\Lambda = 0.7$.

\section{The \textit{XMM-Newton} survey in the Lockman Hole}
\label{LH}
\subsection{The Lockman-SpReSO project}

The Lockman-SpReSO project is an optical spectroscopic survey performed with the Gran Telescopio Canarias (GTC) using the OSIRIS Multi-Object Spectroscopy (MOS) mode. It targets FIR sources observed by the \textit{Herschel} space telescope in the LH field. This is an optimal region in the sky with the lowest Galactic line-of-sight column density \citep[$N_{\mathrm{H}} = 5.7 \times 10^{19} \, \mathrm{cm}^{-2}$,][]{Lockman1986,Brunner2008}. 

The Lockman-SpReSO survey is composed of 1,144 objects observed within the central $24 \times 24\, \mathrm{arcmin}^2$ of the LH field. The parent far-infrared sample is primarily defined from the Herschel/PACS Evolutionary Probe (PEP) survey \citep{Lutz2011}, based on deep PACS observations at 70, 100, and 160\,$\mu$m \citep{Poglitsch2010}. These PACS detections were associated with optical counterparts identified in broad-band OSIRIS imaging, reaching a limiting magnitude of $R_\mathrm{C}\sim24.5$.

The optical and FIR catalogs were cross-matched using a $1.5''$ radius, selected as the best compromise considering the astrometric uncertainties of the PEP catalog \citep{Gonzalez2023}. This procedure resulted in 956 sources with reliable PACS counterparts. Additional FIR photometry was obtained from the HerMES survey \citep{Oliver2010} using the Herschel/SPIRE instrument \citep{Griffin2010} at 250, 350, and 500\,$\mu$m. The SPIRE data provide complementary long-wavelength coverage but were not used to define the initial source selection.

Finally, a compilation of multiwavelength photometric and spectroscopic information made by   \citet{Fotopoulou2012} from the Lockman Hole field was incorporated, providing UV to near-infrared coverage (GALEX to \textit{Spitzer}/IRAC). For a detailed description of the target selection, spectroscopic observations, and catalog preparation, see \citet{Gonzalez2023}.

It is worth mentioning that several studies have recently been published from the Lockman-SpReSO project. For example, \citet{Gonzales2024a} examined galactic outflows and inflows in FIR-selected galaxies using Mg \textsc{ii}, Mg \textsc{i}, and Fe \textsc{ii} spectral lines, while \citet{Gonzales2024b} analysed the general properties of FIR-selected star-forming galaxies (SFGs) in the catalog. Analyses conducted by \citet{Negrete2025} and Assefa et al. (in preparation) have provided detailed characterizations of Type-I and Type-II AGNs, respectively. Moreover, Herrera-Endoqui et al. (in preparation) are conducting an extensive study of the host galaxy properties of AGNs, including a further study of those residing in the Compton-thick regime.  These complementary studies further highlight the rich multi-wavelength nature of the Lockman-SpReSO dataset.

\subsection{The \textit{XMM-Newton} observations}
\label{previous}
The \textit{XMM-Newton} observations of the Lockman Hole (XMM-LH) consist of 18 pointings taken from April 2000 to December 2002, with a total exposure time of $\sim 1.2$ Ms. After filtering out periods of high background, 637 ks of clean exposure were retained using the EPIC PN detector \citep{Brunner2008}. The deepest central region observed by \textit{XMM-Newton} covers an area of 0.196 $\mathrm{deg}^2$ and reaches flux limits of $1.9 \times 10^{-16}$ and $9 \times 10^{-16} \, \mathrm{erg} \, \mathrm{cm}^{-2} \, \mathrm{s}^{-1}$ in the $0.5-2\, \mathrm{keV}$ and $2-10\, \mathrm{keV}$ bands, respectively. This makes the XMM-LH the second most sensitive X-ray survey to date, surpassed only by the \textit{XMM-Newton} survey in the \textit{Chandra} Deep Field South (XMM-CDFS) with a flux limit of $6.6 \times 10^{-16} \, \mathrm{erg} \, \mathrm{cm}^{-2} \, \mathrm{s}^{-1}$ \citep{Ranalli2013} and above the \textit{XMM-Newton} Ultra Narrow Deep Field survey (XMM-UNDF) which reaches $9.8 \times 10^{-16} \, \mathrm{erg} \, \mathrm{cm}^{-2} \, \mathrm{s}^{-1}$ in the hard band \citep{Elias2021}. 

Early work on the XMM-LH was carried out by \citet{Mainieri2002}, who analysed the first 100 ks observation and identified 98 X-ray sources, 67 of them with optical spectroscopic classifications. They found that most Type-II AGNs showed significant absorption with column densities up to $N_{\rm H} \sim 10^{23}\,\mathrm{cm^{-2}}$, while some exhibited little or no absorption, highlighting discrepancies between optical and X-ray classifications. However, they did not find clear evidence for the presence of Compton-thick AGNs. A subsequent study by \citet{Mateos2005}, based on the 123 brightest sources (those with $>500$ counts), reported soft excess emission in $\sim$15 per cent of sources and Fe K$\alpha$ lines in eight AGNs, but again found no strong evidence for CT-AGNs.  However, they reported a 23 per cent discrepancy of optical Type-II AGNs appearing unabsorbed in X-rays. \citet{Brunner2008} published an X-ray source catalog for the central region of the LH (within a 15' radius), identifying 409 sources with high significance $(>3.9 \sigma)$. Using X-ray color-color diagrams, they estimated that approximately 6 per cent of the sources could be CT-AGNs, providing the first constraints on the surface density of CT-AGNs in the LH. 

The main advantages of our Lockman-SpReSO sample compared with earlier analyses are: First, our sample is  well complemented by optical spectroscopic follow-up with GTC, providing precise spectroscopic redshifts for our X-ray spectral analysis up to $z \sim 5$ \citep{Gonzalez2023,Gonzales2024a}. Second, the current multiwavelength coverage of the field enables broadband Spectral Energy Distribution (SED) modelling \citep[e.g. ][and Herrera-Endoqui et al. in prep]{Gonzalez2023}, which provides valuable insights into the AGN nuclear geometry, including torus angular width and inclination angle, information that was not accessible in earlier works. Third, since FIR emission is a distinctive feature of CT-AGNs \citep{KilerciEser2020}, our X-ray sample cross-matched with \textit{Herschel} FIR counterparts is particularly well suited to identify and characterize heavily obscured AGNs that may have been missed in previous studies. Moreover, recent advances in X-ray spectral techniques, such as the use of physically motivated clumpy torus models combined with Monte Carlo approaches \citep[e.g. XCLUMPY,][]{Tanimoto2019}, allow for a more robust characterization of highly obscured AGNs.

\subsection{Statistical Cross-Correlation }

A cross-correlation analysis was performed using a statistical approach. We employed the likelihood ratio (LR), a common method used in X-ray surveys to identify multiwavelength counterparts \cite[e.g.,][]{Brusa2007,Civano2012,Miyaji2024}. LR is defined in Equation \ref{LReq} as the ratio of two statistical probability densities, the probability of a correct detection $dp(r|true)$ and the probability of a false-positive detection $dp(r|false)$ due to background fluctuations.

\begin{equation} \label{LReq}
    \mathrm{LR}(r) = \frac{dp(r|true)q(m)}{dp(r|false)}
\end{equation}

\noindent where $q(m)$ is the probability distribution of having a true counterpart, as a function of magnitude. Following previous studies of \citet{Pineau2011,Ranalli2013,Elias2021}, we perform the cross-correlation and estimate LR, using the software Aladin with the \texttt{xcorr} plugin developed by \cite{Pineau2011}. They did not use explicitly the quantity $q(m)$, rather LR is calculated using positional uncertainties and the local angular density distribution ($\lambda$) to identify the most likely X-ray counterparts\footnote{For a more detailed description of the plugin and the estimation of the local density, see Appendix B of \cite{Pineau2011}}. The correct detection probability density $dp(r|true)$ is  defined in Equation \ref{true} as: 

\begin{equation}\label{true}
    dp(r|true) = re^{-r^2 / 2}dr
\end{equation}

\noindent where $r = d/ \sqrt{\sigma^2_X + \sigma^2_O}$, with $d$ being the angular distance separating both sources, and $\sigma_\mathrm{X}$ and $\sigma_\mathrm{O}$ representing the X-ray and Optical/IR positional errors, respectively. The second probability density of a false-positive detection is defined in equation \ref{false} as: 

\begin{equation} \label{false}
dp(r|false) = 2\lambda rdr
\end{equation}

\noindent where $\lambda = (\sigma^2_X + \sigma^2_O) \times f(\sigma_\mathrm{X}, m_\mathrm{0})$, with $f(\sigma_\mathrm{X}, m_\mathrm{0})$ being the local density estimation \citep{Pineau2011}. Finally, the LR is defined in Equation \ref{LR} as follows: 

\begin{equation} \label{LR}
    LR(r) = \frac{e^{-r^2 / 2}}{2\lambda}
\end{equation}

The standard X-ray positional error, $\sigma_\mathrm{X} = \sqrt{sys_\mathrm{err}^2 + radec_\mathrm{err}^2}$, is calculated as a function of the \textit{XMM-Newton} systematic ($sys_\mathrm{err}$) and statistical ($radec_\mathrm{err}$) positional errors. For our analysis, we adopted a mean systematic error of $sys_\mathrm{err} = 0.43\, \mathrm{arcsec}$. This value was reported by \cite{Traulsen2019} in the first \textit{XMM-Newton} serendipitous source catalog from overlapping observations (3XMM-DR7s) and was used during the cross-correlation process with the XMM Ultra Narrow Deep Field survey \citep{Elias2021}. Finally, the statistical positional  error, $radec_\mathrm{err} = \sqrt{\sigma_{\alpha}^2 + \sigma_{\delta}^2}$, is calculated as the quadrature combination of the errors on the image coordinates. For our X-ray catalog, we obtained a mean positional error of $\sigma_\mathrm{X} = 0.6 \pm 0.17\, \mathrm{arcsec}$. For the FIR sample, we adopted a positional uncertainty of $0.2\, \mathrm{arcsec}$, as reported by \citet{Gonzalez2023} in the construction of the Lockman-SpReSO catalog.

We cross-correlated the FIR catalog with the latest stacked X-ray data release, 4XMM-DR13s\footnote{The full catalog can be downloaded at \url{https://xmmssc.aip.de/cms/catalogues/4xmm-dr13s/}} \citep{Traulsen2020}. This catalog is produced using the standard XMM-Newton EPIC detection pipeline, which performs source detection on individual and overlapping observations obtained at different epochs. Sources are identified through a maximum-likelihood fitting procedure applied in the five standard energy bands ($0.2-0.5$, $0.5-1$, $1-2$, $2-4.5$, and $4.5-12\, \mathrm{keV}$), and only sources with detections corresponding to a significance of $3\sigma$ are considered.

For the cross-correlation process, we first retrieved all X-ray sources within a circular region of 20' radius centred on the Lockman Hole field, obtaining a total of 428 X-ray detections. We then cross-matched both catalogs using the likelihood-ratio method. We found a total of 110 FIR sources with \textit{XMM-Newton} counterparts at a 3$\sigma$ confidence level. Six sources in the sample showed double X-ray associations, representing a 5.4 per cent rate of false-positive detections; for our analysis, we used the counterpart with the highest LR. The median angular separation of our sources from their X-ray counterparts is of $0.67\, \mathrm{arcsec}$. The remaining 318 X-ray sources do not have FIR counterparts in the Lockman-SpReSO catalog. 

Among these 110 sources, 70 have optical spectroscopic measurements with redshifts distribution from $z_\mathrm{spec} = 0.07$ to $5$, the remaining 40 sources have only photometric redshift estimates \citep{Gonzalez2023}. To estimate the effective sensitivity of the X-ray sample, we combined the fluxes of bands 1-3 $(0.2-2\, \mathrm{keV})$ and propagated their uncertainties in quadrature. Adopting a $3\sigma$ detection threshold, the faintest reliable detections correspond to a soft-band flux of $\sim 2\times10^{-16}\,\mathrm{erg\,cm^{-2}\,s^{-1}}$. This value represents the effective flux limit of the X-ray catalog used in this work and is consistent with the reported by \citet{Brunner2008}.

\begin{figure}
\centering
\includegraphics[scale=0.2]{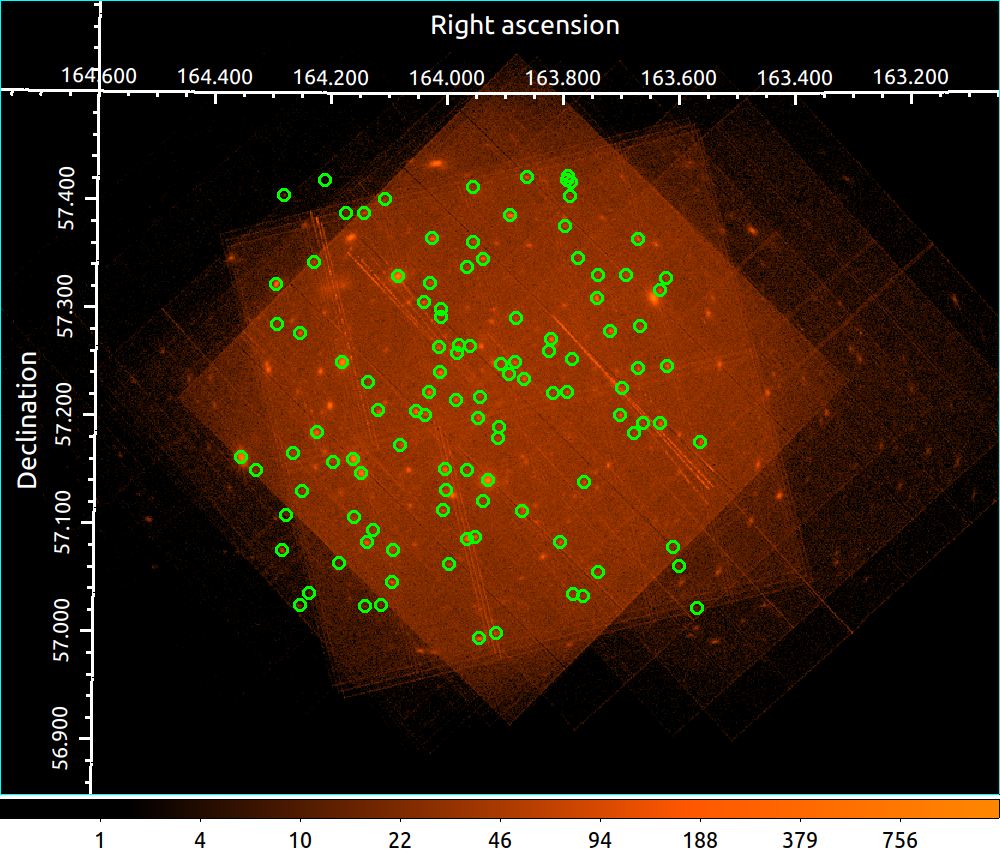}
\caption{PN mosaic image of all observations from the survey in the $0.3-10\, \mathrm{keV}$ band. Green circles indicate the 110 X-ray sources cross-correlated with our FIR catalog. The colour bar is in counts.}
\label{XMM_LH_field}
\end{figure}

Fig. \ref{XMM_LH_field} presents the 18 \textit{XMM-Newton} observations in the $0.3 - 10\, \mathrm{keV}$ band, covering the LH region. The 110 FIR sources with X-ray counterparts are highlighted with green circles. The first ten sources of the sample are listed in Table \ref{cross}, which summarizes the identification of the X-ray counterparts of the FIR sources and the angular separation between the \textit{XMM-Newton} and \textit{Herschel} positions. Whenever optical spectroscopy is available, the table also includes the optical spectroscopic classification for the AGN sources. Type-I AGNs correspond to sources showing broad emission lines with FWHM $>1000\,\mathrm{km\,s^{-1}}$ as reported by \citet{Negrete2025}, while Type-II AGNs correspond to sources without broad emission-line components identified from the available optical spectra.

\begin{table*}
\centering
\caption{Identification of the first 10 FIR sources from the Lockman-SpReSO catalog with X-ray counterparts. Columns 8 and 9 indicate the shortest-wavelength \textit{Herschel} detection band (PACS 100 $\mu$m or SPIRE 250 $\mu$m) and its corresponding flux density in Jansky \citep[see][]{Gonzalez2023}. Column 10 indicates the angular separation between the X-ray and FIR counterparts and column 11 indicates the AGN classification flag based on the X-ray spectral analysis (see Section \ref{Simple_Spec}), where 1 represents AGN and 0 denotes non-AGN. Column 12 reports the optical spectroscopic classification when available. Type-I AGNs correspond to sources showing broad emission lines with FWHM $>1000\,\mathrm{km\,s^{-1}}$ as reported by \citet{Negrete2025}, while Type-II AGNs correspond to sources without broad emission-line components. A dash indicates sources not identified as AGN from optical spectroscopy. The complete table is available online in ASCII format along with this paper.} 
\label{cross}
\begin{tabular}{@{}cccccccccccc@{}}
\toprule
objID & IAUNAME & RA & DEC  & $z$ & $\mathrm{cts}$ & $\log F_{0.3-10\,\mathrm{keV}}$ & FIR flux & FIR band  & Sep & AGN & Type\\
 &  & &   &  &  &   $\mathrm{erg\, s^{-1}\, cm^{-2}}$ &  $\mathrm{Jy}$ & & arcsec & &\\
 (1) & (2) & (3) & (4) & (5) & (6) & (7) & (8) & (9) & (10) & (11) & (12) \\ \midrule
206776 & 4XMMs J105252.7+572859 & 163.2198 & 57.4833 & 0.203 & 858 & $-13.80^{+0.04}_{-0.05}$ & 35($\pm$1.5) & P100 & 0.47 & 1 & II\\
206764 & 4XMMs J105144.7+572808 & 162.9364 & 57.4690 & 3.4 & 1127 & $-13.87^{+0.04}_{-0.05}$ & 5.4($\pm$1.7)E-3 & S250 & 0.35 & 1 & I\\
206720 & 4XMMs J105125.7+573545 & 162.8571 & 57.5960 & 0.073 & 282 & $-14.43^{+0.25}_{-0.13}$ & 9.5($\pm$1.3) & P100 & 1.56 & 0 & -\\
206718 & 4XMMs J105128.0+573503 & 162.8668 & 57.5843 & 0.072 & 973 & $-13.91^{+0.08}_{-0.04}$ & 135($\pm$2.7) & P100 & 1.16 & 0 & -\\
206715 & 4XMMs J105143.6+572938 & 162.9320 & 57.4940 & 0.08 & 229 & $-14.66^{+0.21}_{-0.10}$ & 67($\pm$3.9) & P100 & 1.85 & 0 & -\\
206695 & 4XMMs J105205.8+574121 & 163.0243 & 57.6894 & 0.462 & 2176 & $-13.21^{+0.03}_{-0.03}$ & 8.7($\pm$12)E-4 & S250 & 3.60 & 1 & I\\
206692 & 4XMMs J105223.2+574121 & 163.0970 & 57.6894 & 1.533 & 793 & $-13.84^{+0.06}_{-0.05}$ & 1.6($\pm$1.4)E-3 & S250 & 0.55 & 1 & I\\
206690 & 4XMMs J105206.4+574109 & 163.0270 & 57.6860 & 1.53* & 1366 & $-13.42^{+0.05}_{-0.04}$ & 4.9($\pm$7.5)E-4 & S250 & 0.28 & 1 & - \\
206684 & 4XMMs J105322.3+574009 & 163.3430 & 57.6694 & 0.88 & 192 & $-14.03^{+0.18}_{-0.18}$ & 4.6($\pm$1.3) & P100 & 1.16 & 1 & II\\
206679 & 4XMMs J105230.0+573913 & 163.1253 & 57.6539 & 1.437 & 4270 & $-13.24^{+0.02}_{-0.01}$ & 0.01($\pm$0.002) & S250 & 0.32 & 1 & I\\
\bottomrule
\end{tabular}
\begin{flushleft}
\footnotesize 
$*$ Photometric redshift\\
\end{flushleft}
\end{table*}

\section{X-ray spectral analysis}
\label{Xspectral}

Since the XMM-LH survey consists of 18 pointings conducted at different epochs, we stacked the background-subtracted X-ray spectrum from each individual PN observation for every object in our sample. For our analysis, we used the \textsc{xspec} program (version 12.13.1). We manually selected the source and background regions using the task \texttt{evselect}, defining circular areas with a radius of 15 and 30 arcseconds, respectively. These regions correspond to approximately 75 per cent of the encircled energy fraction. Subsequently, we combined all spectra using the task \texttt{epicspeccombine}\footnote{\url{https://xmm-tools.cosmos.esa.int/external/sas/current/doc/epicspeccombine/index.html}}, thereby enhancing the signal-to-noise ratio and minimizing statistical uncertainties. 

\begin{figure}
\centering
\includegraphics[scale=0.2]{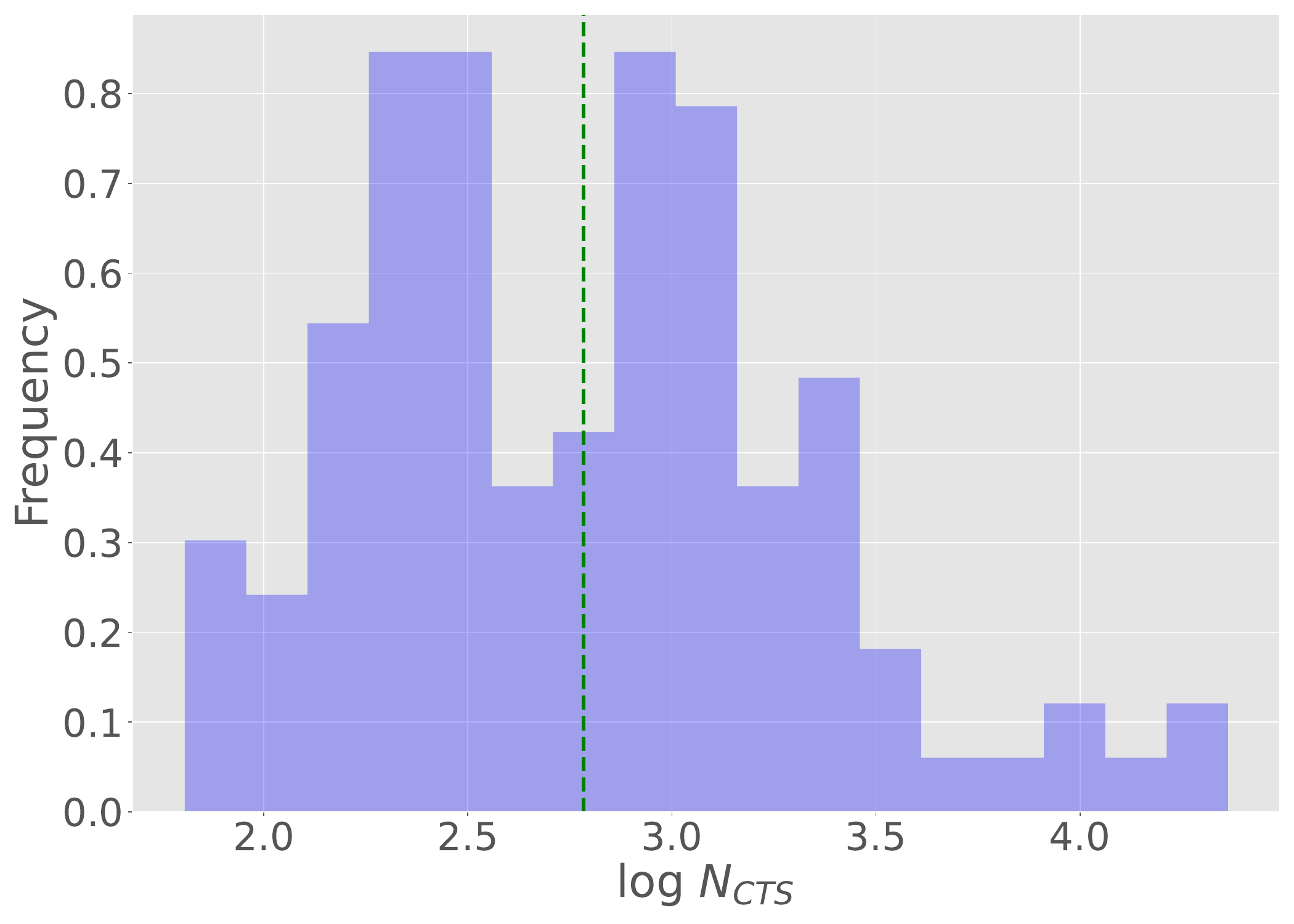}
\caption{Photon count distribution in the $0.3-10\, \mathrm{keV}$ band for the complete X-ray catalog, with a median (green dashed line) of 529.}
\label{CTS}
\end{figure}

Based on the quality of the final combined spectra, specifically the total photon counts $(N_\mathrm{cts})$ in the broadband $(0.3-10\, \mathrm{keV})$, we binned the spectra with 15 to 30 counts per bin to ensure adequate statistical quality for the use of $\chi^2$ statistics. Fig. \ref{CTS} shows the broadband $N_\mathrm{cts}$ density distribution for the sample, ranging from 60 counts for the faintest object to 20,000 counts for the brightest. The green dashed line indicates the median value of 529 X-ray counts.  

\subsection{Simple Spectral fitting and AGNs identification}
\label{Simple_Spec}

We began by fitting our spectra using a simple unabsorbed (model-1) and an intrinsically absorbed  (model-2) power-law models. 

\begin{enumerate}
    \item[1.] \texttt{tbabs*powerlaw}
    \item[2.]  \texttt{tbabs*powerlaw*zphabs}
\end{enumerate}

The Galactic absorption, represented by the component \texttt{tbabs}, was fixed at a column density of $N_{\mathrm{H}} = 5.72 \times 10^{19} \, \mathrm{cm}^{-2}$ in the line of sight of the X-ray field \citep[e.g. as reported by][]{Brunner2008}. The intrinsic absorption, representing the contribution of absorbing material in the circumnuclear region and the host galaxy, was modelled using the \texttt{zphabs} component. The primary power-law emission, characterized by the photon index $(\Gamma)$, was modelled with the \texttt{powerlaw} component. 

Based on the results of the spectral fitting with models 1 and 2, we classified as AGNs those bright objects with absorption-corrected X-ray luminosities (derived from model-2) greater than $ 3\times 10^{42}\, \mathrm{erg\, s^{-1}}$ in the $2-10\, \mathrm{keV}$ rest-frame energy band\footnote{The X-ray luminosity estimated using model-2 shows no significant variation when compared to estimates obtained with models 3, 4, and 5.}, or those with hard spectra (derived from model-1) characterized by $\Gamma < 1$ \citep[e.g.,][and references therein]{Xue2011,Luo2017,Elias2021}.
The luminosity criterion identifies AGNs based on the extreme X-ray emission due to SMBH accretion, which far exceeds those produced by stellar processes in normal galaxies \citep{Wang2013,Hickox2018}. A hard X-ray spectrum, used as a secondary criterion, indicates either strong obscuration or reflection-dominated emission, both of which are characteristics of AGN activity. To mitigate the degeneracy between $\Gamma$ and intrinsic absorption $N_{\mathrm{H}}$, we used the $\Gamma_{m1}$ values estimated without accounting for intrinsic obscuration (i.e., derived from model-1). 
 
For sources where $\Gamma$ could not be reliably constrained (e.g. due to low-count statistics), we adopted a conservative range of $\Gamma =0.5-2.5$  during the fitting. In total, we identified 94 AGNs that satisfied at least one of these criteria. In column 11 of Table \ref{cross} we report the boolean AGN classification, based on the X-ray spectral analysis. A value of 1 indicates a source classified as an AGN, while 0 denotes a non-AGN source. From this point onward, our analysis will focus exclusively on this X-ray AGN population.  

\begin{figure}
\centering
\includegraphics[scale=0.25]{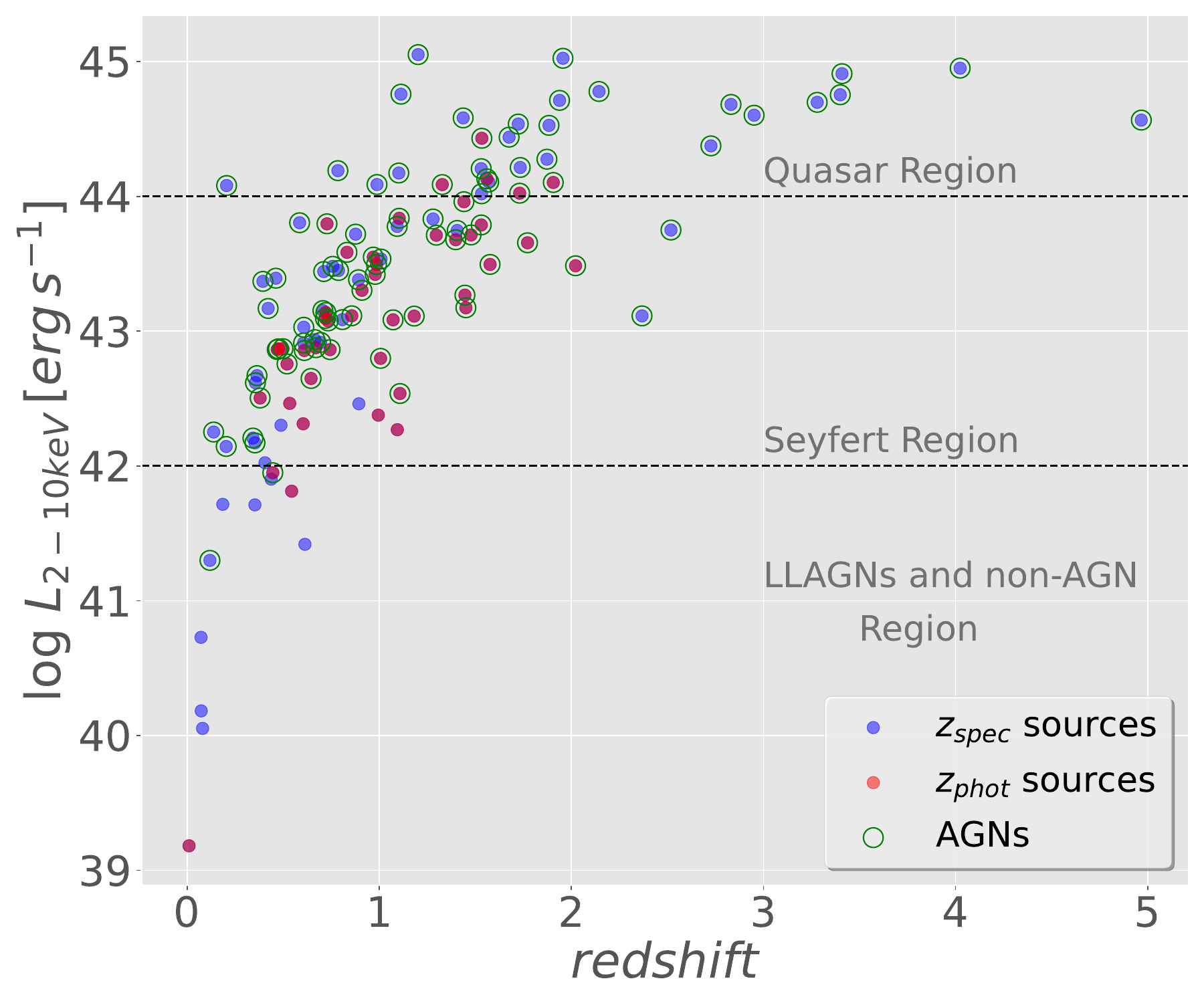}
\caption{$2-10\, \mathrm{keV}$ absorption-corrected rest-frame luminosity versus redshift. Sources classified as AGNs are highlighted with green circles. Blue and red circles represent sources with spectroscopic ($z_{\mathrm{spec}}$) and photometric ($z_{\mathrm{phot}}$) redshifts, respectively. The luminosity regimes for quasars, Seyfert galaxies, and low-luminosity AGN (LLAGN) are indicated with dotted lines.} 
\label{zLx}
\end{figure}

Table \ref{statistic1} summarizes the best-fitting parameters obtained from simple spectral analysis, such as the photon index (column 3) and X-ray luminosity (column 9) used for the AGN classification. Fig. \ref{zLx} shows the $2-10\, \mathrm{keV}$ rest-frame luminosity distribution of our sample, corrected for absorption, as a function of redshift. Red and blue circles represent sources with spectroscopic (54) and photometric (40) redshifts, respectively. Green circles indicate sources classified as AGNs. Based on their luminosities,  we classified 32 AGNs as quasars (34 per cent) with $L_{2-10} > 10^{44} \mathrm{erg\, s^{-1}}$, 60 as Seyfert galaxies (64 per cent) with $L_{2-10} = 10^{42 - 44} \mathrm{erg\, s^{-1}}$, and 2 as Low-luminosity AGNs with  $L_{2-10} < 10^{42} \mathrm{erg\, s^{-1}}$ (2 per cent) \citep[e.g.][]{Elias2021}.   

Our AGN sample is overall consistent with the list reported by \citet{Gonzales2024b}, who applied a set of multiwavelength selection criteria to 409 sources with spectroscopic redshift measurements from the Lockman-SpReSO catalog, identifying AGN activity in 17\% of the sample. A total of 69 AGNs were reported in their analysis, of which 53 are analysed in this work, while the remaining 16 objects do not present X-ray spectral counterparts in the \textit{XMM-Newton} observations.

\subsection{Complex Spectral Fitting}

\begin{table*}
\centering
\caption{ Summary table of the spectral fitting results for the first 10 AGNs in our sample. Column 1 lists the Object ID, while column 2 presents the reduced chi-square values ($\chi^2/\mathrm{dof}$, where $\mathrm{dof}$ are the degrees of freedom). Spectral parameters derived from the best-fitting model (column 10) are reported in columns 4 to 9. The complete table is available online in ASCII format along with this paper.}
\label{statistic1}
\begin{tabular}{@{}ccccccccccc@{}}
\toprule
objID & $\chi^2/\mathrm{dof}$ & $\Gamma_{m1}^{a}$ & $\Gamma$  & $\log N_{\mathrm{H}}$ & $EW$ & $kT$ & R &  $\log L_{2-10\, \mathrm{keV}}$ & model\\ 
 &  &  &  &  $\mathrm{cm}^{-2}$ & $\mathrm{keV}$ & $\mathrm{keV}$ &  & $\mathrm{erg\, s^{-1}}$  & &\\ 
 (1) & (2) & (3) & (4) & (5) & (6) & (7) & (8) & (9) & (10)  \\ \midrule
206764 & 60.29/60 & $1.70_{-0.12}^{+0.12}$ & $1.72_{-0.14}^{+0.21}$ & $21.00_{-0.63}^{+0.96}$ & $<0.079$ & 2 & 0 & $44.75_{-0.03}^{+0.03}$ & 1 \\
206710 & 144.36/122 & $2.21_{-0.02}^{+0.02}$ & $2.11_{-0.04}^{+0.06}$ & $<20$ & $0.067_{-0.027}^{+0.038}$ & 0.12 & -4.18 & $44.77_{-0.01}^{+0.01}$ & 4 \\
206695 & 58.54/58 & $1.91_{-0.07}^{+0.08}$ & $1.91_{-0.07}^{+0.12}$ & $<20.47$ & $0.29_{-0.29}^{+0.27}$ & 0.13 & -6.68 & $43.39_{-0.05}^{+0.04}$ & 1 \\
206692 & 50.93/44 & $2.21_{-0.18}^{+0.20}$ & $2.21_{-0.18}^{+0.20}$ & $<20$ & $<0.29$ & 0.11 & -9.47 & $44.02_{-0.06}^{+0.08}$ & 1 \\
206690 & 45.97/54 & $1.77_{-0.09}^{+0.10}$ & $1.77_{-0.09}^{+0.17}$ & $<21.15$ & $0.074_{-0.062}^{+0.18}$ & 0.01 & -1.15 & $44.43_{-0.04}^{+0.04}$ & 1 \\
206684 & 24.76/25 & $1.53_{-0.42}^{+0.51}$ & $1.52_{-0.41}^{+0.57}$ & $<21.23$ & $1.10_{-0.88}^{+1.88}$ & 0.01 & -10 & $43.08_{-0.19}^{+0.18}$ & 1 \\
206679 & 85.66/78 & $1.94_{-0.05}^{+0.05}$ & $2.08_{-0.09}^{+0.09}$ & $21.13_{-0.32}^{+0.19}$ & $0.038_{-0.030}^{+0.10}$ & 0.52 & -1.36 & $44.58_{-0.02}^{+0.02}$ & 2 \\
206672 & 41.26/37 & $1.68_{-0.25}^{+0.28}$ & $1.58_{-0.34}^{+0.60}$ & $21.79_{-0.37}^{+0.73}$ & $<0.088$ & 0.12 & -10 & $44.37_{-0.05}^{+0.05}$ & 4 \\
206667 & 74.02/66 & $1.91_{-0.07}^{+0.08}$ & $1.91_{-0.07}^{+0.08}$ & $<20$ & $<0.061$ & 0.18 & -0.44 & $44.53_{-0.03}^{+0.02}$ & 1 \\
206666 & 69.22/60 & $1.61_{-0.09}^{+0.09}$ & $1.61_{-0.09}^{+0.11}$ & $<21.15$ & $<0.087$ & 0.01 & -1.99 & $44.71_{-0.02}^{+0.02}$ & 1 \\ \bottomrule
\end{tabular}
\begin{flushleft}
\footnotesize 
$^a$ Photon index derived from model-1 \texttt{tbabs*powerlaw}.\\
\end{flushleft}
\end{table*}

To explore the presence of additional physical components in our AGN spectra, such as neutral Fe K$\alpha$ line emission, soft X-ray excess, and reflection, we increased the complexity of our spectral models as follows:

\begin{enumerate}
    \item[3.] \texttt{tbabs*(powerlaw*zphabs + zgauss)}
    \item[4.] \texttt{tbabs*((powerlaw + zbbody)*zphabs + zgauss)}
    \item[5.] \texttt{tbabs*(powerlaw*zphabs + zgauss + pexrav)}
\end{enumerate}

The third model includes a narrow gaussian emission line (\texttt{zgauss}) to measure the equivalent width of the Fe K$\alpha$ line. Since 33 AGNs in our sample have photometric redshifts, a slight miscalculation of $z_{\mathrm{phot}}$ could lead to an inaccurate measurement of the Fe K$\alpha$ line intensity \cite[for more details on the spectroscopic and photometric redshift estimations, see][]{Gonzalez2023}. Therefore, for these sources with $z_{\mathrm{phot}}$, we used a conservative Fe K$\alpha$ energy interval of $6.1-6.6$ keV \cite[e.g.,][]{Iwasawa2020}. To test the contribution of "soft excess" emission, we employed model-4, which includes a blackbody component (\texttt{zbbody}) with a photon temperature $kT$. This emission is typically observed in the spectra of some bright, unobscured AGNs below 2 keV, usually around $0.1-0.2$ keV \citep[e.g.,][]{Crummy2006}. The blackbody model is added to represent the soft excess, whose possible physical origins are discussed later in Section \ref{Soft}. 

The last model  (model-5) includes a cold-reflection component (\texttt{pexrav}), typically observed in the spectra of obscured AGNs. This component likely originates from reflection from high column density material (e.g. $\sim 10^{23} \mathrm{cm}^{-2}$) in the inner region of the dusty torus or the outer part of the disc. For convenience, the slope and normalization of the \texttt{pexrav} component are linked to those of the primary power-law. This setup allows us to focus only on the reflection scaling factor ($R$), which is free to vary between $-10$ and $0$. For instance, a reflection strength of $R \approx 0$ corresponds to no reflection, while $R = -1$ indicates that the reflected emission contributes equally to the primary power-law emission. We considered the reflection component to be unconstrained when $R$ was not well fixed, for instance, when $R \approx -10$. Also, the high-energy cutoff was set at 100 keV and the inclination angle ($i$) of the reflecting disc is fixed to 45$\degree$, representing a moderate inclination in agreement with typical AGN orientations \citep[e.g.,][]{Ricci2011}.

The reliability of each model was evaluated using two statistical tests. First, we applied an F-test with a 95 per cent confidence level (i.e. $f_{mi} < 0.05$). Then, we used the Akaike Information Criterion (AIC), defined as $AIC = 2k + \chi^{2}$, where $\chi^{2}$ is the chi-squared statistic value, and $k$ is the number of parameters. To measure the improvement of a model when adding a new component, we used Equation \ref{rho} to estimate the inverse of the relative likelihood ($\rho_{x,i}$) using the AIC criterion, as described by \cite{Hebbar2019,Krongold2021,Elias2024}.

\begin{equation}
 \label{rho}
    \rho_{x_i} = {\left(\exp \left(\frac{AIC(x_0,x_i) - AIC(x_0)}{2} \right)\right)}^{-1}
\end{equation}

To identify the best-fitting model, we selected the model with the lowest $f_{mi}$ value and the highest $\rho_{x_i}$\footnote{In cases where the F-test cannot be applied, such as when the added component is multiplicative \citep{Protassov2002}, the best model was selected based only on $\rho_{x_i}$. We also considered the complexity of the model and evaluated whether the improvement in the fit justified the use of a more complex model over a simpler one.}. For instance, source 206672 shows no significant improvement when intrinsic absorption is included using model-2, with test statistics of $\rho_{m2} \sim 3$ and $f_{m2}\sim 0$. In contrast, the significance increases notably after including a gaussian at low energies to model the soft excess contribution in model-4, with $\rho_{m4} \sim 253$ and $f_{m4}\sim 0.05$. Additionally, to complement our analysis, we performed visual inspections of the residual (e.g., Fig. \ref{XID4}). For example, although models 4 and 5 appear to provide a better statistical fit, the residual analysis combined with the constraint of physically realistic parameter values indicates that the best spectral fitting for source 206672 is achieved with model-4. The results of the spectral statistical tests are listed in Table \ref{statistic2}.  In Appendix \ref{xspec_summary} in Table \ref{xspec_table}, we included a summary of all spectral models used during this work with a brief description of each parameter.

\begin{table}
\centering
\caption{Summary table of the F-tests and the Akaike criterion, for the first 10 AGNs in our sample. Column 2 refers to the comparison between the fits obtained with models 1 and 2, column 3 is for models 2 and 3, column 4 is for models 3 and 4, and column 5 is obtained by comparing the complex models 3 and 5. The complete table is available online in ASCII format along with this paper.}
\label{statistic2}
\begin{tabular}{@{}ccccc@{}}
\toprule
objID & $\rho_{m2}/f_{m2}$ & $\rho_{m3}/f_{m3}$ & $\rho_{m4}/f_{m4}$ & $\rho_{m5}/f_{m5}$ \\ 
(1) & (2) & (3) & (4) & (5) \\ \midrule
206764 & 2.79/0.83 & 1.49/0 & 9.51/0.8 & 2.82/0.8 \\
206710 & 2.72/0 & 45.25/0.39 & 6.7E20/8.9E-14 & 5.1E13/1.4E-09 \\
206695 & 2.72/0 & 7.65/0.97 & 7.76/0.96 & 4.71/0.31 \\
206692 & 2.72/0 & 1.83/0 & 102.5/0.13 & 5.68/0.29 \\
206690 & 2.72/0 & 5.99/0 & 7.39/1 & 2.76/0.86 \\
206684 & 2.72/0 & 4.79/0 & 10.07/0.8 & 5.03/0.33 \\
206679 & 492.61/0.003 & 2.76/0 & 44.52/0.21 & 4.02/0.41 \\
206672 & 2.72/0 & 1.75/0 & 253.33/0.05 & 15.92/0.09 \\
206667 & 2.72/0 & 1.84/0 & 16.71/0.51 & 2.93/0.73 \\
206666 & 2.72/0 & 1.77/0 & 7.35/0 & 5.86/0.27 \\ \bottomrule
\end{tabular}
\end{table}

\section{Results of the Spectral Analysis}
\label{SpecResuts}

\subsection{Reduced $\chi^2$ and Photon Index Distributions}

\begin{figure*}
\centering
\includegraphics[scale=0.25]{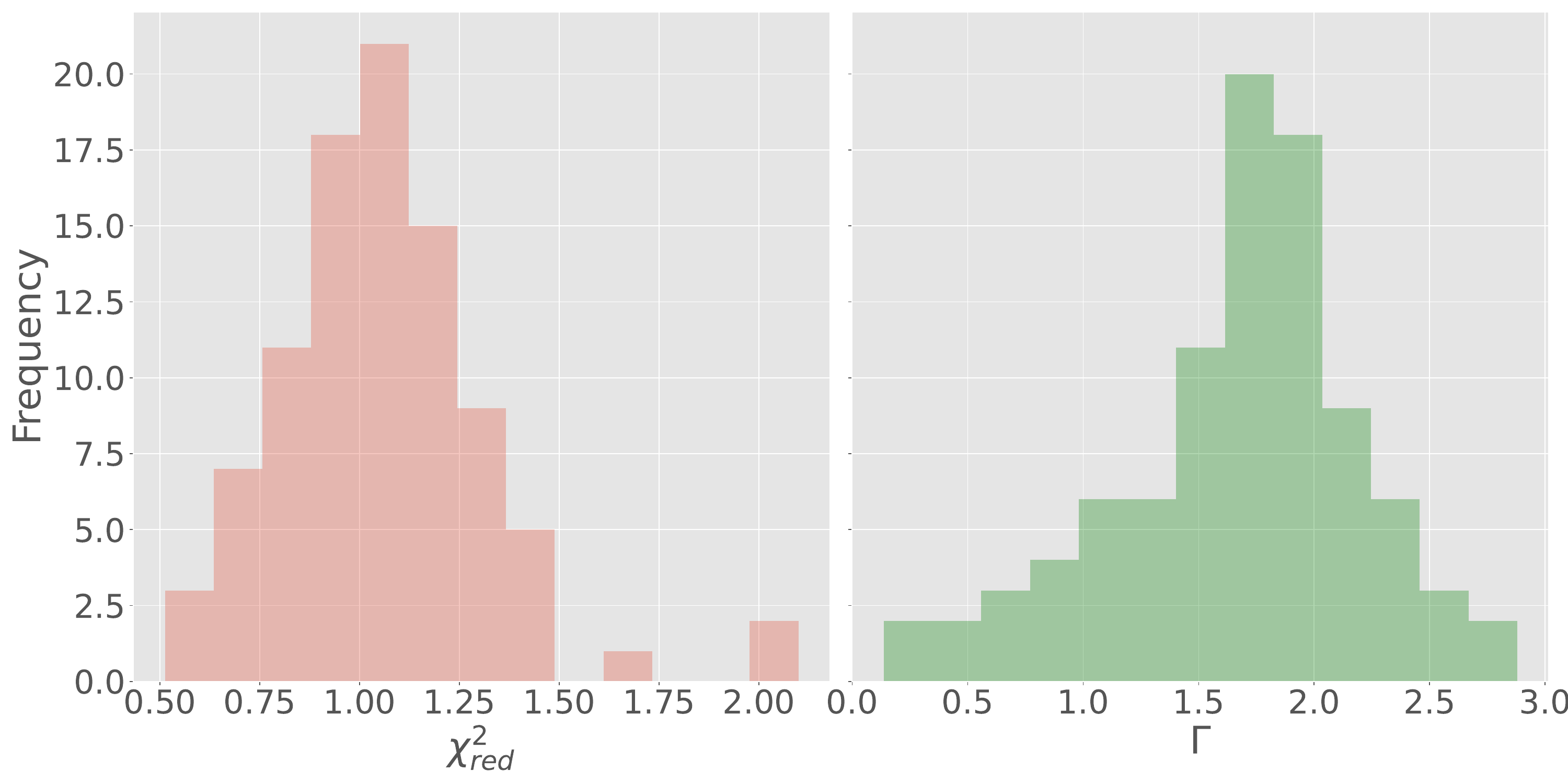}
\caption{Distribution of the spectral fitting parameters for the AGN sample. \textit{Left panel}: distribution of the reduced $\chi^2$ values obtained from the X-ray spectral fits, showing that most sources cluster around $\chi^2_{\rm red} \sim 1$, indicating that the adopted spectral models provide an overall good description of the data. \textit{Right panel}: distribution of the photon index $\Gamma$.}
\label{Chi_gamma}
\end{figure*}

To quantify the reliability of the spectral analysis performed in Section \ref{Xspectral}, we examine the distribution of the reduced chi-square values ($\chi^2_{\rm red}$), presented in the left panel of Fig. \ref{Chi_gamma}. $\chi^2_{\rm red}$ is defined as the ratio between the chi-square statistic ($\chi^2$) and the number of degrees of freedom, and provides a diagnostic of how well the adopted model describes the observed data. We observed that most sources cluster around $\chi^2_{\rm red} \sim 1$, indicating that the adopted spectral models provide a statistically acceptable description for the majority of the sample.

Another parameter that we can use is the distribution of photon indices,  presented in the right panel of Fig. \ref{Chi_gamma}. We obtain an average photon index of $\langle \Gamma \rangle = 1.77 \pm 0.54$, with a median value of $\Gamma = 1.83$, consistent with typical AGN X-ray spectra \citep[e.g.][]{Marchesi2016,Liu2017,Elias2024}. The distribution of $\Gamma$ is wider than those found in brighter AGNs with higher spectral quality in the BASS survey-based analysis \citep{Ricci2017}, where most intrinsic $\Gamma$ is distributed between 1.2 and 2.4. A fraction of sources in our analysis show flatter spectra, which may be due to obscured AGNs being recognized as a flatter power-law due to limited photon statistics or the presence of additional spectral complexity. 

For instance, in Fig. \ref{Chi_gamma} we excluded the outlier source objID-206776, whose spectrum could not be adequately described using the adopted spectral models. In Section \ref{new_spec}, we investigate the spectral complexity of this source in more detail and highlight the existence of sources with spectral features beyond the scope of our baseline models.

\subsection{Soft excess} 
\label{Soft}
A common feature observed in around half of the Narrow-Line Seyfert 1 (NLS1) galaxies is the presence of ``soft excess'' in the X-ray spectrum at energies between 0.1 and 2 keV. Although the physical origin of this component remains unclear, the most commonly proposed models attribute the soft excess either to Comptonization in a warm, optically thick corona or to reflection from an ionized accretion disc \citep[e.g.,][]{Crummy2006,Sobolewska2007,Gliozzi2020}. Because the soft X-ray band is highly sensitive to absorption by the hydrogen column density, the LH with the lowest Galactic $N_{\mathrm{H}}$ is an exceptional field to identify and study the soft excess emission in the AGN population. 

Based on our statistical tests, we identified 26 sources that present a significant improvement in the spectral fit when a blackbody component is included to model the soft excess (i.e., $f_{m4} < 0.05$ and  $\rho_{m4} > 100$). However, seven of these sources were excluded due to unreliable blackbody temperature normalizations. For example, an additional high temperature blackbody component can become degenerate with a steeper power-law index for a low signal-to-noise spectrum.  To mitigate potential degeneracies between the blackbody temperature ($kT$) and other spectral components, such as the power-law continuum, we performed a visual inspection and compared the residuals of models 1 and 2 with those of model-4. As a result, we confirmed the presence of a soft excess in approximately 11 per cent of the sample (10 AGNs). The average blackbody temperature for this subset is $\bar{kT} = 0.12 \pm 0.02\, \mathrm{keV}$ (column 7 in Table \ref{statistic1}). An exception is source objID-206654, which shows a significantly higher blackbody temperature ($kT=0.93\, \mathrm{keV}$). In half of the ten sources, the contribution from reflected emission is negligible, as indicated by the unconstrained reflection strengths ($R \approx -10$), suggesting that the soft excess in these objects is more likely dominated by warm Comptonization rather than ionized disc reflection. In contrast, the remaining sources show moderate reflection components ($R \sim -1$ to $-4$), indicating that reflected emission may contribute to the observed soft X-ray excess.

\begin{figure*}
\centering
\includegraphics[scale=0.25]{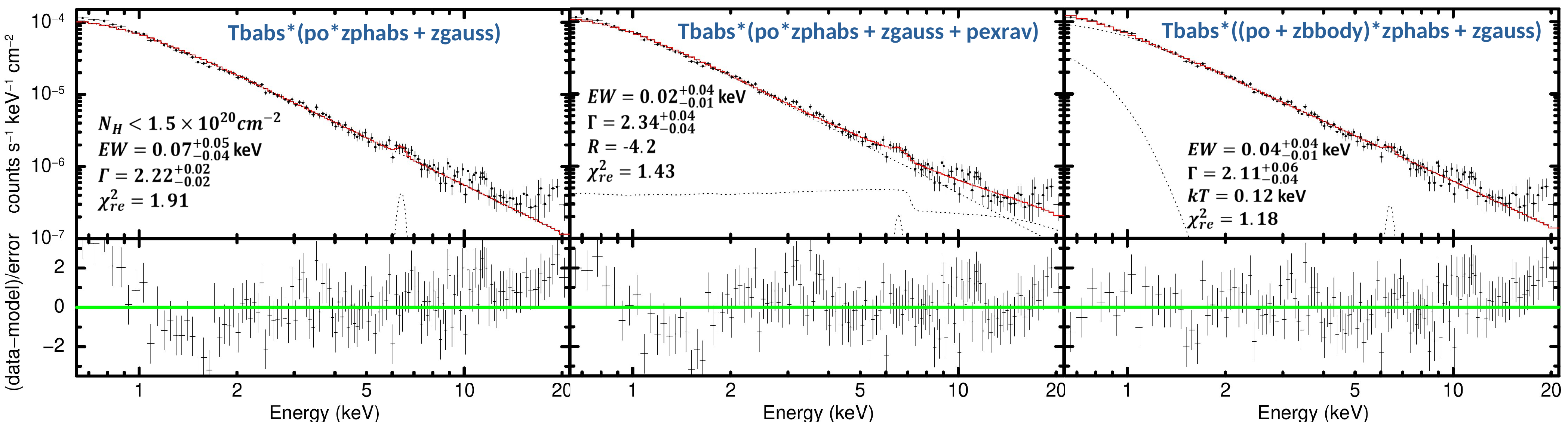}
\caption{Fitted X-ray spectrum and residuals of source 206710 at $z = 1.113$. The left panel shows the spectrum fitted with a simple absorbed power-law and a neutral iron line (model-3); the central panel includes a reflection component (model-5);  and the right panel incorporates an additional blackbody component to account for the soft excess emission (model-4).}
\label{XID4}
\end{figure*}

Fig. \ref{XID4} presents an example of the detected soft excess in the rest-frame spectrum (and residuals, black dots) of source 206710 at $z=1.113$. The left, central, and right panels show the spectral fits (red lines) using models 3, 5, and 4 (black dotted lines), respectively. The observed soft emission is best described in the right panel using model-4, which includes a blackbody component with a blackbody temperature of $kT = 0.12$ whose contribution decreases above $\sim 1 \mathrm{keV}$. This model produces a reduced chi-square value of $\chi^2/\mathrm{dof} = \chi^2_\mathrm{red} = 1.18$, which is the best-fitting compared to model-3 with $\chi^2_\mathrm{red} = 1.91$, and model-5 with $\chi^2_\mathrm{red} = 1.43$. 

It is important to mention that the soft excess was modelled using a phenomenological blackbody component. This approach is commonly adopted in X-ray spectral studies of AGNs when the available statistics do not allow the use of more physically motivated models with a larger number of parameters \citep[e.g.][]{Corral2011,Elias2024}. Possible physical interpretations of the soft excess include warm Comptonization in an optically thick corona or relativistic reflection from an ionized accretion disc \citep[e.g.][]{Crummy2006,Petrucci2018}. Nevertheless, the available data do not allow robust constraints on more complex models. 

Since our sample is selected from far-infrared detected sources, a contribution from star-formation-related processes to the soft X-ray emission is also expected, particularly in the low-luminosity AGN regime. In these systems, thermal emission from hot gas, X-ray binaries, and circumnuclear star-forming regions may partially contribute to the observed soft excess component \citep[e.g.][]{Ranalli2003}. However, the AGN sample analysed in this work is primarily composed of luminous Seyfert galaxies and quasars with $L_{2-10\,\mathrm{keV}} > 3\times10^{42}\,\mathrm{erg\,s^{-1}}$, where the intrinsic AGN emission is expected to dominate the X-ray spectrum. Only two sources are classified as low-luminosity AGNs with $L_{2-10\,\mathrm{keV}} < 10^{42}\,\mathrm{erg\,s^{-1}}$, for which star-formation-related emission may contribute more significantly to the observed soft excess. A further discussion on the relation between the X-ray spectral properties and the star-forming nature of our FIR-selected AGN sample is presented in Section \ref{sfr}.

\subsection{Obscured population}

From the spectral analysis, we measured the hydrogen column density of our AGN sample. We identified 74 sources ($79$ per cent) as unobscured AGNs, with $N_{\mathrm{H}} < 10^{22}\, \mathrm{cm}^{-2}$. Note that 53 of them have only upper limits (i.e., $N_{\mathrm{H}} < 10^{20}\, \mathrm{cm}^{-2}$). The remaining  20 sources ($21$ per cent) were classified as obscured AGNs, with $N_{\mathrm{H}} \ge 10^{22}\, \mathrm{cm}^{-2}$. Five of them are sub-classified as Compton-thick candidates, with $N_{\mathrm{H}} \gtrsim 5\times 10^{23}\,  \mathrm{cm}^{-2}$ and upper errors consistent with $N_{\mathrm{H}} \sim 10^{24}\,  \mathrm{cm}^{-2}$. 

\begin{figure}
\centering
\includegraphics[scale=0.28]{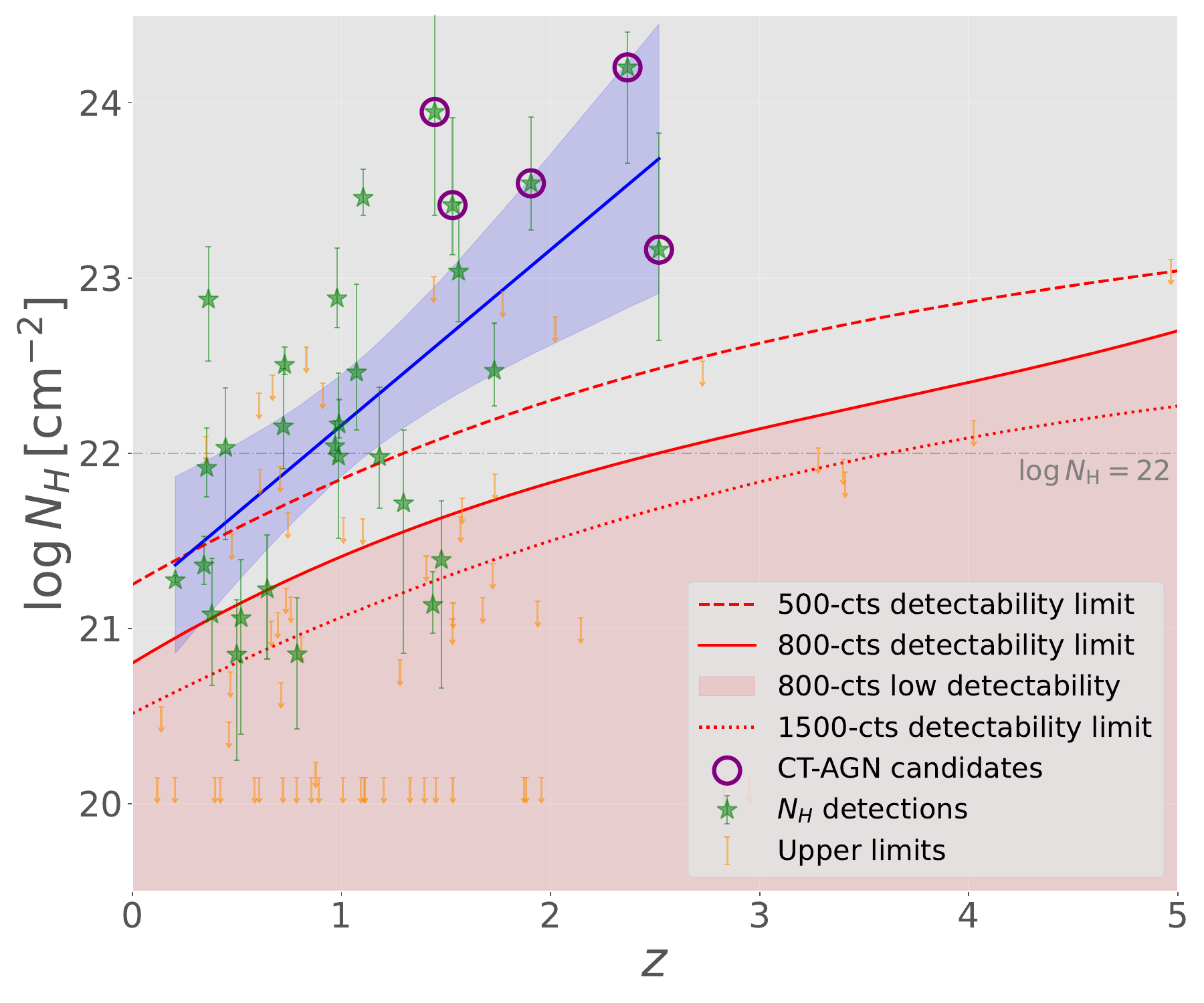}
\caption{Hydrogen column density distribution as a function of redshift for our AGN sample. Green stars, yellow arrows, and purple circles indicate sources with robust $N_{\mathrm{H}}$ measurements, only $N_{\mathrm{H}}$ upper-limits, and Compton-thick AGN candidates, respectively. The shaded blue region represents the $1\sigma$ confidence interval of the linear regression fit (blue line) for the sources with robust $N_{\mathrm{H}}$ measurements. The red curves show the corresponding 50 per cent detectability boundary, for sources with about 500 (dashed-line), 800 (solid-line) and 1500 (dotted-line) X-ray counts in the $0.3 - 10\, \mathrm{keV}$ band.  The shaded red region represents the 800 counts low or no detectability $N_{\mathrm{H}}$ region.}
\label{NH_z}
\end{figure}

Fig. \ref{NH_z} presents the intrinsic hydrogen column density distribution as a function of redshift of the AGN sample. Green stars are those objects with robust $N_{\mathrm{H}}$ measurements at the 90 per cent confidence level (i.e. $\rho > 10$). The data reveal a positive correlation described by Equation \ref{logNHz}, with $N_{\mathrm{H}}$ increasing with redshift, as highlighted by the linear fit and its associated $95$ per cent confidence interval (blue shaded region). The candidate CT-AGN population is located at high redshifts ($z > 1.4$) marked with purple circles. Yellow arrows are those sources with $N_{\mathrm{H}}$ upper-limits and are not considered for the linear regression.

\begin{equation}
    \label{logNHz}
    \log N_{\mathrm{H}} = (1.0 \pm 0.5)z + (21.2 \pm 0.6)
\end{equation}

To quantify the range of intrinsic column densities that can be robustly recovered in the XMM-LH field as a function of redshift, we performed 40,000 spectral simulations using the \texttt{fakeit} command in \textsc{xspec}. The red curves in Fig. \ref{NH_z} show the 50 per cent detectability limits derived for representative simulated spectra  with around 500,  800, and 1500 counts in the $0.3 - 10\, \mathrm{keV}$ band, that represent the faint (dashed-line), average (solid-line) and bright (dotted-line) sources in our sample,  respectively. For each redshift, these limits are defined as the minimum intrinsic column density above which at least 50\% of the simulated spectra satisfy our absorption-detection criteria at the 90\% confidence level (i.e. statistically significant and  well bounded lower limit). A full description of the simulation procedure and the derivation of these detectability curves is provided in Appendix \ref{detectability}.  

This Figure shows that the detectability threshold depends strongly on source counts. In particular, the separation between the 500 and 800 count curves is substantial, indicating that for the faintest sources ($\lesssim$500 counts), moderate absorption ($\log N_{\mathrm{H}} \lesssim 22\,\mathrm{cm^{-2}}$) becomes increasingly difficult to constrain reliably, especially at higher redshift. The shaded red region below the 800 count curve marks the parameter space where intrinsic absorption is not expected to be robustly measurable for a typical source in our sample.

This redshift-obscuration dependence is consistent with previous results from deep X-ray surveys. For example, \citet{Liu2017}, using the 7 Ms \textit{Chandra} Deep Field-South, and \citet{Iwasawa2020}, with the \textit{XMM-Newton} deep survey of the same field, both reported a systematic increase in the fraction of obscured AGNs with redshift. \citet{Liu2017} found that the obscured AGN fraction follows the relation $f_{\mathrm{obs}} \approx 0.42 \times (1+z)^{0.60}$, and their analysis also confirmed that more luminous AGNs tend to be less obscured. Similar results were reported by \citet{Iwasawa2020}, based on a homogeneous X-ray spectral analysis of 185 bright sources up to $z \sim 3.8$.

These studies support the idea that AGNs at earlier cosmic epochs tend to be more heavily obscured, likely due to the higher gas and dust content in their host galaxies and surrounding environments \citep[e.g.][]{Gilli2022}. Our results also show an increasing obscured fraction with redshift. Although, since our sample is not based on a well-defined flux-limited selection, it may be biased by selection effects as observed in the different detectability curves of Fig. \ref{NH_z}. For instance, at high $z$, it is expected to detect the brightest AGNs with lower column densities (as predicted by the receding torus model). In addition, rest-frame harder X-rays are more likely to enter the observed band, further affecting obscuration estimates. Thus at higher redshift, a larger fraction of obscured AGNs enter the \textit{XMM-Newton} band. This mimics the increased obscured fraction with redshift. However, the increasing obscured AGN fraction with redshift is observed even after correcting for this effect \citep[e.g.][]{Ueda2014,Peca2023}. Finally, since our sources are FIR-selected, some host galaxies are dominated by star-forming activity (see  Section \ref{sfr}), which could introduce additional biases. Indeed, \citet{Gonzales2024b} reported that 78 per cent of the initial spectroscopic sample of 340 non-AGN host galaxies in the Lockman-SpReSO catalog are of the starburst type.

\subsection{Optical and X-ray AGN Classifications }  

We compared our X-ray classifications with the optical spectroscopic classifications of AGNs in the Lockman-SpReSO sample, reported by \citet{Negrete2025} (Type-I) and Assefa et al. (in preparation, Type-II). We identified 44 sources in common, 27 classified as optical Type-I and 17 as Type-II. Overall, we found a 73 per cent agreement between the optical and X-ray classifications, consistent with previous X-ray surveys (e.g., \citealt{Brusa2010,Ordov2017}). However, while all optical Type-I AGNs in our sample are unobscured in X-rays, a significant discrepancy arises for Type-II sources. Specifically, only 5 out of 17 (30 per cent) optical Type-II AGNs show X-ray obscuration with $N_{\mathrm{H}} > 10^{22}\, \mathrm{cm^{-2}}$, whereas the remaining 12 (70 per cent) display little or no X-ray absorption. Similar mismatches between optical and X-ray classifications were also reported in early analyses of the XMM-LH by \citet{Mainieri2002} and \citet{Mateos2005}. 

The origin of these discrepancies remains debated. One explanation is host-galaxy dilution: since our sources are FIR-selected, some may be dominated by star-forming activity, which can bias the optical classification as stellar or star-forming emission outshines AGN signatures, particularly at low AGN luminosities or high redshifts \citep{Brusa2010,Marchesi2016}. Another possibility is a non-uniform, clumpy torus structure, where the line of sight to the broad-line region is obscured while the more compact X-ray-emitting region remains visible, producing divergent optical and X-ray classifications.

\subsection{Compton-thick AGNs} 

\subsubsection{XCLUMPY model} 
\label{xclum}

For heavily obscured sources approaching or exceeding the Compton-thick regime ($N_{\mathrm{H}} \gtrsim 10^{24}\, \mathrm{cm}^{-2}$),  a more detailed spectral analysis is required, since the model \texttt{phabs} alone is not well suited to manage Compton-thick obscuration.

In the previous section, we identified five Compton-thick candidates based on their preliminary column density estimates. To gain deeper insight into these sources, we conducted a more complex spectral analysis using the XCLUMPY\footnote{\url{https://github.com/AtsushiTanimoto/XClumpy}} model \citep{Tanimoto2019}, a sophisticated modelling designed to characterize the clumpy torus structure in AGNs through Monte Carlo simulations. This model adopts a toroidal geometry consistent with the CLUMPY infrared model \citep{Nenkova2008a}, where clumps follow a power-law distribution in the radial direction and a normal distribution in the axial direction. 

The XCLUMPY model is presented in Equation \ref{xc}, composed of the following three main components: 

 \begin{itemize} 

\item \texttt{phabs*cabs*zcutoffpl}: Represents the transmitted continuum through the torus. 

\item \texttt{xclumpy\_R}: Models the reflection continuum. 

\item \texttt{xclumpy\_L}: Accounts for fluorescence line emission. 

\end{itemize}

Both the reflected and fluorescent components are parameterized by the equatorial hydrogen column density ($N_{\mathrm{H}}^\mathrm{Equ}$), the torus angular width ($\sigma_\mathrm{tor}$), and the inclination angle $(i)$. These parameters are intrinsically linked to the properties of the primary continuum, including the photon index and cutoff energy. 

\begin{equation}
    \label{xc}
    \begin{split}
        \texttt{phabs}*( & \texttt{zphabs}*\texttt{cabs}*\texttt{zcutoffpl} + \texttt{constant}*\texttt{zcutoffpl} \\
        & + \texttt{atable\{xclumpy\_R.fits\}} \\
        & + \texttt{atable\{xclumpy\_L.fits\}} )
    \end{split}
\end{equation}

The \texttt{zcutoffpl} component models the intrinsic continuum as a power-law with an exponential cutoff. 
In Eq. \ref{xc}, the first term in the parentheses represents the primary continuum transmitted through the torus and thus is subject to photoelectric absorption (\texttt{zphabs}) and Compton scattering (\texttt{cabs}) with the line-of-sight column density $N_\mathrm{H}^\mathrm{Los}$.  The second term is the scattered component where \texttt{const} is the scattered fraction. The XCLUMPY model is composed of two components: Reflection continuum (\texttt{xclumpy\_R.fits}) and emission lines (\texttt{xclumpy\_L.fits}). The power-law indices and normalizations of all components are tied together, and the high-energy cutoff is fixed at 100\,keV. From the equatorial column density, we can derive the line-of-sight column density $N_\mathrm{H}^\mathrm{Los}$ using Equation \ref{Nlos}, as follows: 

\begin{equation}
\label{Nlos}
    N_\mathrm{H}^\mathrm{Los} = N_\mathrm{H}^\mathrm{Equ} \left [ \exp \left (- \frac{(i - 90\degree)^2}{\sigma^2_\mathrm{tor}} \right) \right] 
\end{equation}

We modelled the X-ray spectra of our Compton-thick AGN candidates using the XCLUMPY model, following the approach of \citet{Miyaji2019}. For each source, we performed a Markov Chain Monte Carlo (MCMC) analysis with a chain length of 40,000 steps to explore the parameter space (\texttt{chain} command in \textsc{xspec}). The best-fit parameters were taken as the median of the posterior distributions, while the uncertainties were estimated from the 5th and 95th percentiles (i.e. $\sim$90 per cent confidence intervals). In cases where the photon index could not be constrained, $\Gamma$ was fixed to 1.8.

We define strong CT-AGN candidates as sources with median column densities above $10^{24}\,\mathrm{cm}^{-2}$ and lower confidence interval consistent with the Compton-thick regime. Sources with column densities around this threshold but with uncertainties extending significantly below $10^{24}\,\mathrm{cm}^{-2}$ are classified as tentative or borderline CT-AGN candidates.

\subsubsection{XCLUMPY fitting results}

Our analysis identifies one strong Compton-thick AGN candidate, objID-206545, with both equatorial and line-of-sight column densities exceeding $10^{24}\,\mathrm{cm}^{-2}$. Two sources, objID-206463 and objID-206603, are best classified as borderline CT-AGN candidates, as their best-fit column densities are consistent with the Compton-thick regime but their uncertainties extend below $10^{24}\,\mathrm{cm}^{-2}$. The remaining sources, objID-206524 and objID-206642, are heavily obscured but consistent with Compton-thin absorption, with $N_\mathrm{H}^{\mathrm{Los}} \sim 5 \times 10^{23}\,\mathrm{cm}^{-2}$.

\begin{figure*}
\centering
\includegraphics[scale=0.24]{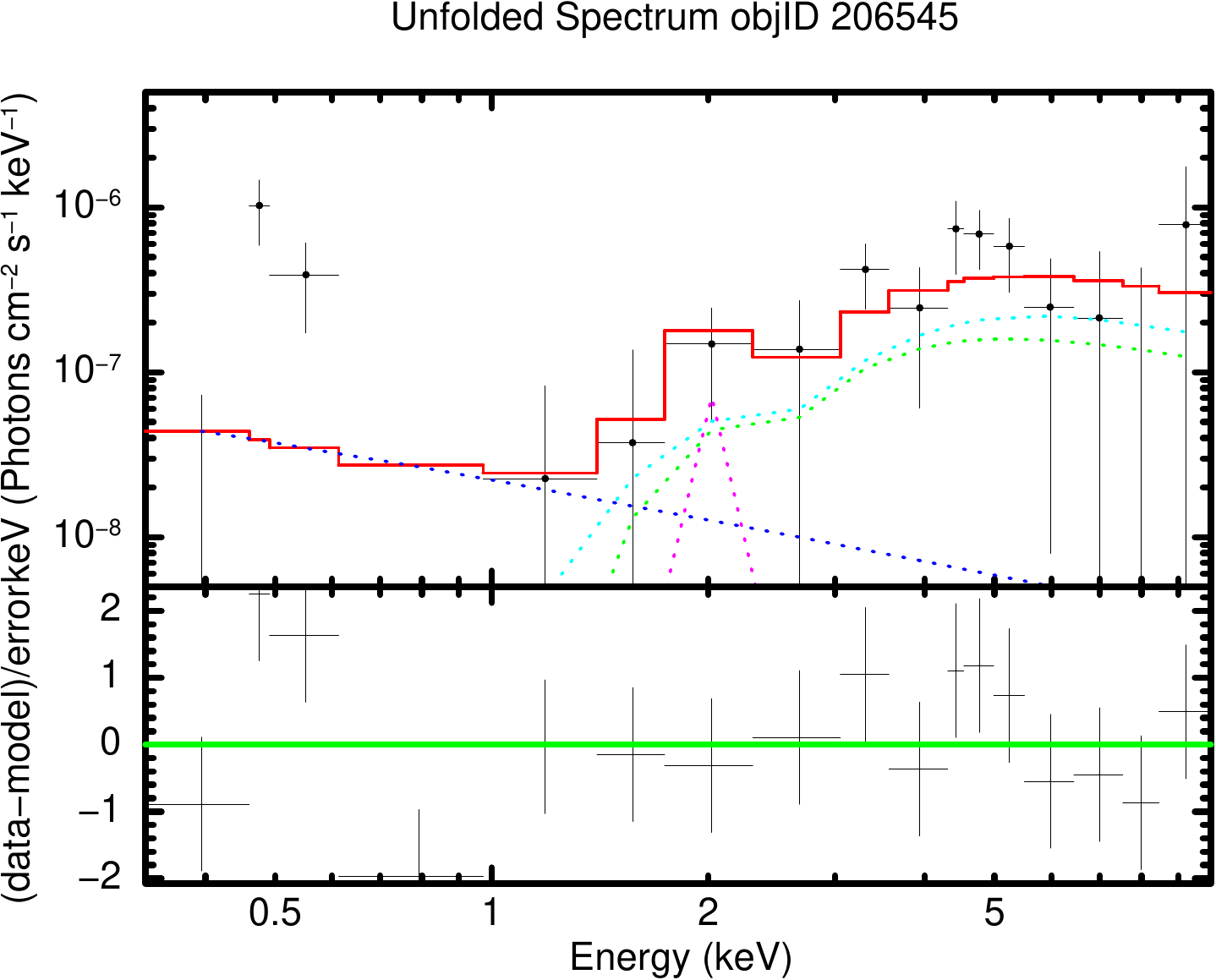}
 \includegraphics[scale=0.24]{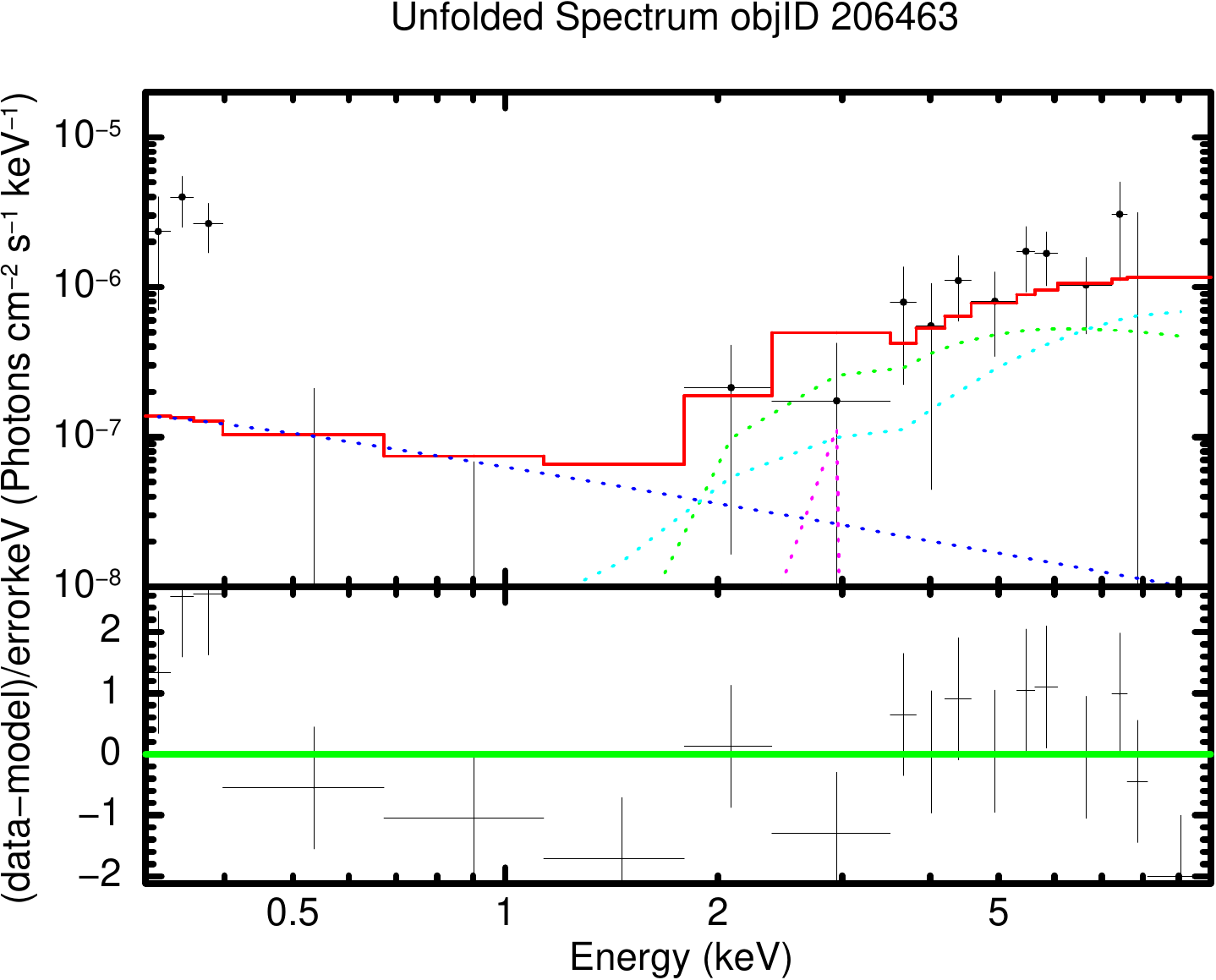}
 \includegraphics[scale=0.24]{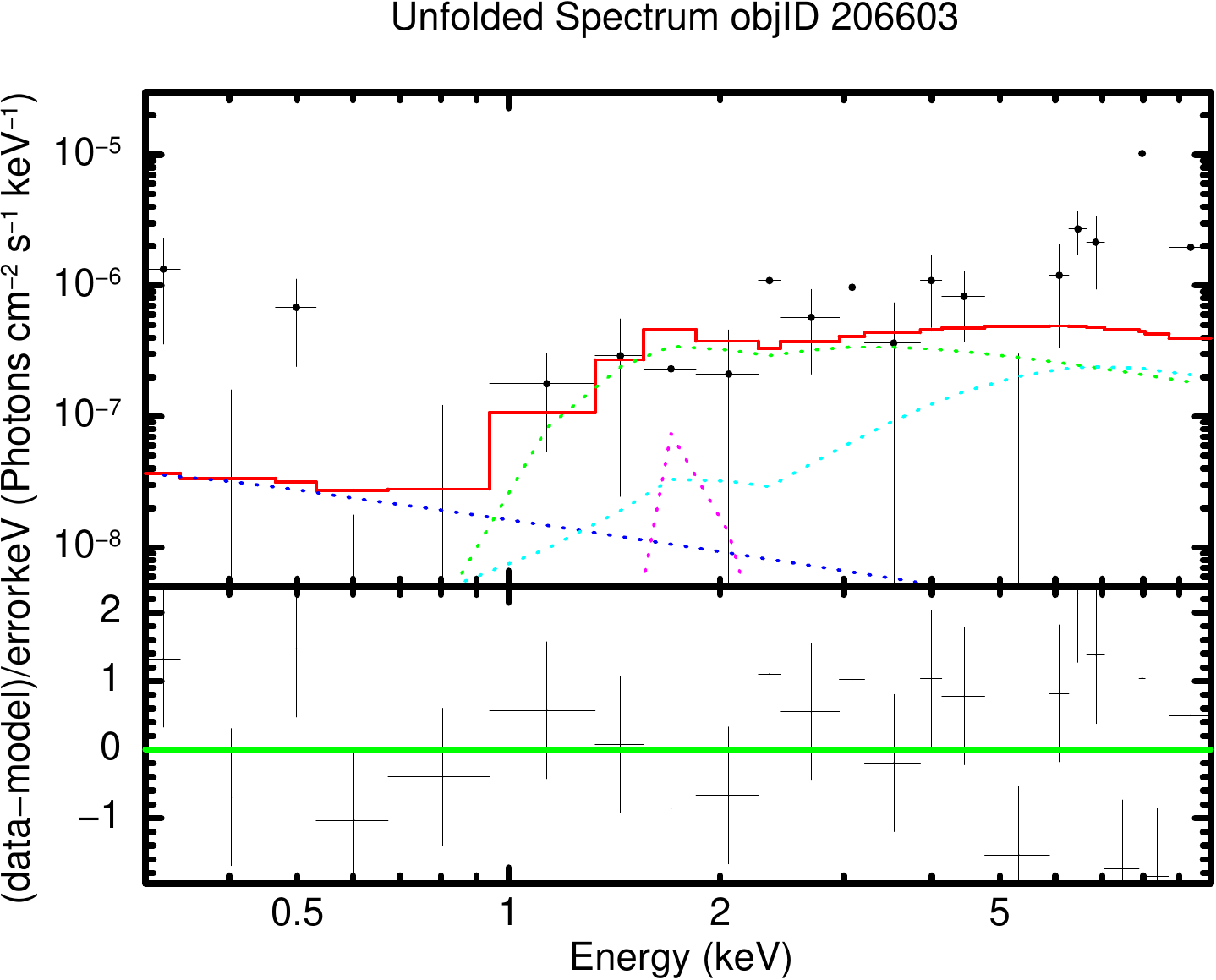}
 \includegraphics[scale=0.21]{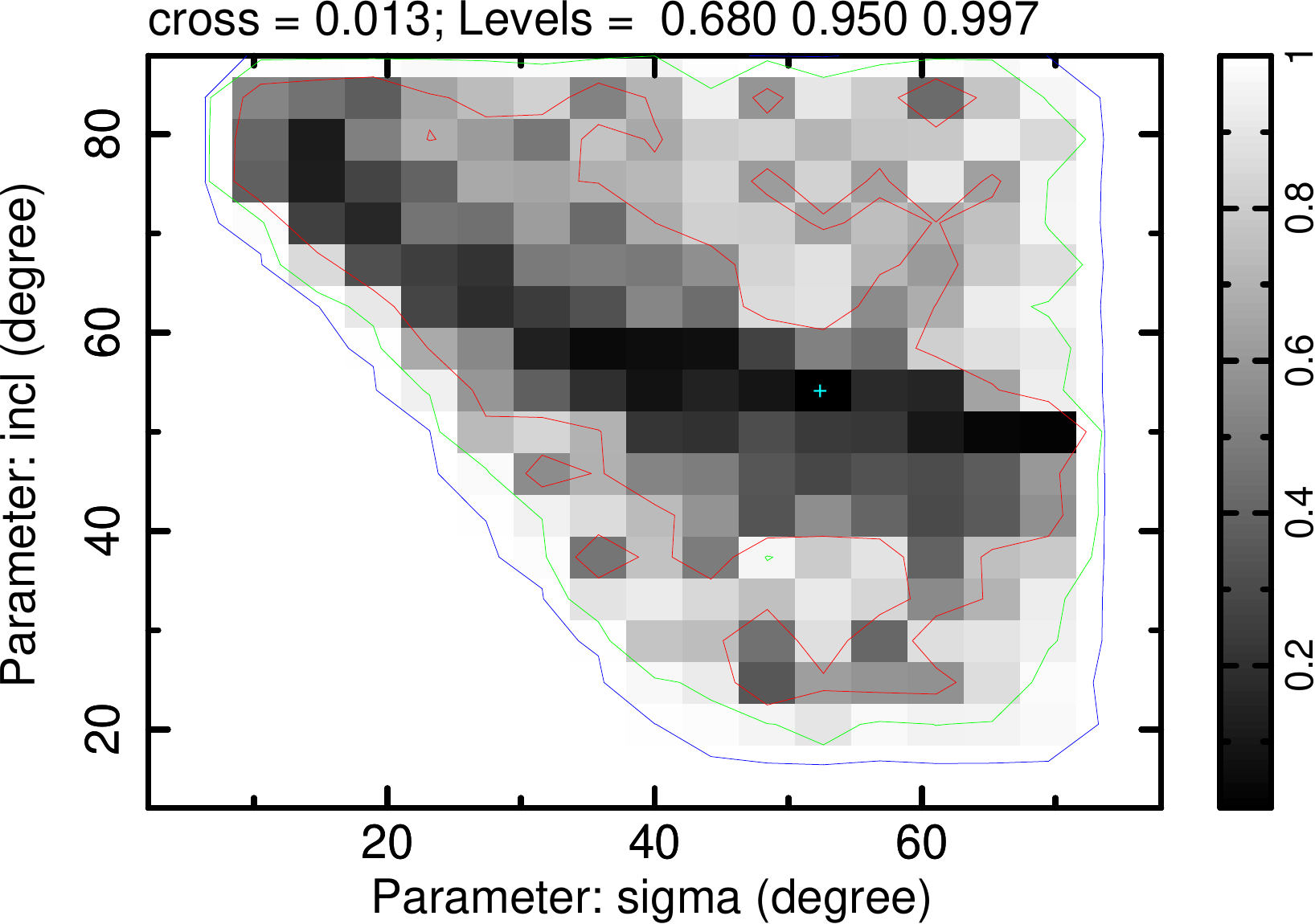}
 \includegraphics[scale=0.218]{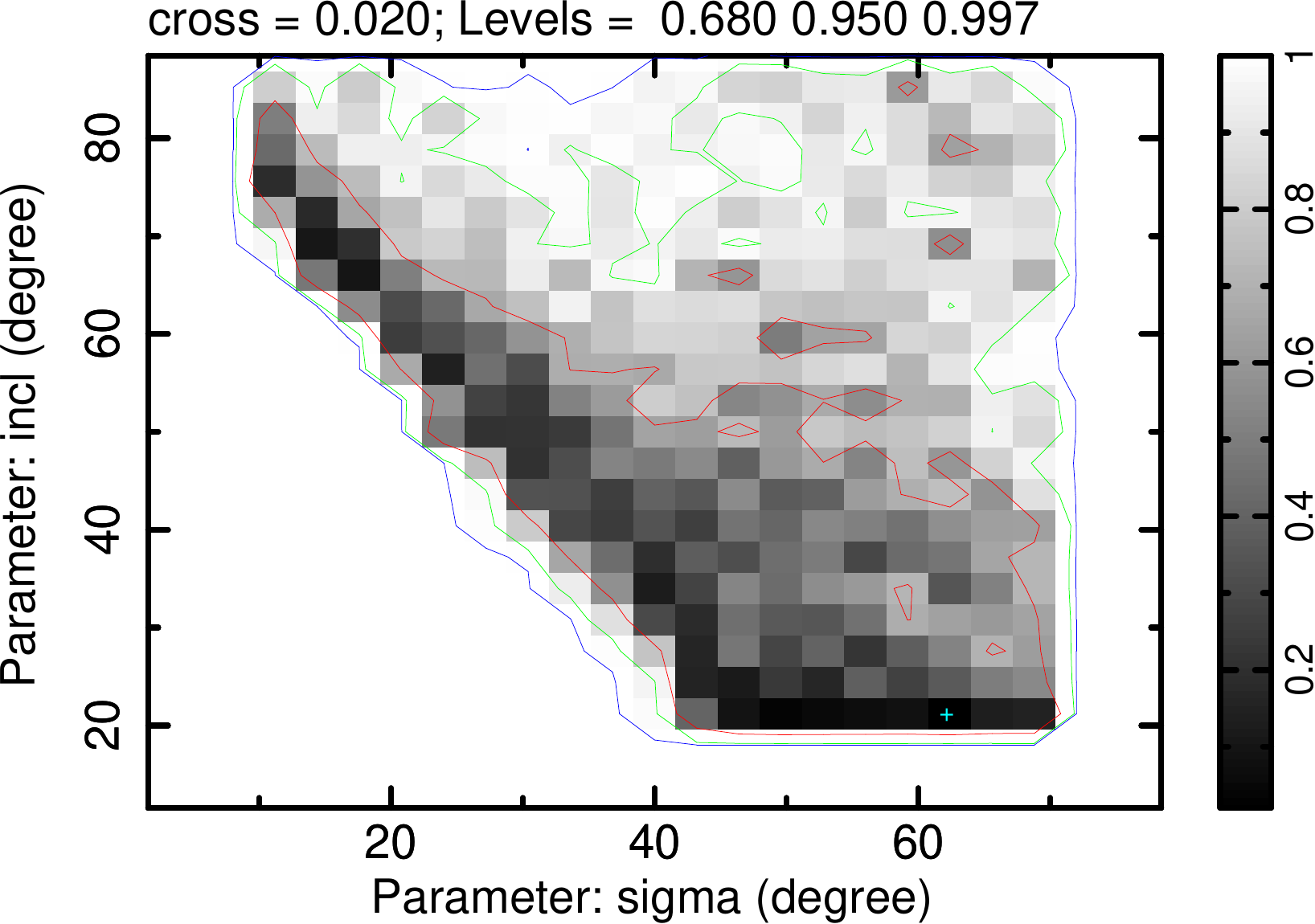}
 \includegraphics[scale=0.21]{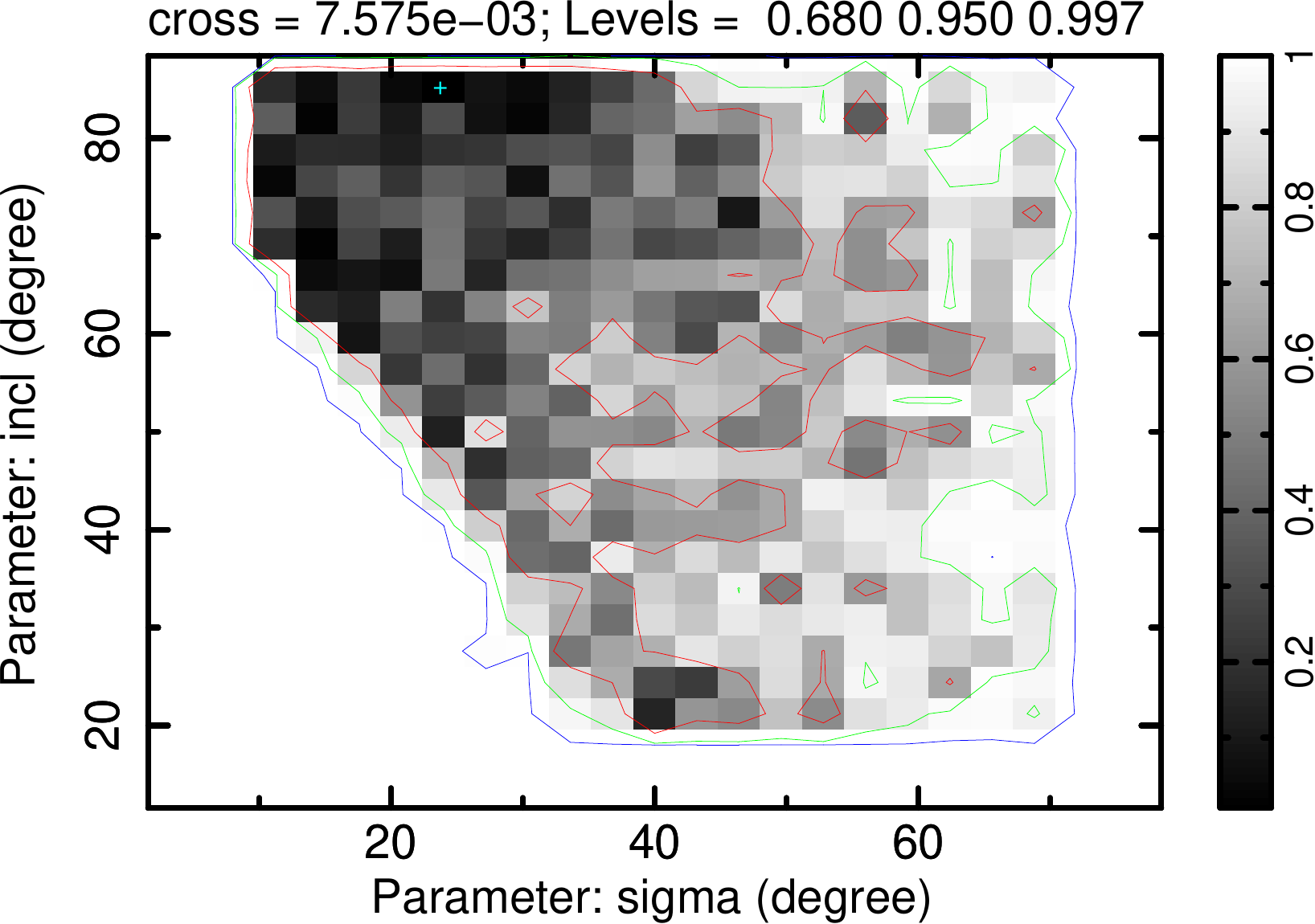}
    \caption{Upper-panels: Unfolded X-ray spectra and residuals of the three strongest CT-AGN candidates modelled with XCLUMPY. The black points with error bars represent the binned data, as described in Section \ref{Xspectral}. The red lines correspond to the best-fitting models. Individual spectral components are overplotted: the intrinsic X-ray emission (blue dot-dashed line), reflected continuum (green dot-dashed line), Fe K$\alpha$ emission line (pink dot-dashed line), and the scattered component (cyan dot-dashed line). The  model provides an acceptable fit to the data ($\chi^2_{\mathrm{red}} \approx 1$), with both the equatorial and the line-of-sight column densities consistent with the Compton-thick regime $N_{\mathrm{H}} \gtrsim 10^{24}\, \mathrm{cm^{-2}}$. Lower-panels: Contour plots of the torus inclination angle versus the torus angular width from the MCMC XCLUMPY model. The contours represent the joint confidence intervals at 68.3 per cent (blue), 90 per cent (green), and 99 per cent (red). For visual clarity, the spectra shown in this Figure were rebinned for display purposes only.}
\label{CTAGNs}
\end{figure*}

Fig. \ref{CTAGNs} in the upper-panels shows the unfolded X-ray spectra of the strongest CT-AGN candidates modelled with XCLUMPY. Among them, objID-206545 in the left-panel represents our strongest CT-AGN candidate, with equatorial and line-of-sight column densities of $\log\,N_\mathrm{H}^\mathrm{Equ} = 24.5^{+0.41}_{-0.43}\, \mathrm{cm}^{-2}$ and $\log\, N_\mathrm{H}^\mathrm{Los} = 24.2^{+0.41}_{-0.35}\ \mathrm{cm}^{-2}$, respectively. The fit provides an acceptable description of the data, with a reduced chi-square of $\chi^{2}_{red} \approx 1$. The lower-panels show the MCMC joint confidence contours for the torus angular width and the inclination angle. For instance, for objID-206545 we obtain a torus angular width of $\sigma_\mathrm{tor} = 45.4^{+22.4}_{-29.6}\deg$ and an inclination angle of $i = 58^{+26}_{-31}\deg$. These values suggest a line of sight intersecting the obscuring torus, consistent with the high level of X-ray obscuration inferred from the spectral fit. Among the five sources analysed, only objID-206642 and objID-206524 show marginal detections of the Fe K$\alpha$ emission line, estimated using the model \texttt{TBabs(powerlaw*zphabs + zgauss)}. For the remaining sources, the Fe K$\alpha$ equivalent widths could not be robustly constrained given the current data quality, and only upper limits could be derived, as reported in column 10 of Table \ref{CT_table}.

Furthermore, since the strongest CT-AGN candidates, objID-206603 and objID-206545, have secure spectroscopic redshifts of $z_\mathrm{spec}=2.519$ and $z_\mathrm{spec}=2.368$, respectively,  we include their optical spectra in Fig. \ref{A1} of Appendix \ref{Optical}. The spectra show clear narrow emission lines, such as Ly$\alpha$ ($\lambda1216$ \AA), [N\,{\sevensize V}] ($\lambda1240$ \AA), and C\,{\sevensize IV} ($\lambda1549$ \AA), supporting the reported redshift estimate and the reliability of the X-ray spectral analysis. Additionally, in Appendix \ref{Appendix_SED} the Figures \ref{SEDs1} and \ref{SEDs2} present the spectral energy distributions of the three strongest CT-AGN candidates\footnote{The spectral energy distributions were derived from the multiwavelength photometric analysis of the Lockman-SpReSO field by Herrera-Endoqui et al. (in preparation), using optical-to-far-infrared photometry.}.

All three sources present infrared emission consistent with dust reprocessing by an active nucleus. The AGN contribution derived from the CIGALE SED fitting, quantified through the AGN fraction ($f_{\rm AGN}$), i.e. the fraction of the total infrared luminosity attributed to the AGN component, is highest for the strongest CT-AGN candidate, objID-206545, with $f_{\rm AGN}=0.33 \pm 0.09$. The two borderline CT-AGN candidates show lower AGN contributions, with $f_{\rm AGN}=0.18 \pm 0.08$ and $0.11 \pm 0.04$ for objID-206463 and objID-206603, respectively. We also note that objID-206603 is hosted by the galaxy with the highest star-formation rate in the sample ($\sim 10^3\,M_\odot\,\mathrm{yr^{-1}}$). These results indicate that the infrared emission of these systems is affected by both AGN activity and host-galaxy star formation, as discussed in Section \ref{sfr}.

We tested the impact of fixing the torus inclination and torus angular width to $i = 60^\degree$ and $\sigma_\mathrm{tor} = 45^\degree$, respectively, on the estimated column densities. In general, this configuration resulted in slightly lower equatorial column densities and modest increase in the line-of-sight column densities. Under this configuration, the first two sources (objID-206524 and objID-206642) remain consistent with heavily obscured but Compton-thin AGNs with line-of-sight column densities of $\log N_\mathrm{H}^{\mathrm{Los60}} \sim 23.7\, \mathrm{cm^{-2}}$, while the remaining three sources maintain column densities consistent with the Compton-thick regime ($\log N_\mathrm{H} \gtrsim 24 \, \mathrm{cm^{-2}}$), although with significant uncertainties in some cases. These results indicate that, while the inferred column densities depend on the assumed torus geometry, the classification of the sources as strong and borderline CT-AGN candidates remains broadly robust. A summary of the spectral fitting results obtained with the XCLUMPY model is presented in Table \ref{CT_table}.

\begin{table*}
\caption{Best-fit spectral parameters for the  Compton-thick AGN candidates derived using the XCLUMPY model. The reported column densities and angles correspond to the median of the MCMC analysis with a chain length of 40,000 steps, while the uncertainties represent the 5th and 95th percentiles (i.e. $\sim$90 per cent confidence intervals). Equatorial ($N_\mathrm{H}^\mathrm{Equ}$) and line-of-sight ($N_\mathrm{H}^\mathrm{Los}$) column densities are listed  in columns 3 and 5, respectively. Columns 4 and 6 report the corresponding column densities assuming fixed torus parameters of  $i= 60^\degree$ and $\sigma_\mathrm{tor}=45^\degree$. The photon index ($\Gamma$), torus angular aperture ($\sigma$), inclination angle ($i$), and iron line Equivalent width  are also presented in columns, 7, 8, 9, and 10, respectively.}
\label{CT_table}
\begin{tabular}{@{}ccccccccccc@{}} 
\toprule
objID & $z$ & $\chi^2/\mathrm{dof}$ & $\log\, N_\mathrm{H}^\mathrm{Equ}$ & $\log\, N_\mathrm{H}^\mathrm{Equ60}$ & $\log\, N_\mathrm{H}^\mathrm{Los}$ & $\log\, N_\mathrm{H}^\mathrm{Los60}$ &$\Gamma$ & $\sigma_\mathrm{tor}$ & $i$ & EW$^c$ \\ 
 & & & $\mathrm{cm^{-2}}$ & $\mathrm{cm^{-2}}$ & $\mathrm{cm^{-2}}$ & $\mathrm{cm^{-2}}$ & & deg & deg & keV \\ 
 & (1) & (2) & (3) & (4) & (5) & (6) & (7) & (8) & (9) & (10)\\ \midrule
206524 & $1.10^a$ &41.3/41 & $24.3^{+0.61}_{-0.52}$ & $23.9^{+0.08}_{-0.10}$ & $23.7^{+0.09}_{-0.10}$ & $23.7^{+0.08}_{-0.10}$ & $1.95^{+0.19}_{-0.22}$ & $40.8^{+24.6}_{-19.2}$ & $38.4^{+37.49}_{-16.49}$ & $0.12^{+0.08}_{-0.08}$ \\
206642 & $1.53^a$ &50.4/45 & $24.4^{+0.54}_{-0.60}$ & $23.9^{+0.20}_{-0.19}$ & $23.7^{+0.23}_{-0.20}$ & $23.7^{+0.20}_{-0.19}$ & $1.8^b$ & $40.1^{+23.16}_{-21.95}$ & $38.4^{+38.81}_{-17.53}$ & $0.22^{+0.08}_{-0.26}$ \\
206463 & $1.45^a$& 50.5/28 & $24.5^{+0.40}_{-0.46}$ & $24.4^{+0.33}_{-0.29}$ & $24.1^{+0.33}_{-0.30}$  & $24.2^{+0.33}_{-0.30}$ & $1.8^b$ & $47.2^{+19.91}_{-30.91}$ & $47.8^{+33.34}_{27.48}$ & < 1.2 \\
206603 & 2.519 & 38.0/30 & $24.7^{+0.23}_{-0.68}$ & $24.7^{+0.30}_{-0.65}$ & $24.3^{+0.55}_{-0.82}$  & $24.5^{+0.30}_{-0.65}$ & $1.8^b$ & $34.9^{+30.91}_{-17.19}$ & $62.6^{22.31}_{35.45}$ & < 0.6 \\
206545 & 2.368& 37.9/38 & $24.5^{+0.41}_{-0.43}$ & $24.5^{+0.43}_{-0.33}$ & $24.2^{+0.41}_{-0.35}$  & $24.3^{+0.43}_{-0.33}$ & $1.8^b$ & $45.4^{+22.42}_{-29.61}$ & $58.0^{25.81}_{31.02}$ & < 1.0 \\ \bottomrule
\end{tabular}
\begin{flushleft}
\footnotesize 
$^a$ Photometric redshift\\
$^b$ parameter fixed during the fit\\
$^c$ Estimated using the model \texttt{TBabs(powerlaw*zphabs + zgauss)}
\end{flushleft}
\end{table*}

We also compared our sample with the hardness-ratio selection proposed by \citet{Brunner2008}, who identified candidate Compton-thick AGNs using the criteria $HR1 < -0.1$ and $HR2 > 0.1$, where $HR1$ and $HR2$ compare the count rates between the $0.5-2.0$, $2.0-4.5$, and $4.5-10.0$ keV bands. From the 13 sources reported by \citet{Brunner2008} satisfying these conditions, we found five sources in common (objID-206776, objID-206649, objID-206401, objID-173466, and objID-206715). Interestingly, none of these objects are classified as CT-AGN candidates based on our spectral fitting. Instead, several of them exhibit spectrally complex X-ray continua that cannot be adequately described using simple absorbed power-law models.

For instance, one of them, objID-206776, was analysed individually in Section \ref{new_spec} using the relativistic reflection model \texttt{relxill}, resulting in a significantly improved fit and suggesting that its spectrum is likely dominated by reflection components. This may indicate that the hardness-ratio selection of \citet{Brunner2008} preferentially identifies AGNs with complex reflection-dominated spectra rather than sources that can be described solely by neutral absorption.

\subsection{Fe K$\alpha$ line emission}

The fluorescent Fe K$\alpha$ emission line at 6.4 keV, associated with reprocessed X-ray radiation in cold material (e.g., the molecular torus or accretion disc), is a characteristic X-ray spectral feature observed in many bright AGNs. A well-established anti-correlation exists between the iron line equivalent width ($EW$) and the X-ray luminosity, known as the X-ray Baldwin effect (or Iwasawa-Taniguchi effect). Originally identified in unobscured AGNs by \citet{Iwasawa1993}, this trend is often interpreted as a consequence of a luminosity-dependent decrease in the torus covering factor \citep{Bianchi2007,Ricci2014}.

In our X-ray spectral analysis, we modelled the Fe K$\alpha$ line using a narrow Gaussian component (\texttt{zgauss}). Following the methodology of \citet{Iwasawa2020}, the line was fixed at 6.4 keV in the rest-frame for sources with spectroscopic redshifts and allowed to vary within $6.1-6.6$ keV for those with photometric redshifts. For 49 AGNs, we obtained reliable $EW$ measurements with $EW > 0.02 \,\mathrm{keV}$ with errors at 90 per cent of confidence level. In cases where the \texttt{error} command failed to return uncertainties on the $EW$, we employed an MCMC analysis using the \texttt{chain} command with a length of 50,000 steps. For the remaining sources, upper limits on the $EW$ were reported. Since a large fraction of our sample has measured Fe K$\alpha$ equivalent widths, the following regression analysis was performed excluding sources with only upper limits.

\begin{figure}
\centering
\includegraphics[scale=0.4]{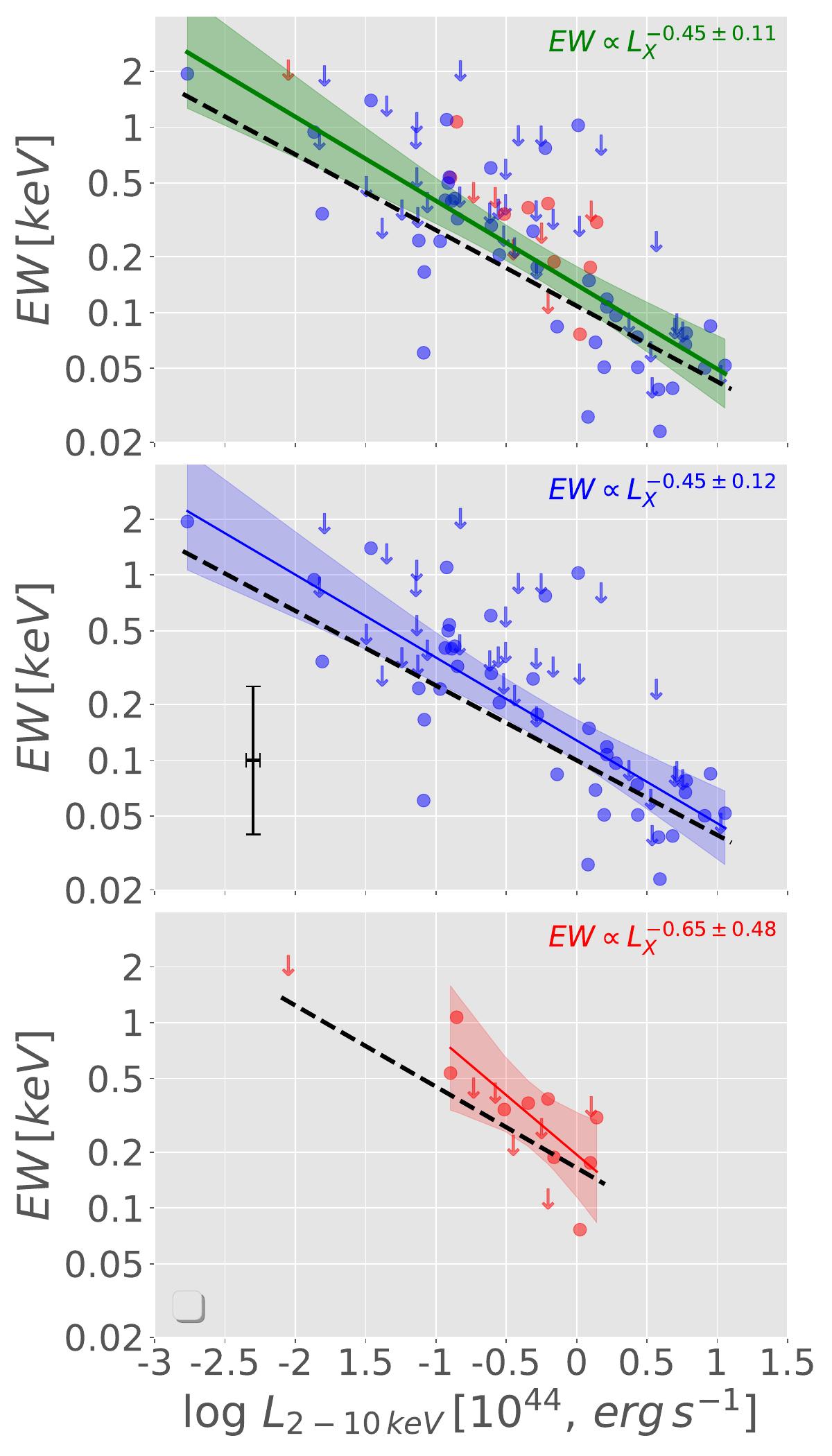}
\caption{Distribution of the Fe K$\alpha$ equivalent width as a function of hard X-ray luminosity for both obscured (red circles) and unobscured (blue circles) AGN populations. The upper panel shows the full sample, while the central and lower panels display the unobscured and obscured subsamples, respectively. The solid lines represent the linear regression fits for each sample without upper-limits, with the shaded regions indicating the 95 per cent confidence intervals in the regression fits. The black dotted lines represent the linear regression considering upper limits. The average error bar at 90 per cent of confidence for the whole sample is reported in the central panel.}
\label{Lx_ew}
\end{figure}

The upper panel in Fig. \ref{Lx_ew} shows the distribution of $EW$ as a function of the obscuration-corrected X-ray luminosity in the $2-10$ keV band of our sample. Unobscured and obscured AGNs are represented by blue and red circles, respectively. The green solid line indicates a linear regression fit using weighted least squares, while the green shaded region corresponds to the 95 per cent confidence interval. Central and lower panels show the individual distributions of both populations, unobscured and obscured AGNs, respectively.

For unobscured AGNs, we find a strong and statistically significant anti-correlation, described by the relation presented in Equation \ref{ew_Lx_unobs}, with a Pearson correlation coefficient of $r = -0.77$ and a $p$-value of $7.35 \times 10^{-9}$, indicating a significance above the 99 per cent confidence level.

\begin{equation}
\label{ew_Lx_unobs}
\log(EW) = (-0.45 \pm 0.12) \log\left(\frac{L_{2-10\,\mathrm{keV}}}{10^{44}\,\mathrm{erg\,s}^{-1}}\right) - (0.90 \pm 5.4)
\end{equation}

We note that this anti-correlation is also observed in the obscured AGN population, although these sources present systematically higher $EW$ values at a given luminosity. The corresponding linear relation is presented in Equation \ref{ew_Lx_obs}, with the same Pearson correlation coefficient ($r = -0.77$) and a statistically significant $p$-value of 0.016 (95 per cent confidence level). 

\begin{equation}
\label{ew_Lx_obs}
\log(EW) = (-0.65 \pm 0.48) \log\left(\frac{L_{2-10\,\mathrm{keV}}}{10^{44}\,\mathrm{erg\,s}^{-1}}\right) - (0.70 \pm 21),
\end{equation}

Finally, we test the contribution of the $EW$ upper limits in our linear regression, by applying a survival analysis using the \texttt{ASURV} software\footnote{\url{https://github.com/rsnemmen/asurv}} \citep{Feigelson1985,Isobe1986}, which has been extensively tested with censored data in multiwavelength studies of AGNs \citep[e.g.,][]{Steffen2006,Lusso2010,Castro2014,Castro2025}. For our analysis, we employed the EM (Estimate and Maximize) algorithm for linear regression of bivariate data. The EM method estimates the regression coefficients under the assumption of normally distributed residuals \citep{Wolynetz1979}. Our results are shown as black dotted lines in each panel of Fig. \ref{Lx_ew}, maintaining a consistent trend with those obtained using only detections, though slightly shifted toward the lower boundary of the confidence interval region.

Both distributions show comparable slopes, where $EW$ decreases as $\propto L_{2-10\,\mathrm{keV}}^{-0.5}$. These results suggest that the physical origin of the EW-luminosity anti-correlation in both unobscured and obscured AGNs may arise from the same mechanism. The similar slopes of both anti-correlations indicate that obscuration does not alter the fundamental driver. This could be attributed to the receding torus effect, higher luminosities cause dust sublimation at larger radius, reducing the torus covering fraction and thus diminishing reprocessed Fe K$\alpha$ emission. 

In a few objects, we measure relatively large Fe K$\alpha$ equivalent widths ($EW \sim 1\,\mathrm{keV}$), which are often associated with reflection-dominated spectra in heavily obscured or Compton-thick AGNs \citep{Lanzuisi2015}. However, the fitted column densities and photon indices for these sources do not indicate a clear Compton-thick regime. One possible explanation is that the limited photon statistics prevent a robust decomposition between the reflection continuum and the Fe line emission, causing part of the reflection component to be modelled as a strong Gaussian line.

We tested the inclusion of a reflection component using the \texttt{pexrav} model; however, in most cases the reflection strength is not well constrained. Therefore, the large $EW$ values should be interpreted with caution and may reflect spectral complexity that cannot be fully resolved in some low count statistic sources.

\section{Dusty Torus Structure and fraction of obscured AGNs} 
\label{torus}

\subsection{Torus angular width $\sigma_\mathrm{tor}$}

The torus angular width $(\sigma_\mathrm{tor})$ and the dust covering factor $(f_{\mathrm{cov}})$ are critical parameters for probing the geometry and physical properties of AGN torus structure. $\sigma_\mathrm{tor}$ is defined as the angle between the equatorial plane and the torus edge, while $f_{\mathrm{cov}} =\cos(90\degree - \sigma_\mathrm{tor})$ represents the fraction of the AGN obscured by the dusty torus \citep[e.g.][]{Nenkova2008a,Ricci2013}. 

For our analysis, we used the $\sigma_\mathrm{tor}$ derived by Herrera-Endoqui et al. (in prep.) through SED multicomponent decomposition analysis for the complete Lockman-SpReSO survey in the UV to FIR wavelength regime, using the code CIGALE \citep{Boquien2019}.  Briefly, they used their implementation of the Nenkova clumpy dusty torus emission models \citep{Nenkova2008a,Nenkova2008b} to fit the AGN contribution \citep[see e.g.,][]{Miyaji2019,Yamada2023}. They estimated the $\sigma_\mathrm{tor}$ from the fits to the SEDs of the AGNs using priors of 20, 40 and 60 degrees in their models. It is important to mention that Herrera-Endoqui et al. (in prep.) define their AGNs as those sources having an AGN contribution of at least 20 per cent of the total IR emission, and as expected, their sample of AGNs might not fully overlap with ours.

\begin{figure}
\centering
\includegraphics[scale=0.3]{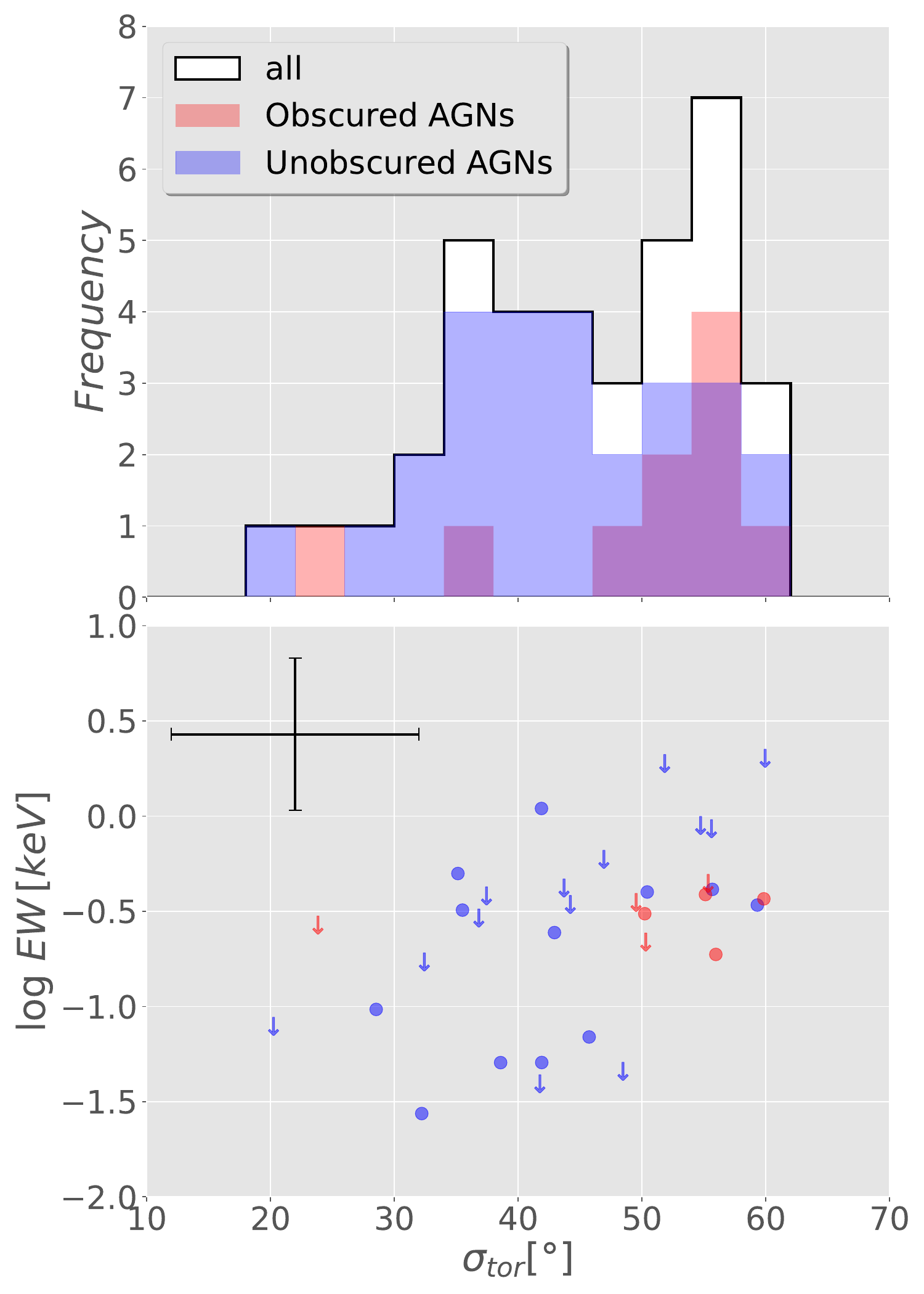}
\caption{Upper-panel: angular width histogram. Lower-panel: distribution of the Fe K$\alpha$ equivalent width as a function of the torus angular width for both obscured (red circles) and unobscured (blue circles) AGNs. We included the average error bar at 90 per cent of confidence for the whole sample.}
\label{sigma_tor}
\end{figure}

For a total of 36 sources in our AGN sample, reliable torus angular width estimates were obtained from the work by Herrera-Endoqui et al. Among them, 10 are classified as obscured and 26 as unobscured AGNs. We compared the values of $\sigma_\mathrm{tor}$ with other X-ray spectral properties, including the iron line equivalent width and the X-ray luminosity. The upper panel of Fig. \ref{sigma_tor} shows the distribution of torus angular widths. The obscured AGN population (red histogram) displays systematically higher values of $\sigma_\mathrm{tor}$, mostly clustered between $50\degree$ and $60\degree$, with a median of $\mu = 52.3\degree \pm 12.9\degree$. In contrast, the unobscured AGNs (blue histogram) present a broader distribution, centred at $\mu = 43.2^\degree \pm 12.2^\degree$. This result is consistent with expectations from the clumpy torus model, where Type-II AGNs preferentially exhibit larger torus angular widths \citep{Elitzur2008,Sazonov2015}.

In the lower panel of Fig. \ref{sigma_tor}, we test the correlation between the torus angular width and the iron K$\alpha$ equivalent width. We found a positive correlation between these two parameters. This result supports the physical interpretation of the Iwasawa-Taniguchi effect \citep[IT, ][]{Iwasawa1993},  according to which, larger $\sigma_\mathrm{tor}$ corresponds to a greater covering fraction, which enhances the reprocessed X-ray emission, and results in a higher $EW$. Moreover,  although the IT effect is observed in both unobscured and obscured AGN population (Fig. \ref{Lx_ew}), we found a flat slope $EW-\sigma_\mathrm{tor}$ trend for those obscured AGNs (red circles). Among them six (75 per cent) are highly obscured ($N_{\mathrm{H}} > 10^{23}\, \mathrm{cm}^{-2}$) with two CT-AGN candidates. This suggests that at high column densities, the torus becomes optically thick, making the spectrum reflection-dominated. In this regime, $EW$ saturates and becomes less sensitive to the torus angular width changes.

\subsection{Dust covering factor $f_{\mathrm{cov}}$ and X-ray Luminosity}

\begin{figure}
\centering
\includegraphics[scale=0.25]{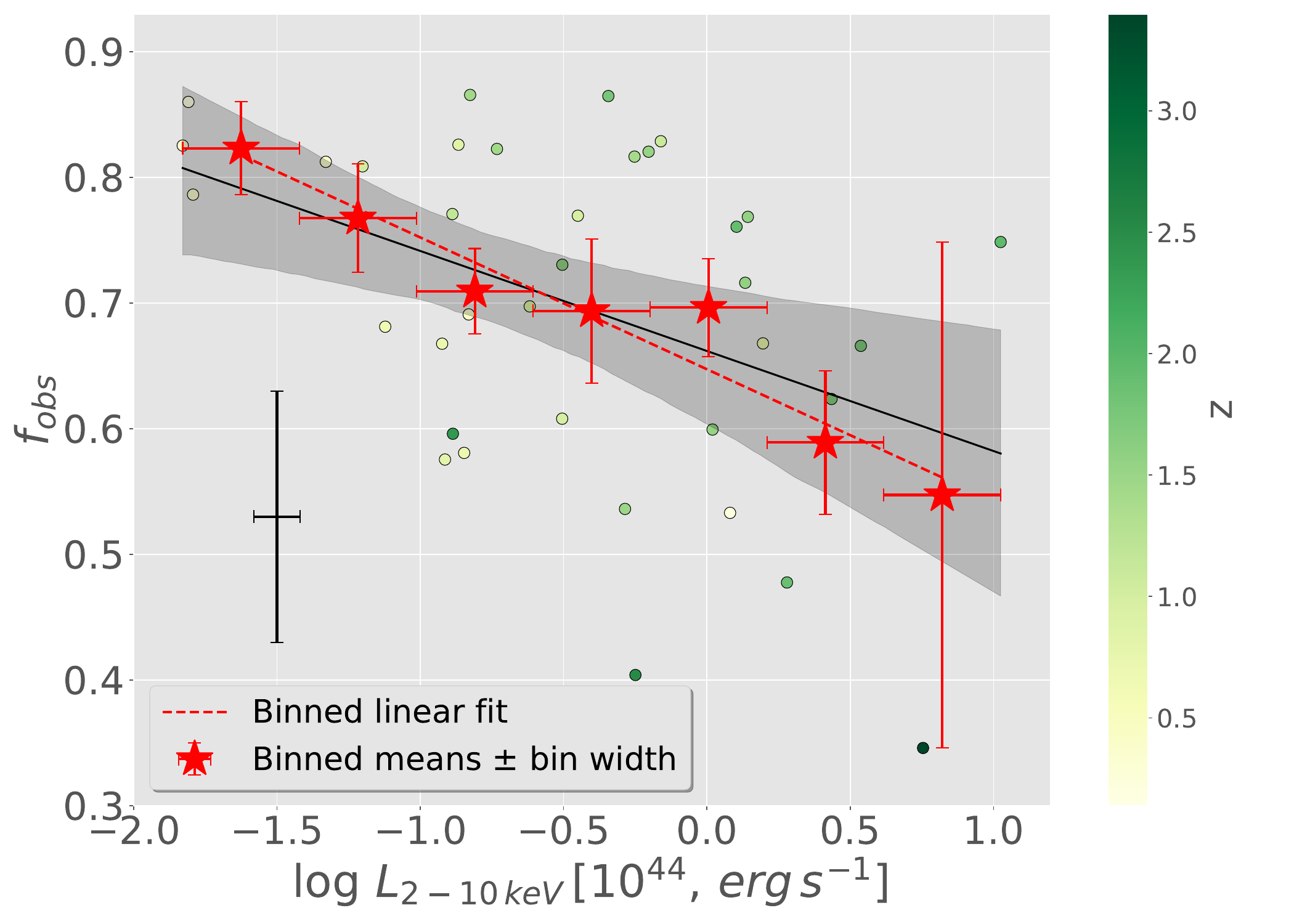}
\caption{Distribution of the dust covering factor as a function of hard X-ray luminosity. The black line represents the linear regression fit, with the shaded region indicating the 95 per cent confidence interval. The red line shows the regression obtained by dividing the sample into seven equally distributed luminosity bins, represented by red stars. The colour bar indicates the redshift of each source. The average error bar of each individual source at 90 per cent of confidence is also included (black cross).}
\label{fobs_Lx}
\end{figure}

We investigate how the fraction of obscured AGNs evolves as a function of X-ray luminosity. Fig. \ref{fobs_Lx} shows the dust covering factor $f_{\mathrm{cov}}$, plotted against the hard X-ray luminosity for our sample. A high-dispersion anti-correlation is observed between the two parameters, as indicated by the linear fit (black line) and its associated 95 per cent confidence interval (shaded dark region) estimated using a bootstrap method with 1000 resampling runs. The average errors at 90 per cent of confidence of each individual object are represented with the black cross in the lower-left part of the Figure.

This relation is strengthened by dividing the sample into equally separated log-scale luminosity bins distributed from $10^{42}$ to $10^{45}\, \mathrm{erg \, s^{-1}}$ (red stars). For instance, with 7 bins we found a tighter linear regression fit, presented with the red dashed-line and described by Equation \ref{f_eq}. We obtained a strong Spearman correlation coefficient of $r = -0.9$ with a statistically significant $p$-value of 0.0005, corresponding to a 99 per cent confidence level. The x-axis error bar represents the luminosity bin width, while the y-axis error bars correspond to the standard deviation of the mean ($\sigma_{f_{\mathrm{cov}}}/\sqrt{n}$)\footnote{where $\sigma_{f_{\mathrm{cov}}}$ is the standard deviation of the $f_{\mathrm{cov}}$ values in the bin, and $n$ is the number of sources in that bin.}. 

From Equation \ref{f_eq}, we can infer that the expected obscured fraction at a luminosity of $10^{44}\, \mathrm{erg\, s^{-1}}$ is approximately 0.65, increasing to about 0.8 at $\sim 10^{42}\, \mathrm{erg\, s^{-1}}$. These results support a luminosity-dependent obscuration scenario, in agreement with the receding torus model.

\begin{equation}
\label{f_eq}
    f_{\mathrm{cov}} = (-0.1 \pm 0.01) \log\left(\frac{L_{2-10\, \mathrm{keV}}}{10^{44}\, \mathrm{erg\, s^{-1}}}\right)+ (0.65 \pm 0.01)
\end{equation}

We tested the robustness of the observed distribution by applying various binning strategies, including increasing the number of bins and varying bin widths. As a result, we consistently identified a flattened trend (i.e., slope $\approx 0$) in the range $L_{2-10\, \mathrm{keV}} = 10^{43}$ - $10^{44} \, \mathrm{erg\, s^{-1}}$, even though it is not statistically significant. A similar behaviour was reported by \citet{Lusso2013} based on the XMM-COSMOS data. They found a plateau or a very weak luminosity dependence of the covering factor within this luminosity range \citep[see also][]{Stalevski2016}.

\section{Discussion}
\label{Discussion} 

\subsection{CT-AGN fraction}

During our $N_{\mathrm{H}}$ estimations, we identified 5 out of 94 AGNs (5.3 per cent) as Compton-thick candidates. A more detailed X-ray spectral analysis using the XCLUMPY model (Section \ref{xclum}) indicates that one of these sources is best classified as a strong CT-AGN candidate, while two additional sources are best classified as borderline or tentative CT-AGN candidates, with column densities at or close to $N_{\mathrm{H}} \sim 10^{24}\, \mathrm{cm}^{-2}$ within the uncertainties. This subset represents a small fraction of 3.2 per cent of our total sample, consistent with those reported in other deep X-ray surveys. For example, \citet{Iwasawa2020} found a similar CT-AGN fraction in the \textit{XMM-Newton} CDFS survey, while \citet{Signorini2023} reported 3.3 per cent in the J1030 \textit{Chandra} survey. Moreover,  \citet{Lanzuisi2018} using the \textit{Chandra} COSMOS survey, showed that the CT-AGN fraction increases with redshift, following the relation $f_{\mathrm{CT}}= 0.11(1 + z)^{1.11}$. 

It is important to note that our analysis focuses on the subset of X-ray sources with far-infrared counterparts selected from the \textit{Herschel} Lockman-SpReSO catalog. FIR emission is a key feature observed in heavily obscured AGNs, as shown by \citet{KilerciEser2020}, who analysed the spectral energy distributions of 338 obscured hard X-ray AGNs, including 52 CT-AGNs. They found that CT-AGNs typically exhibit a strong FIR emission, attributed to reprocessed radiation from heavily obscuring dust structures. 

Therefore, selecting X-ray sources with strong infrared counterparts may help the identification of heavily obscured AGNs. However, infrared selection alone does not guarantee a Compton-thick classification, and additional multiwavelength diagnostics are required. In particular, heavily obscured AGNs are generally expected to present enhanced mid-infrared emission due to dust reprocessing \citep[e.g.][]{Laloux2023}, while a low X-ray-to-infrared luminosity ratio can be used as an indicator of heavy obscuration \citep[e.g.][]{Stern2015,Alexander2008}.

The combination of deep X-ray data and FIR selection provides an advantageous framework for identifying heavily obscured AGN candidates, particularly sources that may be underrepresented in optical or soft X-ray selected samples.

\subsection{Clumpy torus structure and receding model}

Throughout this work, we have confirmed that X-ray luminosity, directly linked to the SMBH mass, is one of the main mechanisms shaping the structure of the dusty torus, as proposed by the receding torus model \citep{Elitzur2006, Ricci2013}. Our results support this scenario through several observed results, including the decrease in both the Fe K$\alpha$ equivalent width (Fig. \ref{Lx_ew}) and the fraction of obscured AGNs (Fig. \ref{fobs_Lx}) as functions of X-ray luminosity. These results suggest that the size of the dusty torus ($R_\mathrm{tor}$), and hence the covering factor, is directly dependent on the AGN luminosity, primarily driven by radiation pressure and dust sublimation. 

In the case of the Iwasawa-Taniguchi relation (EW-$L_X$), the neutral iron line emission originates from reprocessed X-ray radiation in the torus. As luminosity increases, the inner radius of the torus expands due to stronger radiation pressure, and the structure becomes more compact, reducing the torus contribution to the observed reflection features. For example, \citet{Suganuma2006}  and  \citet{Tristram2011} suggest that the inner radius of the dusty torus scales with luminosity as $R_{\mathrm{in}} \propto L^{0.5}$. This relation was confirmed using near-infrared reverberation mapping and mid-infrared interferometry with the VLTI, respectively.

In our analysis, we obtained a strong anti-correlation between the obscured fraction and the hard X-ray luminosity, described by the expression $f_{\mathrm{cov}}\propto L_X^{-0.1 \pm 0.01}$ (Equation  \ref{f_eq}). Similar results were reported by \citet{Hasinger2008}, who found a steeper slope of -0.23 by using a sample of Type-II AGNs. They also showed that the obscured AGN fraction increases with redshift.

We note that our sample reveals a flattening variation of $f_{\mathrm{cov}}$ within the luminosity range of $L_{2-10\, \mathrm{keV}} = 10^{43}$ - $10^{44} \, \mathrm{erg\, s^{-1}}$, consistent with the results reported by \citet{Lusso2013} and \citet{Stalevski2016}. They attribute the observed weak luminosity dependence of the obscured fraction to the transition from a thin to an optically thick torus regime, where the infrared emission and the covering factor become largely insensitive to further increases in AGN luminosity due to dust self-absorption and reprocessing saturation. An alternative explanation could be that, at intermediate luminosities, the rate of dust sublimation driven by a strong radiation field is effectively balanced by the accretion-driven inflow of material, maintaining a stable torus inner radius and an unchanged covering factor. However, \citet{Elitzur2006} proposed that below a critical bolometric luminosity ($L_\mathrm{bol} < 10^{42}\, \mathrm{erg\, s^{-1}}$), the outflow of dusty clouds becomes insufficient to sustain the torus structure. Supporting this study, \citet{gonzalez2017} found that a significant fraction of low-luminosity AGNs (LLAGNs) with $2-10$ keV X-ray luminosities below $10^{41}\, \mathrm{erg\, s^{-1}}$ show weak or absent torus signatures in mid-infrared observations, suggesting that the torus structure dissipates under low-accretion regimes. 

\subsection{Sources with complex spectral features}
\label{new_spec}

A small number of sources in our sample present spectral features that are not adequately reproduced by the baseline models adopted in this work, and therefore require a more detailed source-by-source analysis. A clear example is source 206776, classified as an optical Type-II AGN at $z = 0.203$ (see Table \ref{cross}), whose spectrum could not be satisfactorily fitted with our standard spectral models. The best fit obtained with the baseline models produced a poor fit statistic ($\chi^2_{\rm red} \approx 4$), a relatively flat photon index ($\Gamma \approx 0.8$), and a low column density ($\log N_{\mathrm{H}} < 20\,\mathrm{cm^{-2}}$), despite presenting a large Fe K$\alpha$ equivalent width ($EW \sim 0.94\,\mathrm{keV}$). This combination of spectral properties suggests that the source may be affected by a significant reflection component that is not properly captured by the simpler models.  In particular, for low- and moderate-count spectra, part of the reflected continuum emission may be misidentified as an enhanced Fe K$\alpha$ line when using simple absorbed spectral models with a limited number of components.

\begin{figure}
\centering
\includegraphics[scale=0.35]{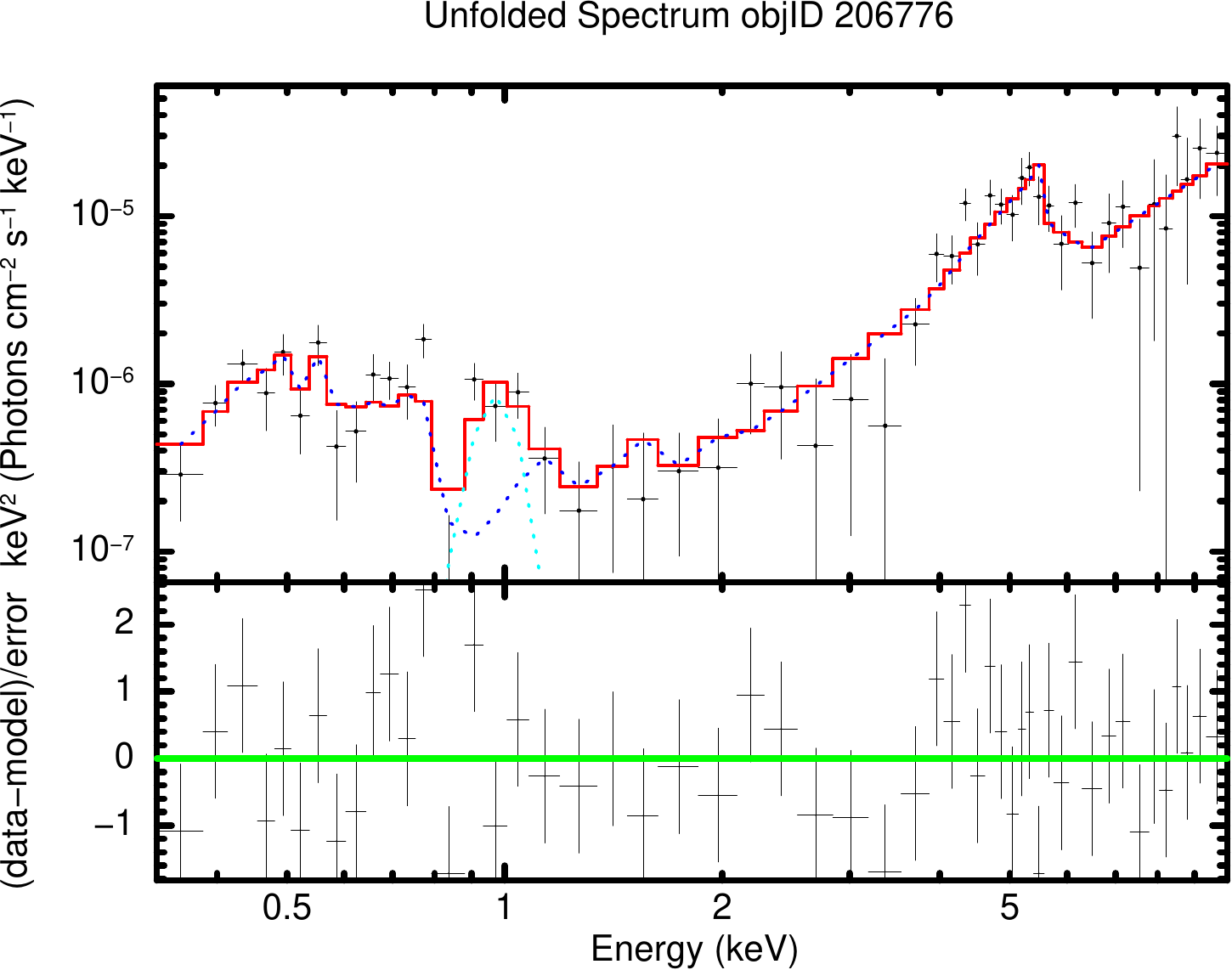}
\caption{ Unfolded $E\times F(E)$ spectrum of source 206776 together with the best-fit model $\texttt{TBabs(powerlaw + relxill + gaussian)}$ and residuals.}
\label{rexill}
\end{figure}

Since this source is among the brightest in our sample (with $\sim 850$ counts in the $0.3-10\, \mathrm{keV}$ band), we performed an additional exploratory analysis using a more physically motivated model that includes relativistic reflection from the accretion disc \citep[\texttt{relxill},][]{Garcia2014,Dauser2016}. We find that the spectrum shown in Fig. \ref{rexill} can be reasonably reproduced with the model $\texttt{TBabs(powerlaw + relxill + gaussian)}$, producing an acceptable fit statistic of $\chi^2_{\rm red} \approx 1.2$.

The best-fitting parameters indicate a hard primary continuum ($\Gamma \sim 1.16$) and a moderately ionized accretion disc ($\log \xi \sim 2.7$), with an iron abundance slightly above solar ($A_{\mathrm{Fe}} \sim 3.5$). Additionally, a Gaussian component centred at $\sim 0.97\,\mathrm{keV}$ significantly improves the fit, accounting for residual emission that may be associated with a blend of $\mathrm{Fe}-\mathrm{L}$ and Ne lines. Although the available data do not allow robust constraints on parameters such as the emissivity profile or the black hole spin, the inclusion of a relativistic reflection component provides a more physically consistent description of the observed spectral features. Other sources in the sample may present similar complexities; however, a systematic analysis with more sophisticated models is beyond the scope of the present work.

Therefore, the spectral parameters derived for some complex sources, particularly the Fe K$\alpha$ equivalent widths and photon indices, should be interpreted with caution. In particular, a few low-luminosity sources in the sample present large upper limits on the Fe K$\alpha$ equivalent width together with moderate upper limits on the intrinsic absorption, properties commonly associated with obscured or reflection-dominated AGNs. However, due to their relatively low X-ray luminosities, these sources do not satisfy the AGN classification criteria adopted in Section \ref{Simple_Spec}. Furthermore, no clear optical spectroscopic AGN classification is currently available for these sources. Therefore, some of these objects may still host obscured or low-luminosity AGN activity that cannot be robustly confirmed with the current data quality.

\subsection{Star formation and X-ray emission}
\label{sfr}

\begin{figure}
\centering
\includegraphics[scale=0.25]{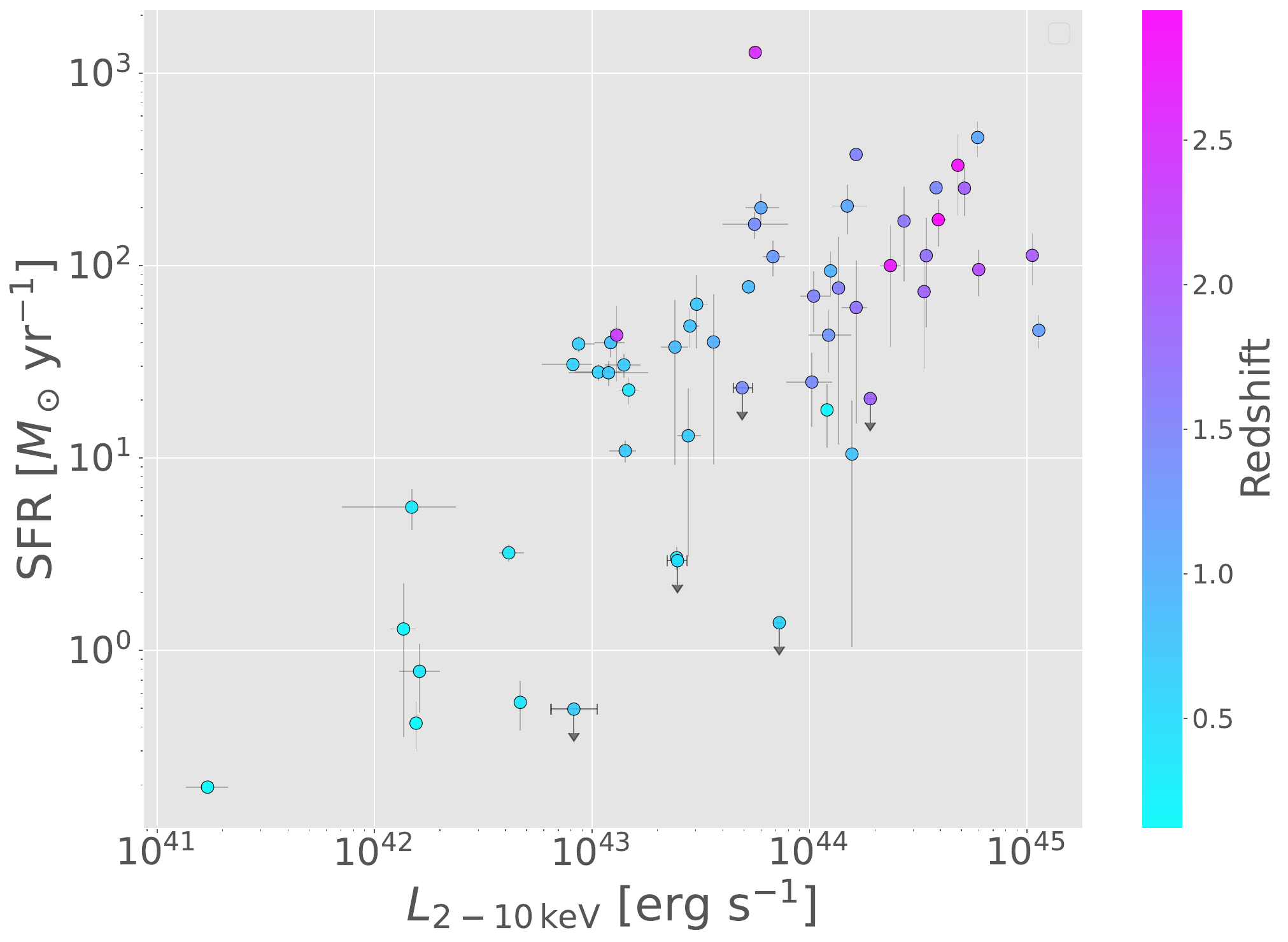}
\caption{Star-formation rate derived from the CIGALE SED fitting as a function of the absorption-corrected $2-10\,\mathrm{keV}$ X-ray luminosity for the AGN sample. Dark arrows are used for those sources with SFR upper-limits.}
\label{Star}
\end{figure}

One of the main characteristics of the AGN population analysed in this work is that it is selected from a parent far-infrared sample from the Lockman-SpReSO project. Since FIR emission traces dust-rich systems with enhanced star-formation activity, this selection may favour the identification of obscured AGNs hosted by actively star-forming galaxies. To explore the star-forming properties of the AGN population, in Fig. \ref{Star} we show the relation between the star-formation rate (SFR) derived from the CIGALE SED fitting reported by \citet{Gonzales2024b} and the absorption-corrected $2-10\,\mathrm{keV}$ X-ray luminosities estimated in this work for 54 AGNs with spectroscopic redshifts. The AGN population spans a broad range of SFRs, indicating that these systems are hosted by galaxies with significant ongoing star-formation activity. 

This Figure shows that the most X-ray luminous AGNs at higher redshifts tend to be associated with host galaxies exhibiting higher SFRs, some of them with complex X-ray spectra, as discussed in Section \ref{new_spec}, consistent with heavily obscured or reflection-dominated systems. For instance, we found that four AGNs that show soft excess emission are hosted by galaxies with high star-formation rates ($\mathrm{SFR} \gtrsim 100\,M_\odot\,\mathrm{yr^{-1}}$), all of them with high X-ray luminosities ($L_{2-10\,\mathrm{keV}} \gtrsim 10^{44}\,\mathrm{erg\,s^{-1}}$). One of them is objID-206710 (Fig. \ref{XID4}), which presents one of the highest star-formation rates with $\mathrm{SFR} \approx 460\,M_\odot\,\mathrm{yr^{-1}}$. 

Another remarkable case is the borderline CT-AGN candidate objID-206603, hosted by the galaxy with the highest star-formation rate in the sample ($\mathrm{SFR} \sim 10^3\,M_\odot\,\mathrm{yr^{-1}}$). This result suggests that gas- and dust-rich environments may contribute both to the nuclear obscuration and to part of the soft X-ray emission observed in some sources. Although the limited sample size prevents a robust statistical analysis, these trends are consistent with the dusty and actively star-forming nature of the parent FIR-selected population.

Despite the FIR selection, the overall X-ray properties of the AGN population are broadly consistent with those reported in previous deep X-ray surveys such as COSMOS \citep{Lanzuisi2018} and CDFS \citep{Iwasawa2020}. The photon index distribution ($\langle \Gamma \rangle = 1.77 \pm 0.54$), obscured AGN fraction ($f_\mathrm{obs} = 21\%$, $N_\mathrm{H} > 10^{22}$ cm$^{-2}$), soft excess fraction (11\%, $kT \approx 0.12$ keV), and Fe K$\alpha$ equivalent widths follow typical trends observed in X-ray-selected AGN samples. This suggests that FIR-selected AGNs do not constitute a fundamentally different X-ray population, although the FIR selection enhances the detection of dusty and highly star-forming host galaxies and heavily obscured systems.

\section{Summary and Conclusions}
\label{summary} 

In this work, we have conducted a detailed X-ray spectral study of the AGN population identified in the Lockman-SpReSO project.  We applied a set of absorbed power-law models with X-ray spectra obtained from the combination of 18 \textit{XMM-Newton} observations of the Lockman Hole field. We measured their main spectral properties and studied how $L_{2-10\, \mathrm{keV}}$, $N_{\mathrm{H}}$ and $f_{\mathrm{cov}}$ evolve with redshift. We studied the dusty torus structure and divided our sample into obscured and unobscured sources, based on their hydrogen column density. Our main results are summarized as follows: 

\begin{itemize}
    \item From a statistical cross-correlation analysis, we identified a total of 110 X-ray sources detected in the LH field, with \textit{Herschel} FIR  emission. 64 per cent of them with spectroscopic redshift  distributed at a broad redshift range $z = 0.07 - 5$. 
    
    \item A total of  94 sources (85 per cent) were identified as AGNs,  based purely on their X-ray properties, including the X-ray luminosity $(Lx > 3\times 10^{42}\, \mathrm{erg\, s^{-1}})$ and the hardness of the spectral photon index $(\Gamma < 1)$. The sample was classified based on their luminosity as  LLAGNs,  Seyfert galaxies and quasars.  
    
    \item We confirmed the presence of a soft excess in a total of ten AGNs. If we model the soft excess with a blackbody, their average fitted temperature becomes $kT = 0.12 \pm 0.02\, \mathrm{keV}$, consistent with typical soft excess temperatures observed in AGNs. 
    
    \item We observed a trend suggesting an increase of the fraction of obscured AGNs toward higher redshifts. 21 per cent of our sample are classified as obscured AGNs with $N_{\mathrm{H}} > 10^{22}\, \mathrm{cm}^{-2}$. 
    
    \item We identified five heavily obscured sources with column densities approaching the Compton-thick regime. A more detailed analysis using the XCLUMPY torus model indicates that one of these (objID-206545) is best classified as a strong Compton-thick AGN candidate, with both equatorial and line-of-sight column densities exceeding $10^{24}\,\mathrm{cm}^{-2}$. Two additional sources (objID-206463 and objID-206603) are classified as borderline CT-AGN candidates, with line-of-sight column densities consistent with, or close to, $N_\mathrm{H}^{\mathrm{Los}} \sim 10^{24}\,\mathrm{cm}^{-2}$, although their uncertainties extend below this threshold. Their AGN infrared fractions derived from the CIGALE SED fitting range from $f_{\rm AGN}=0.11$ to $0.33$, indicating that both AGN activity and host-galaxy star formation contribute to their infrared emission. The remaining sources are better described as heavily obscured but Compton-thin AGNs. 
    
    \item  Our comparison between optical and X-ray classifications revealed an overall agreement of $\sim$73 per cent, with all optical Type-I AGNs confirmed as unobscured in X-rays. However, a substantial fraction of optical Type-II AGNs (70 per cent) appeared unabsorbed in X-rays, a discrepancy that may be linked to host-galaxy dilution.
    
    \item We detected significant Fe K${_\alpha}$ emission in 49 AGNs and confirmed the X-ray Baldwin effect as an anti-correlation between $EW$  and X-ray luminosity, observed in both obscured and unobscured sources.

    \item By combining X-ray spectral parameters with SED measurements, we explored the torus geometry and covering factor of the AGN population. We found that obscured AGNs show larger torus angular widths $(\sigma_\mathrm{tor} \sim 50\degree)$, while unobscured AGNs show a broader distribution centred at $\sigma_\mathrm{tor} = 43.2 \pm 12.2$. Additionally, we identified a positive correlation between the torus angular width and the Fe K$\alpha$ equivalent width.

    \item We observed a strong anti-correlation between dust covering factor and X-ray luminosity $(\rho \approx -0.9)$, supporting the receding torus scenario in which radiation pressure regulates the torus structure. We also identified a possible plateau in the covering factor at intermediate luminosities $(L_{2-10\, \mathrm{keV}} \sim 10^{43}-10^{44}\,\mathrm{erg\,s^{-1}})$, consistent with a balance between dust sublimation and replenishment processes, in agreement with previous works.

    \item Overall, our results indicate that FIR-selected AGNs represent a population preferentially associated with dusty and actively star-forming host galaxies, while preserving the general X-ray spectral properties observed in deep X-ray AGN surveys. 
      
\end{itemize}

Our results reinforce the scenario in which AGN obscuration is closely linked to both the geometry of the dusty torus and the properties of the host galaxy. The increasing fraction of obscured and CT-AGN candidates toward higher redshifts is consistent with a larger incidence of gas- and dust-rich environments in the early universe, where black hole growth and galaxy evolution are expected to be more active. In addition, the observed anti-correlation between covering factor and X-ray luminosity supports the receding torus scenario, while the possible plateau observed at intermediate luminosities may indicate a balance between dust sublimation and replenishment processes within the torus. Overall, our results demonstrate the importance of combining deep X-ray observations with FIR-selected samples to identify faint and heavily obscured AGN populations in the Lockman Hole field.

\section*{Acknowledgements} 
MEC and MHE acknowledge support from SECIHTI through a postdoctoral fellowship within the program ``Estancias Posdoctorales por México''. This work is partially supported by UNAM-DGAPA PAPIIT IN-114423 and IT-101226. YK acknowledges support from UNAM-PAPIIT grant IN102023. ICG and MHE acknowledge financial support from DGAPA-UNAM grant IN-119123 and CONAHCYT grant CF-2023-G-100. HMHT acknowledges CONAHCYT project CF-G-543 entitled ``Arqueología y filogenética de dinosaurios galácticos: formación y evolución de galaxias masivas apagadas''. HMHT acknowledges support from CONAHCYT project CF-2023-G-1052 entitled ``Sinergia y Retos del Censo LSST del Observatorio Vera Rubin para la Astrofísica, la Ciencia de datos, la Química y otras disciplinas''. MC acknowledges support by grant PID2022-136598NB-C33 funded by MCIN/AEI/10.13039/501100011033 and by ``ERDF A way of making Europe''.  JAD acknowledges support from UNAM-PAPIIT grant IN116325. MIR acknowledges the support of the Spanish Ministry of Science, Innovation and Universities through the project PID-2021-122544NB-C43. ICG and EB acknowledge financial support from DGAPA-UNAM grant IN-119123 and SECIHTI grant CF-2023-G-100. J. Cepa and C.P. Padilla acknowledge the support of the Spanish Ministry of Science, Innovation and Universities through the project PID-2021-122544NB-C41.

\section*{Data Availability}
The data underlying this article will be shared on reasonable
request to the corresponding author.
 


\bibliographystyle{mnras}
\bibliography{bibliography} 




\appendix

\section{summary of \textsc{xspec} Spectral models}
\label{xspec_summary}

Summary of the spectral models employed in our analysis with \textsc{xspec}, together with a brief description and their corresponding parameters, including the models:  \texttt{tbabs}\footnote{\url{https://heasarc.gsfc.nasa.gov/docs/software/xspec/manual/node279.html}}, \texttt{powerlaw}\footnote{\url{https://heasarc.gsfc.nasa.gov/docs/software/xspec/manual/node221.html}}, \texttt{zgauss}\footnote{\url{https://heasarc.gsfc.nasa.gov/docs/software/xspec/manual/node181.html}}, \texttt{pexrav}\footnote{\url{https://heasarc.gsfc.nasa.gov/docs/software/xspec/manual/node214.html}}, \texttt{cabs}\footnote{\url{https://heasarc.gsfc.nasa.gov/docs/software/xspec/manual/node248.html}}, \texttt{zcutoffpl}\footnote{\url{https://heasarc.gsfc.nasa.gov/docs/software/xspec/manual/node163.html}}, \texttt{zbbody}\footnote{\url{https://heasarc.gsfc.nasa.gov/docs/software/xspec/manual/node136.html}}, and \texttt{xclumpy}\footnote{\url{https://github.com/AtsushiTanimoto/XClumpy}}.

\clearpage

\begin{table*}
\centering
\caption{Spectral models used in the analysis.}
\label{xspec_table}
\resizebox{\textwidth}{!}{%
\begin{tabular}{@{}p{1.4cm} p{4.0cm} p{5cm} p{5cm}@{}}
\toprule
Name & Description & Equation & Parameters \\ 
\toprule
\texttt{tbabs} &
Photoelectric absorption by the interstellar medium \citep[ISM,][]{Wilms2000}. 
& -- 
& $N_{\mathrm{H}}$: hydrogen column density \\ 
\midrule

\texttt{powerlaw} &
Simple photon power-law. 
& \( A(E) = K E^{-\alpha} \)
& $\alpha$: photon index of power-law (dimensionless)\\
& & & $K$: photons/keV/cm$^2$/s at 1 keV \\ 
\midrule

\texttt{zgauss} &
Redshifted Gaussian emission line. 
& 
\(A(E) = K \frac{\exp\!\left[-(E(1+z)-E_l)^2 / (2\sigma^2)\right]}{(1+z)\sigma\sqrt{2\pi}
\left[1-\mathrm{erf}\!\left(-E_l/(\sqrt{2}\sigma)\right)\right]}\)
& $E_l$: line energy (keV)\\
& & & $\sigma$: width (keV)\\
& & & $z$: redshift\\
& & & $K$: total photons/cm$^2$/s \\ 
\midrule

 &
&  
& $\Gamma$: first power-law photon index, $N_\mathrm{E} \propto E^{\Gamma}$\\
& & & $E_{\mathrm{c}}$: cutoff energy (keV)\\
& & & $rel_{\mathrm{refl}}$: reflection factor\\
& & & $z$: redshift\\
\texttt{pexrav} & Reflected power-law spectrum with an exponential cutoff from a neutral medium \citep{Magdziarz1995}. 
& --
& Abundance of elements heavier than He, iron abundance and cosine of inclination angle.  $K$: photon flux  at 1 keV (photons/keV/cm$^2$/s)  of the cutoff broken  power-law only in the observed frame. \\

\midrule

\texttt{cabs} &
Optically thin Compton scattering. 
&
\( M(E)=\exp[-\eta_\mathrm{H}\sigma_T(E)] \)
& $\eta_\mathrm{H}$: column density\\
& & & $\sigma_T$: Thomson cross-section \\ 
\midrule

\texttt{zcutoffpl} &
Redshifted power-law model with
a high-energy exponential cutoff. 
&
\( A(E)=K E^{-\alpha}\exp(-E/\beta) \)
& $\alpha$: photon index\\
& & & $\beta$: e-folding energy (keV)\\
& & & $z$: redshift\\
& & & $K$: photons/keV/cm$^2$/s at 1 keV\\ 
\midrule

\texttt{zbbody} &
Redshifted blackbody spectrum. 
&
\( 
A(E)=\frac{
K\times8.0525[E(1+z)]^2dE
}{
(1+z)(kT)^4\left[\exp(E(1+z)/kT)-1\right]
}
\)
& $kT$: temperature (keV)\\
& & & $z$: redshift\\
& & & $K = L_{39}/[D_{10}(1+z)]^2$ \\ 
\midrule

\texttt{xclumpy} &
Clumpy torus model for AGN. 
&
\begin{tabular}{@{}l@{}}
\texttt{phabs*(zphabs*cabs*zcutoffpl} \\
\texttt{+ constant*zcutoffpl} \\
\texttt{+ atable\{xclumpy\_R.fits\}} \\
\texttt{+ atable\{xclumpy\_L.fits\})}
\end{tabular}
& \texttt{phabs*(zphabs*cabs*zcutoffpl:} transmitted continuum\\
& & & xclumpy\_R: reflection continuum\\
& & & xclumpy\_L: fluorescence lines \\

\bottomrule
\end{tabular}}
\end{table*}

\clearpage

\section{$N_{\mathrm{H}}$  detectability limit}
\label{detectability}

To estimate how the $N_{\mathrm{H}}$ detectability limit of our survey evolves with redshift, we adopted the following approach. First, we selected an AGN with a well-behaved spectrum located near the centre of the field, where the instrumental response is representative of the survey. We modelled the spectrum (as described in Section \ref{Xspectral}) using an absorbed power-law model, \texttt{tbabs*zphabs*powerlaw}. For this source, we measured a photon index of $\Gamma = 1.77$, consistent with the median spectral slope of the full AGN sample, with $\sim 1600$ counts in the $0.3-10\,\mathrm{keV}$ band, and  a low intrinsic absorption of $\log N_{\mathrm{H}} = 20.85^{+0.33}_{-0.85}\, \mathrm{cm^{-2}}$. This spectrum was then used as the instrumental template for the simulated spectra.

\begin{figure*}
\centering
\includegraphics[scale=0.21]{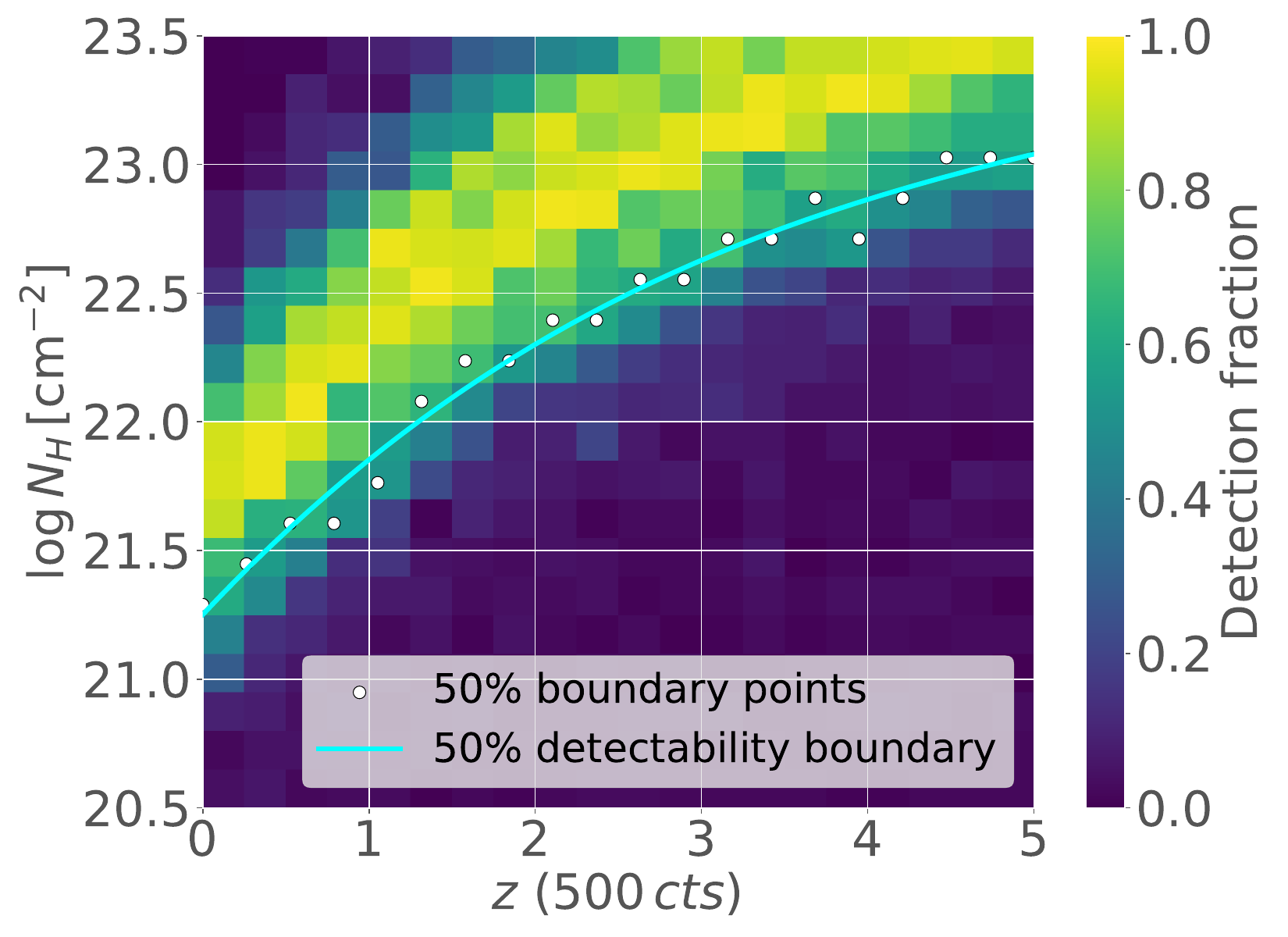}
\includegraphics[scale=0.21]{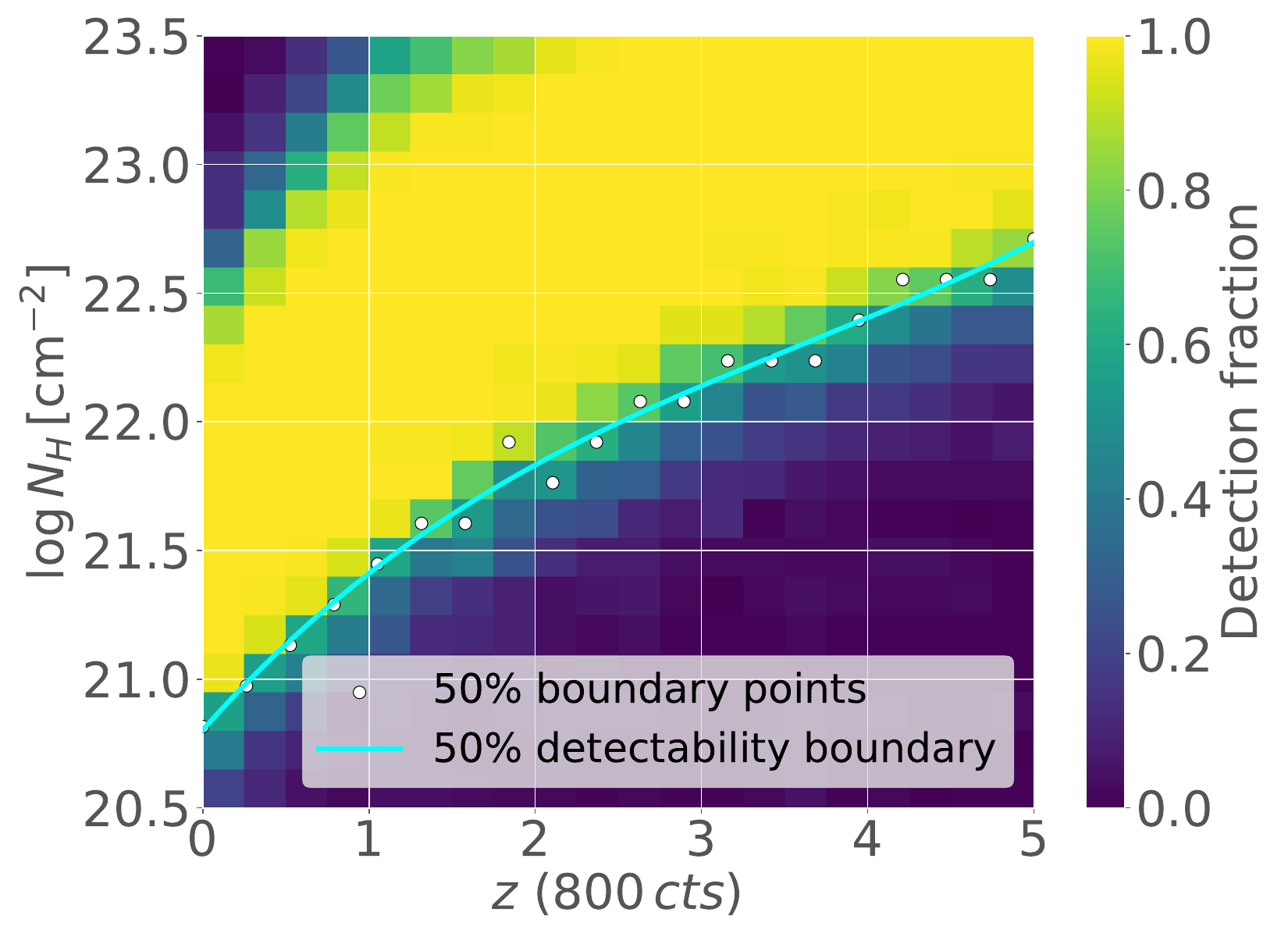}
\includegraphics[scale=0.21]{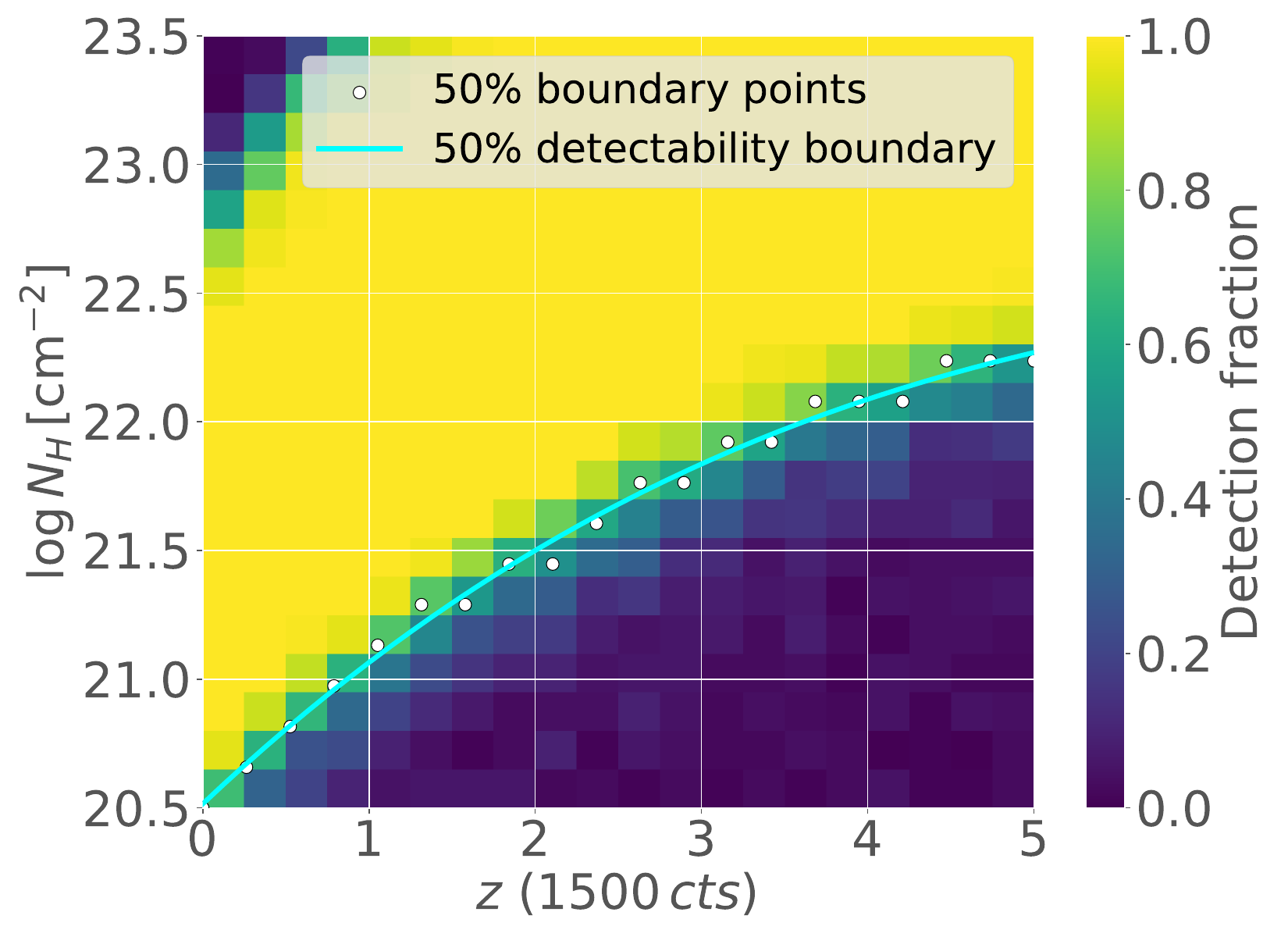}
\includegraphics[scale=0.18]{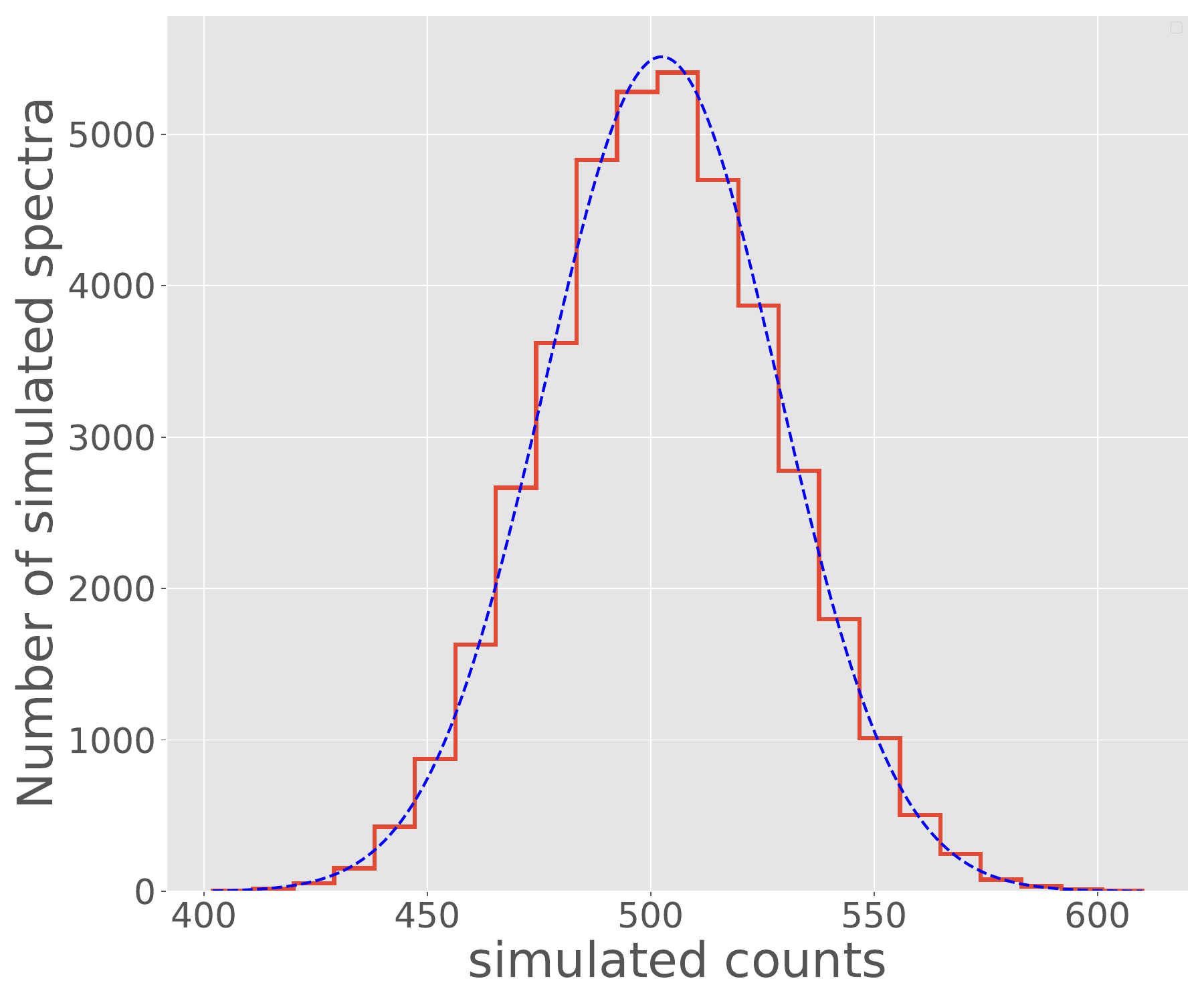}
\includegraphics[scale=0.18]{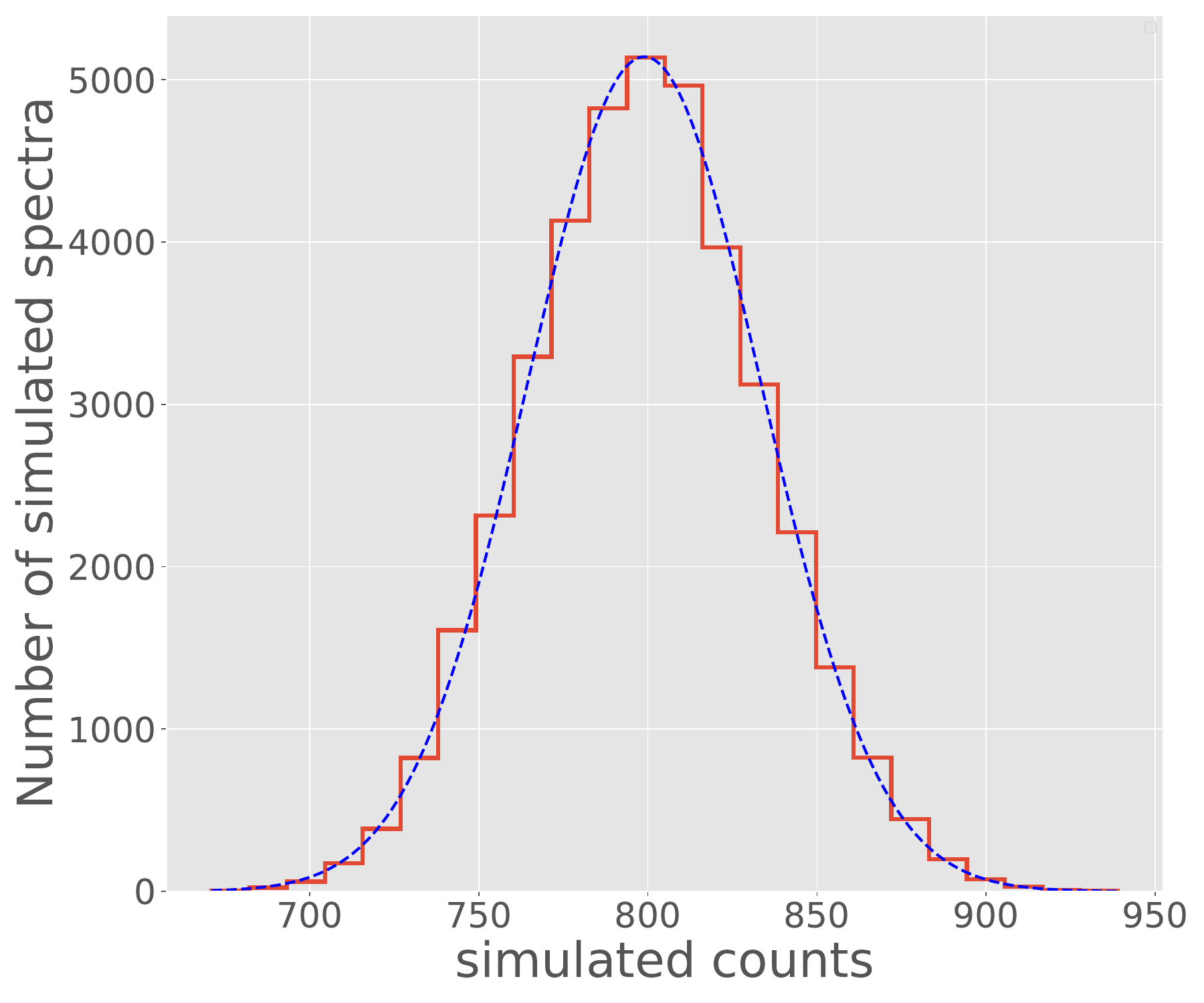}
\includegraphics[scale=0.18]{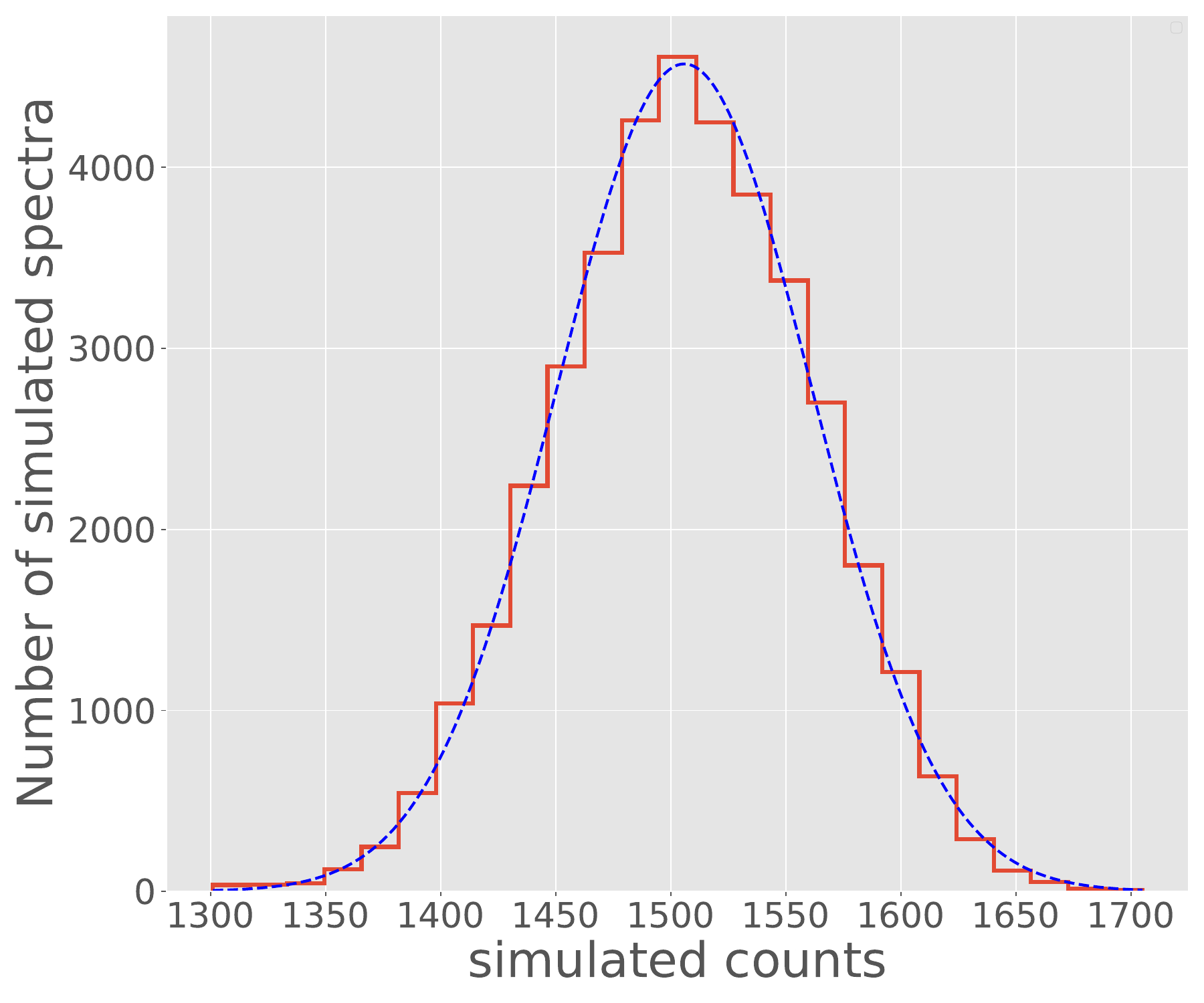}
\caption{Detectability of intrinsic absorption as a function of redshift and source counts. Upper panels: heat maps showing the fraction of simulated spectra in which intrinsic absorption is successfully detected at each $(z, N_{\mathrm{H}})$ grid point, based on the detection criteria. The white circles mark the 50 per cent boundary points, and the cyan curve shows the corresponding 50 per cent detectability boundary, defined as the minimum intrinsic column density above which at least half of the simulated spectra satisfy the absorption-detection criteria. Lower panels: distributions of the total $0.3-10\,  \mathrm{keV}$  counts for the 40,000 simulated spectra in each representative case, centred at 500, 800, and 1500 counts, respectively. The blue dashed curves show the corresponding Gaussian fits.}
\label{heatmap}
\end{figure*}

The simulated spectra were generated using the \texttt{fakeit} command, adopting the same response matrix and background as the original source, while varying the intrinsic absorption and source redshift over the ranges $\log N_{\mathrm{H}}  = 20 - 23.5\, \mathrm{cm^{-2}}$ and $z = 0 - 5$, respectively. For each $(N_{\mathrm{H}}, z)$ combination, we generated 100 simulated spectra, constructing a two-dimensional grid in intrinsic absorption and redshift, for a total of 40,000 simulated spectra. For each spectrum, the power-law normalization was adjusted to produce spectra with approximately a specific number of X-ray counts in the $0.3 - 10\,\mathrm{keV}$ band for an exposure time of 200 ks, including Poisson fluctuations. We first adopted 800 counts, corresponding to the median of the AGN sample, and then repeated the analysis for 500 and 1500 counts to represent faint and bright sources, respectively. The count distributions of these three sets of simulated spectra, together with their Gaussian fits, are shown in the lower-central, lower-left, and lower-right panels of Fig. \ref{heatmap}, respectively.

Each simulated spectrum was then fitted with two nested models: an unabsorbed power-law, \texttt{tbabs*powerlaw}, and an absorbed power-law, \texttt{tbabs*zphabs*powerlaw}. In both cases, the Galactic absorption and photon index were fixed at $N_{\mathrm{H,Gal}} = 5.7 \times 10^{19}\,\mathrm{cm^{-2}}$ and $\Gamma = 1.8$, respectively, while $N_{\mathrm{H}}$ was allowed to vary in the absorbed model. Since the original spectrum was grouped to 30 counts per bin to ensure adequate statistical quality, we used the $\chi^2$ statistic in all fits. We adopted two conservative detection criteria to determine whether intrinsic absorption is detectable: 

\begin{enumerate}
    \item A statistically significant improvement of the absorbed model over the unabsorbed one, defined as $\Delta \chi^2 = \chi^2_1 - \chi^2_2 \geq 2.71$. Where $\chi^2_1$ and $\chi^2_2$ correspond to the best-fit statistics of the unabsorbed and absorbed models, respectively. This threshold corresponds to a 90 per cent confidence level for one additional free parameter.

    \item A positive intrinsic absorption measurement with a well-constrained lower bound at the 90 per cent confidence level, i.e. $N_{\mathrm{H,low}} > 0$, as determined using the \texttt{error} command.
\end{enumerate}

For each grid point, we computed the fraction of simulated spectra in which intrinsic absorption was successfully detected according to our adopted detection criteria. We then defined the detectability boundary as the minimum intrinsic column density for which at least 50 per cent of the simulated spectra at a given $(z, N_{\mathrm{H}})$ grid point satisfy the absorption-detection criteria. The resulting detectability heat maps are shown in the upper panels of Fig. \ref{heatmap}, where the colour bar indicates the detection fraction. The three panels correspond to the results of our simulated spectra with 500 (upper-left), 800 (upper-centre), and 1500 (upper-right) counts.

\section{Optical spectrum of the strongest CT-AGN candidates}
\label{Optical}

\begin{figure*}
\centering
\includegraphics[scale=0.7]{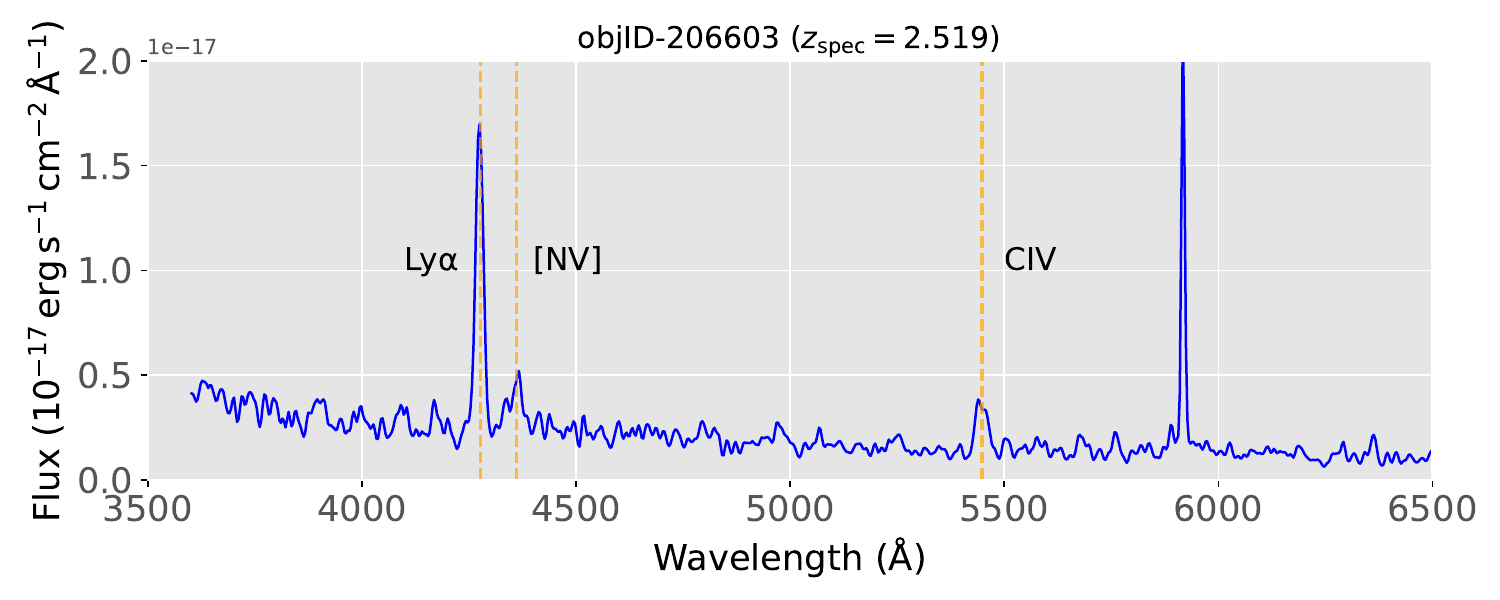}
\includegraphics[scale=0.7]{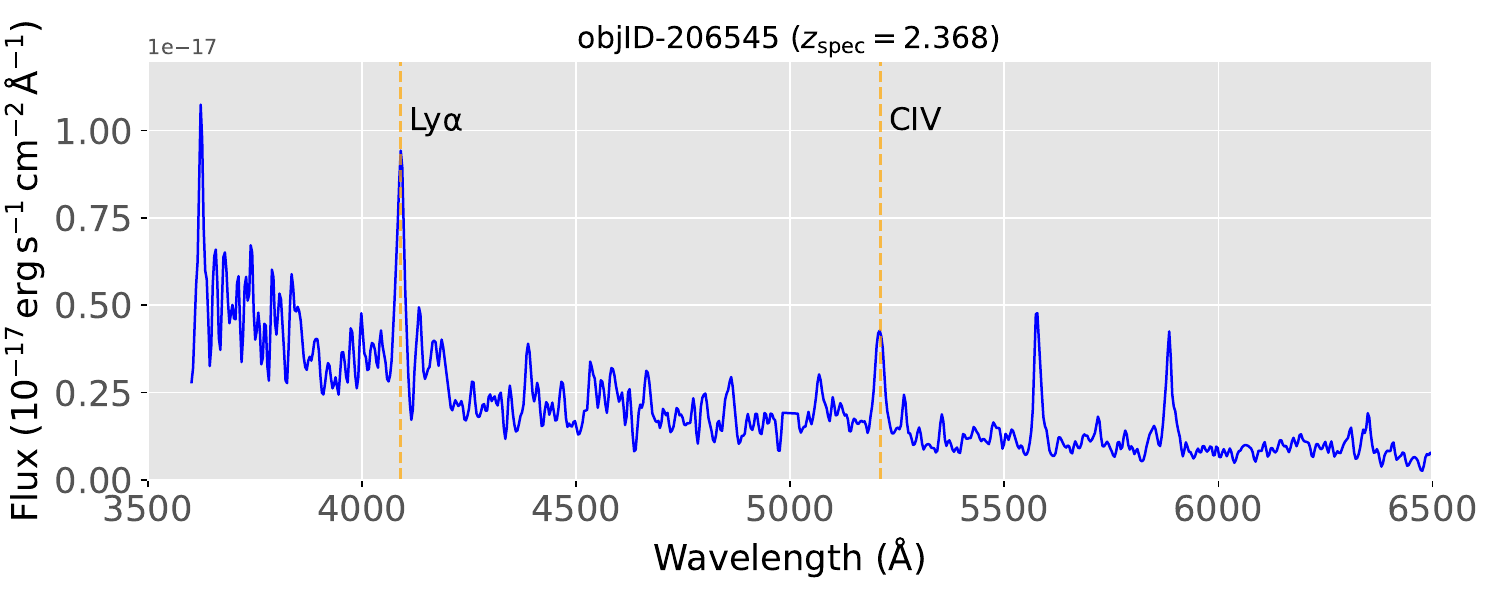}
\caption{Optical spectra of the strongest CT-AGN candidates, objID-206603 (upper-panel) and objID-206545 (lower-panel), obtained with the GTC/OSIRIS R500B grism. The dashed vertical lines indicate the expected positions of the Ly$\alpha$, [N\,{\sevensize V}],  and C\,{\sevensize IV} emission lines at the spectroscopic redshift of $z_\mathrm{spec}=2.519$ and $z_\mathrm{spec}=2.368$, respectively.}
\label{A1}
\end{figure*}

In this appendix, we present the only available optical GTC spectra of our strongest CT-AGN candidates. Fig. \ref{A1} shows the optical spectrum of objID-206603 in the upper-panel, while the source objID-206545 is presented in the lower-panel. The spectra were obtained with the GTC/OSIRIS R500B grism. We detected some emission lines such as Ly$\alpha$ ($\lambda1216$ \AA), [N\,{\sevensize V}] ($\lambda1240$ \AA), and C\,{\sevensize IV} ($\lambda1549$ \AA), consistent with the reported spectroscopic redshift of $z_\mathrm{spec}=2.519$ and $z_\mathrm{spec}=2.368$, respectively.

\section{Spectral energy distributions of the CT-AGN candidates}
\label{Appendix_SED}

In this appendix, we present the spectral energy distributions (SEDs) of the strongest CT-AGN candidates: objID-206545, objID-206463, and objID-206603. The SEDs were derived from the multiwavelength photometric analysis of the Lockman-SpReSO field by Herrera-Endoqui et al. (in preparation), using optical-to-far-infrared photometry. All three sources show prominent infrared emission, particularly objID-206545 consistent with dust reprocessing in heavily obscured AGNs.

\begin{figure*}
\centering
\includegraphics[scale=1.1]{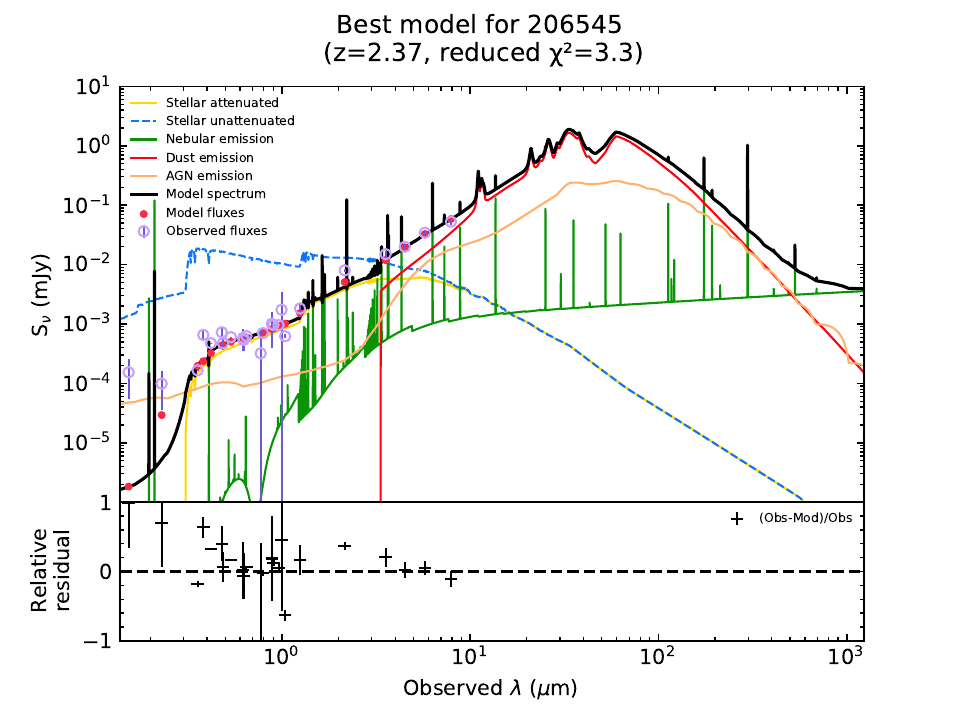}
\caption{Spectral energy distributions of the strongest CT-AGN candidate  objID-206545. The black curves represent the best-fit models obtained from the multiwavelength photometric analysis, while the photometric data points are shown as purple circles.}
\label{SEDs1}
\end{figure*}

\begin{figure*}
\centering
\includegraphics[scale=0.9]{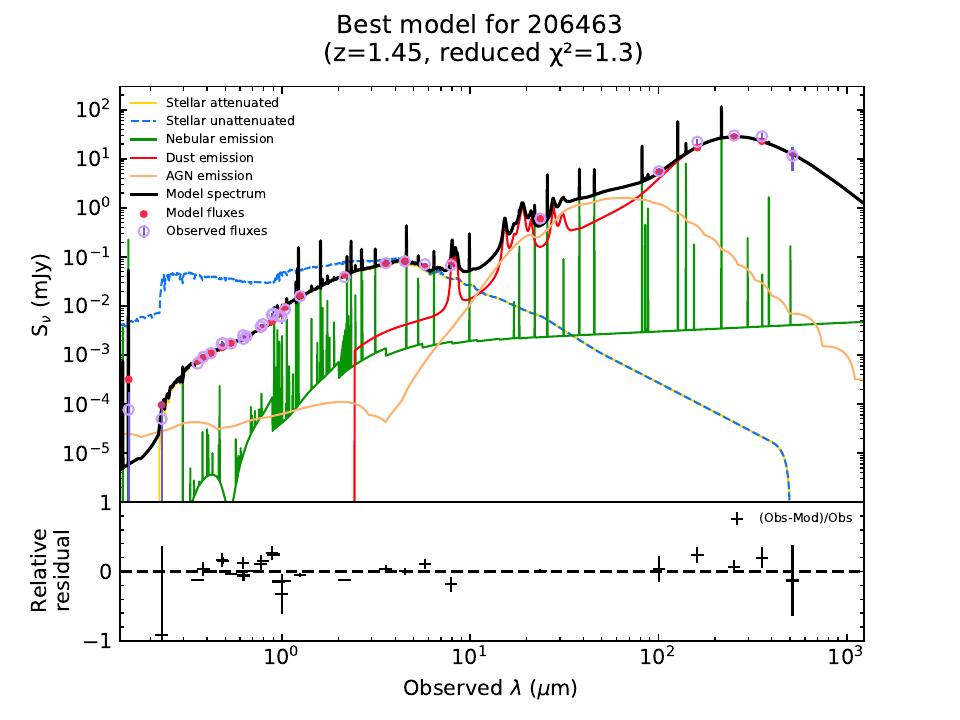}
\includegraphics[scale=0.9]{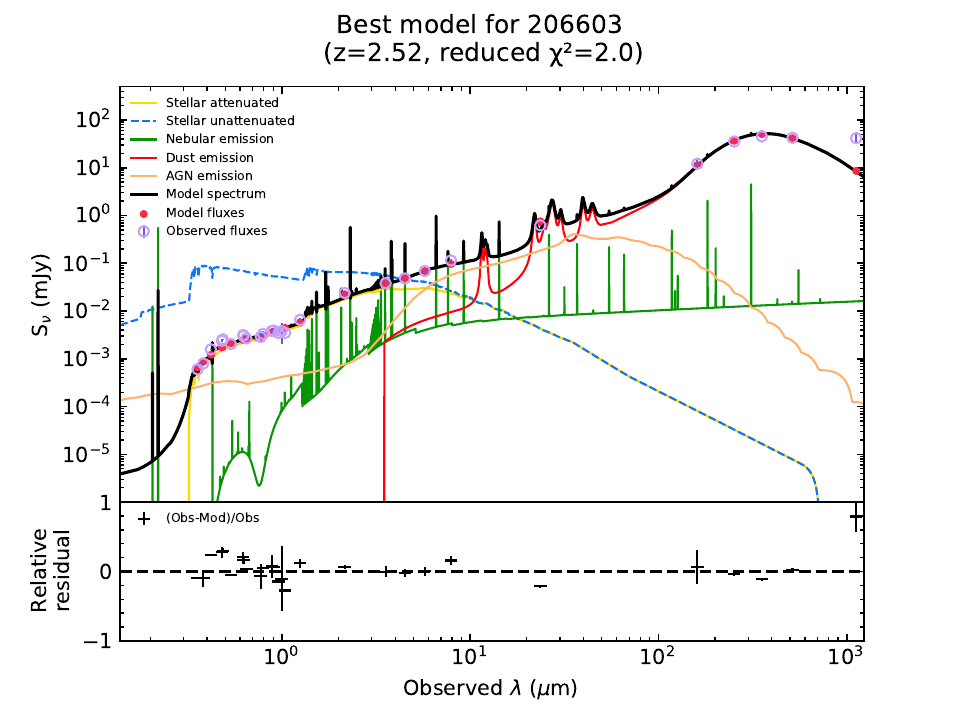}
\caption{Spectral energy distributions of the borderline CT-AGN candidates objID-206463 (upper-panel) and objID-206603 (lower-panel). The black curves represent the best-fit models obtained from the multiwavelength photometric analysis, while the photometric data points are shown as purple circles.}
\label{SEDs2}
\end{figure*}


\bsp	
\label{lastpage}
\end{document}